\documentclass[journal=jctcce,manuscript=article]{achemso}
\usepackage{multirow}
\usepackage{graphicx}
\usepackage{dcolumn}
\usepackage{bm}
\usepackage{enumitem}
\usepackage{hyperref}
\usepackage{bigints}
\usepackage{booktabs}
\usepackage{siunitx}
\usepackage[autostyle]{csquotes}
\usepackage[table,xcdraw]{xcolor}
\definecolor{codegreen}{rgb}{0,0.6,0}
\definecolor{codegray}{rgb}{0.5,0.5,0.5}
\definecolor{codepurple}{rgb}{0.58,0,0.82}
\definecolor{backcolour}{rgb}{0.95,0.95,0.92}
\usepackage[T1]{fontenc}
\usepackage[english]{babel}
\MakeOuterQuote{"}
\usepackage{siunitx}
\usepackage[version=4]{mhchem} 
\usepackage[acronym]{glossaries}

\makeatletter
\newcommand\reaction@[1]{\begin{equation}\ce{#1}\end{equation}}
\newcommand\reaction@nonumber[1]%
{\begin{equation*}\ce{#1}\end{equation*}}
\newcommand\reaction{\@ifstar{\reaction@nonumber}{\reaction@}}
\makeatother

\usepackage{braket}

\newcommand{\mr}{\mathrm}

\newcommand{\ev}[1]{\langle #1 \rangle}

\newacronym{sic}{SIC}{self-interaction correction}
\newacronym{lrdft}{LR-TDDFT}{linear response time-dependent density functional theory}
\newacronym{paw}{PAW}{projector augmented wave}
\newacronym{tdm}{TDM}{transition dipole moment}
\newacronym{dm}{DM}{dipole moment}
\newacronym{1rtdm}{1-RTDM}{one-particle reduced transition density matrix}
\newacronym{1rdm}{1-RDM}{one-particle reduced density matrix}
\newacronym{ao}{AO}{atomic orbital}
\newacronym{mo}{MO}{molecular orbital}
\newacronym{lcao}{LCAO}{linear combination of atomic orbitals}
\newacronym{rsg}{RSG}{real-space grid}
\newacronym{pw}{PW}{plane-wave}
\newacronym{xc}{xc}{exchange-correlation}
\newacronym{ks}{KS}{Kohn--Sham}
\newacronym{doo}{DOO}{direct orbital optimization}
\newacronym{oo}{OO}{orbital-optimized}
\newacronym{mom}{MOM}{maximum overlap method}
\newacronym{si}{SI}{Supporting Information}
\newacronym{cbs}{CBS}{complete basis set}
\newacronym{fc}{FC}{Franck-Condon}
\newacronym{tbe}{TBE}{theoretical best estimate}
\newacronym{homo}{HOMO}{highest occupied molecular orbital}
\newacronym{lumo}{LUMO}{lowest unoccupied molecular orbital}
\newacronym{dft}{DFT}{density functional theory}
\newacronym{aug}{aug}{\mbox{aug-cc-pVDZ+sz}}
\newacronym{daug}{d-aug}{\mbox{d-aug-cc-pVDZ+sz}}
\newacronym{gpaw}{GPAW}{grid-based project augmented wave}

  \newcommand*{\citen}{}
\DeclareRobustCommand*{\citen}[1]{%
  \begingroup
    \romannumeral-`\x 
    \setcitestyle{numbers}%
    \cite{#1}%
  \endgroup
}

\usepackage[authormarkuptext=name,authormarkup=none]{changes}
\definecolor{red}{RGB}{255, 0, 0}
\definecolor{blue}{RGB}{0, 0, 255}
\definecolor{green}{RGB}{0, 192, 0}
\definechangesauthor[name={Authors}, color={red}]{GL}

\title{Excited-state Properties Beyond the Excitation Energy from
Orbital-Optimized Density Functional Calculations I: Dipole Moments of Rydberg States}

\author{Lorenzo Restaino}
\email{lorenzo@hi.is}
\affiliation{Science Institute and Faculty of Physical Sciences, University of Iceland, 107 Reykjavík, Iceland}
\author{Jukka John}
\affiliation{Science Institute and Faculty of Physical Sciences, University of Iceland, 107 Reykjavík, Iceland}
\author{Diego Llorena Prieto}
\affiliation{Science Institute and Faculty of Physical Sciences, University of Iceland, 107 Reykjavík, Iceland}
\author{Yorick Leonard Adrian Schmerwitz}
\affiliation{Max-Planck-Institut f\"ur Kohlenforschung, 45470 M\"ulheim an der Ruhr, Germany}
\author{Elvar Örn Jónsson}
\affiliation{Science Institute and Faculty of Physical Sciences, University of Iceland, 107 Reykjavík, Iceland}
\author{Gianluca Levi}%
\email{gianluca.levi@units.it}
\affiliation{Department of Chemical and Pharmaceutical Sciences, University of Trieste, 34127 Trieste, Italy}
\alsoaffiliation{Science Institute and Faculty of Physical Sciences, University of Iceland, 107 Reykjavík, Iceland}

\begin{document}

\begin{abstract}
Rydberg excited states are challenging to describe due to their highly diffuse character. Orbital-optimized density functional calculations typically provide more accurate values of the excitation energy of Rydberg states than time-dependent density functional theory approaches, but their performance with respect to excited-state properties such as the dipole moment remains largely unexplored as existing benchmarks are limited to the lowest excited state. Here, orbital-optimized density functional calculations with a plane-wave basis set are used to compute the dipole moment of several Rydberg states of a set of small molecules. Plane waves provide a flexible representation of diffuse Rydberg orbitals, whereas commonly used atomic orbitals basis sets are found to yield significant discrepancies in the magnitude and orientation of the dipole moment, even when diffuse functions are included and the excitation energy agrees well with the plane-wave results. The generalized gradient approximation functional PBE used in combination with the plane-wave representation of the orbitals gives good agreement with higher-level coupled-cluster calculations performed with sufficiently diffuse basis sets, when the latter are available. The hybrid functional PBE0 further improves the results, while PBE with globally scaled explicit Perdew-Zunger self-interaction correction generally leads to larger errors and an overestimation of the dipole moment, despite restoring the correct asymptotic $-1/r$ dependence of the effective Kohn--Sham potential.
\end{abstract}

\maketitle

\section{Introduction}
Rydberg states are electronically excited states with highly diffuse electron density. The energy of molecular Rydberg states approximately follows a Rydberg series, converging to the ionization limit. Far from a mere curiosity, these states play a crucial role in many photophysical and photochemical processes~\cite{Mori2012-hb,Sandorfy1999-bv}, including photoionization~\cite{Jochim2017-vn,Merkt1997-na}, spectroscopy~\cite{Softley2004-gl,Kuthirummal2003-pa,Sandorfy1999-bv}, and long-range interactions~\cite{Dunning2024-ck,Deiglmayr2016-ow,Kay2008-gs}. In small molecules such as water, ammonia, and methanol, all low lying excited states have Rydberg character. Because they are highly sensitive to the molecular geometry and the local electrostatic environment, Rydberg states can be valuable probes of the molecular electronic structure~\cite{Kuthirummal2003-pa}. 

The description of Rydberg excited states in electronic structure calculations poses several challenges. Commonly used \gls*{lcao} basis sets struggle to describe the long-range tail of Rydberg states even when diffuse functions are included~\cite{Levi2026-ts,Sigurdarson2023-do,Chrayteh2021-ib,Loos2018-ww,Yang2011-rm,Ciofini2007-hm,Li2006-mq,Muller2001-hq,Kaufmann1989-lb}, and the inclusion of multiple diffuse functions can lead to linear dependence in the basis set. The diffuse character of Rydberg orbitals can introduce other complications in common multi-configurational methods, such as complete active space self-consistent field (CASSCF) and complete active space second-order perturbation theory (CASPT2)~\cite{Janos2024,Kaufold2023-mz,Zobel2017-cq,Veryazov2011-kj,Serrano-Andres1996-rm,Serrano-Andres1996-vk}. An unbalanced description of Rydberg states relative to valence states can result in problems such as state mixing at the CASPT2 level~\cite{Sauri2011-au}. Excluding Rydberg orbitals from the active space can give rise to intruder states, while their inclusion may destabilize the active space and lead to convergence difficulties during the wave function optimization~\cite{Janos2024}. \Gls*{lrdft}~\cite{Casida1995, Runge1984} offers a computationally efficient alternative, but common practical implementations based on the adiabatic approximation often fail for Rydberg excitations. When standard local, semi-local, or global hybrid \gls*{xc} functionals are used, the excitation energy of Rydberg states is typically underestimated~\cite{Seidu2015-fz,Yang2011-rm,Cheng2008,Peach2008,Casida1998}, and the states may spuriously mix with valence or charge-transfer excitations~\cite{Selenius2024-rk, Shu2017}. For molecular Rydberg states, these limitations are associated with deficiencies in the approximate ground-state \gls*{xc} potential, including an erroneous upshift in the molecular region and incorrect long-range asymptotic behavior~\cite{VanMeer2014,Tozer1998}. As a result, the excitation spectrum of the ground-state Kohn--Sham system does not display the correct Rydberg series, a deficiency that carries over to \gls*{lrdft} because these Kohn--Sham excitations constitute its zeroth-order excitations~\cite{Tozer1999-xv}. Nonadiabatic TDDFT methods may improve the description of such excitations, although their development remains at an early stage~\cite{Lacombe2023}.

In light of these challenges, \gls*{oo} density functional calculations~\cite{Restaino2026-yg, Herbert2023, Hait2021}, where excited states are obtained by state-specific variational optimization of the orbitals, provide an attractive alternative. In fully variational \gls*{oo} methods, excited states correspond to stationary points of the energy functional lying higher in energy than the ground state and having nonaufbau orbital occupations. Since such solutions are typically saddle points on the electronic energy surface rather than minima, specialized optimization strategies~\cite{Qin2026, Schmerwitz2026, Bogo2025, Schmerwitz2023, Levi2020-nz, Carter2020, Levi2020farad, Hait2020, Barca2018, Gilbert2008, Cheng2008} are needed to target them without the risk of collapsing to lower-energy solutions, the so-called variational collapse. \Gls*{oo} density functional calculations are computationally more affordable than ab initio multireference approaches while still accounting for orbital relaxation, which has been shown to be important for a correct description of Rydberg excitations~\cite{Levi2026-ts,Vigneshwaran2026,Sigurdarson2023-do,Seidu2015-fz,Yang2011-rm,Cheng2008}. 

\Gls*{oo} density functional methods have been extensively assessed with respect to the vertical excitation energy as well as excited-state geometry for various classes of excitations~\cite{Froitzheim2024, Bogo2024, Selenius2024-rk, Schmerwitz2022, Toffoli2022, Hanson-Heine2013, Maurer2011, Kowalczyk2011}, including Rydberg excited states~\cite{Sigurdarson2023-do,Seidu2015-fz,Yang2011-rm, Cheng2008}. Remarkably, recent studies have shown that even the local density approximation and generalized gradient approximation \gls*{ks}~\cite{Kohn1965, Hohenberg1964} functionals can provide good results for Rydberg states of molecules~\cite{Sigurdarson2023-do,Seidu2015-fz}. In particular, the PBE~\cite{Perdew1997-rt,Perdew1996-il} functional has a relatively low mean absolute error on the excitation energy of $\sim0.2$~eV for Rydberg states of small molecules, despite the fact that the effective potential of local and semilocal functionals lacks the correct $-1/r$ long-range form, where $r$ is the distance from an atom. The inclusion of explicit Perdew--Zunger \gls*{sic}~\cite{Perdew1981-oz}, which restores the correct asymptotic $-1/r$ dependence of the effective potential, was found to improve the excitation energy. More recently, OO studies on Rydberg excitations have shown that the approach can provide potential energy surfaces of Rydberg excited states of small and medium-sized molecules in agreement with higher-level calculations and experiments~\cite{Barreiro-Lage2026, Birgisson2025dmp}.

Assessments of excited state properties beyond the excitation energy and the geometry are scarce. Recently, Paetow and Neugebauer~\cite{Paetow2025-ye} assessed OO methods with respect to the electric dipole moment of valence and charge-transfer excited states of small and medium-sized molecules. They found that OO excited-state dipole moments are generally competitive with \gls*{lrdft}, with the main improvements occurring in cases that are pathological for adiabatic TDDFT, such as doubly excited states\footnote{Although OO methods can target doubly excited states, the single-determinant formulation used here does not explicitly describe their mixing with singly excited configurations. Such mixing can be important near conical intersections and would require a multistate treatment.} and charge-transfer excitations. However, the study was limited to low-lying singlet HOMO-LUMO excitations. Thus, much less is known about the performance of \gls*{oo} density functional calculations for excited-state dipole moments, especially for states above the lowest energy one. This current limitation may be traced back to the fact that standard self-consistent field (SCF) algorithms are not tailored to convergence to saddle points and therefore often struggle to target higher-lying excited states without variational collapse, hampering extensive benchmarks.

Here, \gls*{oo} density functional calculations are assessed with respect to the prediction of dipole moments of Rydberg excited states above the lowest energy excitation using a robust direct orbital optimization approach~\cite{Ivanov2021, Levi2020-nz}. To this end, \gls*{oo} calculations are performed for several singlet and triplet Rydberg excited states with excitation energy up to 10~eV of water (H$_2$O), formaldehyde (CH$_2$O), ammonia (NH$_3$), and methanol (CH$_3$OH). The calculations are carried out using a \gls*{pw} basis set and compared with calculations using commonly employed \gls*{lcao} basis sets with progressively more diffuse functions. Because \glspl*{pw} are delocalized over the simulation cell, they provide a suitable representation for spatially extended electronic states, granted that the cell is large enough to avoid confinement effects. Calculations are also performed with a range of \gls*{xc} functionals, including PBE, the hybrid functional PBE0~\cite{Adamo1999}, and PBE with explicit Perdew-Zunger \gls*{sic}~\cite{Perdew1981-oz} with globally scaled self-interaction correction. The results are benchmarked against higher-level wave function calculations performed with sufficiently diffuse atomic basis sets. 

An accompanying article~\cite{restaino2026arXiv-spectra} assesses the performance of \gls*{oo} density functional calculations for oscillator strengths and optical spectra for a set of molecules with valence and Rydberg states, including those investigated here.

\section{Methodology}
\subsection{Orbital-optimized excited state calculations}
A variety of time-independent, OO density functional approaches for excited states exist, which may be broadly divided into fully variational~\cite{Schmerwitz2026, Schmerwitz2023, Vandaele2022, Hait2021, Levi2020-nz, Kowalczyk2011} and constrained methods~\cite{Lemke2026, Pham2025, Evangelista2013, Gavnholt2008}, the latter employing constraints on the excited-state solutions. In the present work, a fully variational approach is employed, in which the orbitals of a single Slater determinant with nonaufbau occupation are optimized. The resulting states are excited-state solutions of the \gls*{ks} equations and are stationary points of the energy functional. 

To robustly converge these excited-state stationary solutions and avoid variational collapse, we employ a direct optimization approach~\cite{Restaino2026-yg,Schmerwitz2026,Schmerwitz2023,Ivanov2021,Levi2020-nz}.
Here, the orbitals are optimized by directly finding the unitary transformation $\mathbf U$ that makes the energy stationary starting from a set of orthonormal initial orbitals $\boldsymbol{\psi}_0=\{\psi_i^0(\mathbf r)\,|\,1\le i\le N_{\mr{orb}}\}$:
\begin{equation}
    \boldsymbol{\psi}=\boldsymbol{\psi}_0\mathbf U.
\end{equation}
When the unitary matrix is parametrized as an exponential of an anti-Hermitian matrix\cite{Head-Gordon1988}, as commonly done,
\begin{equation}
    \mathbf U = e^{\boldsymbol{\kappa}}, \qquad
    \boldsymbol{\kappa}=-\boldsymbol{\kappa}^{\dagger},
\end{equation}
an excited-state solution is obtained by imposing the stationarity of the energy with respect to the orbital-rotation parameters, together with minimization with respect to the underlying orbital representation~\cite{Ivanov2021},
\begin{equation}
    \underset{\boldsymbol{\psi}}{\mathrm{stat}}\,
    E[\boldsymbol{\psi}]
    =
    \underset{\boldsymbol{\psi}_0}{\mathrm{min}}\,
    \underset{\boldsymbol{\kappa}}{\mathrm{stat}}\,
    E[\boldsymbol{\psi}_0 e^{\boldsymbol{\kappa}}].
\end{equation}
In this work, both \gls*{pw} and \gls*{lcao} basis set representations are employed, using the direct optimization implementation~\cite{Schmerwitz2026, Schmerwitz2023, Ivanov2021, Levi2020-nz} in the \gls*{gpaw} software package~\cite{Mortensen2024-ji, Mortensen2005-nw}. 

All OO calculations performed here are spin-unrestricted. Unrestricted \Gls*{oo} calculations of open-shell singlet excited states provide spin-contaminated solutions consisting of an approximately equal mixture of triplet and pure singlet states. Herein, such solutions are denoted as mixed-spin solutions and indicated by the label M. The energy of open-shell pure singlet states can be obtained through the approximate spin purification formula~\cite{Ziegler1977-xv}
\begin{equation}\label{eq:energy_spin_purification}
    E_{\mathrm{S}} = 2 E_{\mathrm{M}} - E_{\mathrm{T}} \, ,
\end{equation}
where $E_{\mathrm{M}}$ is the energy of the mixed-spin solution, and $E_{\mathrm{T}}$ is the energy of the corresponding triplet state.

\subsection{Self-interaction correction}\label{sec:sic}
In \gls*{ks} density functional theory, the Hartree term is constructed from the total electron density and therefore includes, for each electron, a Coulomb self-interaction. For the exact \gls*{xc} functional, this unphysical contribution is cancelled. However, local and semi-local approximate functionals do not fully provide this cancellation, because the self-interaction is intrinsically nonlocal. 

Here, calculations with \gls*{sic} are performed using the Perdew-Zunger formulation~\cite{Perdew1981-oz}, where the total energy is given by
\begin{equation}\label{eq:sic}
    E_{\mathrm{SIC}}[
    n_1,n_2,\ldots,n_N]
    =
    E_{\mathrm{KS}}[n]
    -
    \alpha \sum_i^N
    \left(
        E_{\mathrm H}[n_i]
        +
        E_{\mathrm{XC}}[n_i, 0]
    \right),
\end{equation}
In eq\ \ref{eq:sic}, $E_{\mathrm{KS}}$ is the Kohn-Sham energy, $E_{\mathrm H}$ is the Hartree energy, $E_{\mathrm{XC}}$ is the \gls*{xc} energy, which is approximated in practice, $n$ is the total density, $n_i$ the density of orbital $i$, and $\alpha$ is a global scaling factor of the self-interaction energy (SIE) term. In SIC, the SIE is subtracted for each occupied orbital and includes the self Hartree and self \gls*{xc} contributions that arise from the orbital density. In this way, the functional becomes explicitly dependent on the individual orbital densities rather than only on the total density. As a result, the SIC energy is no longer invariant under unitary transformations among the occupied orbitals, and the direct optimization needs to include a minimization in the occupied-occupied rotation space~\cite{Ivanov2021, Schmerwitz2024}. When used in excited-state calculations, this yields a fully variational optimization of the \gls*{sic} orbitals for excited states, including both the orbital relaxation associated with the excitation and the localization arising from the orbital-density-dependent correction.

The scaling factor, $\alpha$, in eq\ \ref{eq:sic} accounts for the fact that SIC tends to overcorrect the approximate functionals, and in practice, a scaled down correction, typically by $\alpha=1/2$, is found to provide better results for, e.g., the atomization energy of molecules~\cite{Lehtola2016b}, band gaps of solids~\cite{Gudmundsdottir2015}, and excitation energy of molecular valence excited states~\cite{Ivanov2021}.

\subsection{Electric dipole moment}
For an electronic wave function $\Psi^k$ of a state $k$, the dipole moment is given by the following sum of electronic and nuclear contributions (in atomic units):
\begin{equation}\label{eq:dipole_wavefunction}
  \boldsymbol{\mu}^{k}
  =
  -\left\langle
  \Psi^k
  \left|
  \sum_{p}^{N} \mathbf r_p
  \right|
  \Psi^k
  \right\rangle
  +
  \sum_{a}^{M} \mathcal{Z}_{a}\mathbf R_{a},
\end{equation}
where $\mathbf{r}_p$ is the position of electron $p$, $N$ is the number of electrons, and $\mathcal{Z}_{a}$ and $\mathbf R_{a}$ are the charge and position of nucleus $a$, respectively. For an excited state $k$ obtained in a single-determinant OO calculation, the dipole moment can be evaluated from the state-specific electron density or equivalently from the excited-state orbitals,
\begin{align}\label{eq:dipole_density}
  \boldsymbol{\mu}^{k}
  &=
  -\int \mathbf r\, n^k(\mathbf r)\, d\mathbf r
  +
  \sum_{a}^{M} \mathcal{Z}_{a}\mathbf R_{a}
  \\
  &=
  -\sum_{i}^{N}
  \left\langle
  \psi_i^k
  \left|
  \mathbf r
  \right|
  \psi_i^k
  \right\rangle
  +
  \sum_{a}^{M} \mathcal{Z}_{a}\mathbf R_{a},
  \nonumber
\end{align}
where $\psi_i^k$ are the optimized occupied spin orbitals of state $k$. The OO calculations presented in this work use the \gls*{paw} formalism~\cite{Blochl1994-gk}. The evaluation of the dipole moment using the dipole integrals in the \gls*{paw} formalism is described in the Appendix.

For open-shell singlet excited states represented by a single determinant, the optimized solution is generally spin-mixed. In analogy to the spin purification applied to the energy, eq~\ref{eq:energy_spin_purification}, the dipole moment of  open-shell singlet excited states in unrestricted single-determinant OO calculations is obtained as\cite{Restaino2026-yg,Paetow2025-ye}
\begin{equation}
    \boldsymbol{\mu}_{\mathrm S}
    =
    2\boldsymbol{\mu}_{\mathrm M}
    -
    \boldsymbol{\mu}_{\mathrm T},
    \label{eq:dipole_spin_purification}
\end{equation}
where $\boldsymbol{\mu}_{\mathrm M}$ is the dipole moment of the mixed-spin solution and $\boldsymbol{\mu}_{\mathrm T}$ is that of the corresponding triplet state. This spin-purification formula for the dipole moment was previously used by Paetow and Neugebauer~\cite{Paetow2025-ye} in their 2025 benchmark of OO excited-state dipole moments, which included charge-transfer excited states.

\subsection{Computational Settings}
Ground and \gls*{oo} excited-state calculations are carried out for water, formaldehyde, ammonia and methanol using the \gls*{xc} functionals PBE~\cite{Perdew1997-rt,Perdew1996-il}, PBE0~\cite{Perdew1996-pr}, and PBE with explicit Perdew-Zunger \gls*{sic}~\cite{Perdew1981-oz}, as implemented in the GPAW~\cite{Mortensen2024-ji,Mortensen2005-nw} program v.~25.7.1b1. All calculations make use of the PAW formalism~\cite{Blochl1994-gk} and the frozen-core approximation for the orbitals. Previous studies by Loos et al.~\cite{Loos2018-ww} indicate that the impact of the frozen-core approximation on the excitation energy of small molecules, including those investigated here, is on the order of 0.01~eV for post-Hartree-Fock methods. The \gls*{lcao} calculations use aug-cc-pVDZ+sz (hereafter referred to as aug) and d-aug-cc-pVDZ+sz (hereafter referred to as d-aug) basis sets, where the valence electrons are expanded in atomic basis sets composed of primitive Gaussian-type functions from the aug-cc-pVDZ~\cite{Pritchard2019-vq,Kendall1992-wv,Dunning1989-vi} and d-aug-cc-pVDZ~\cite{Pritchard2019-vq,Woon1994-fz,Kendall1992-wv,Dunning1989-vi} sets, each augmented by a single set of numerical atomic orbitals denoted as “sz”. The contracted functions of type "s" are removed to avoid redundancy with partial waves centered on the atoms in the augmentation regions in the \gls*{paw} formalism. The \gls*{pw} calculations use an energy cutoff of 1200~eV. SIC calculations are performed using the Perdew–Zunger formulation described in Section~\ref{sec:sic} in combination with complex orbitals, with both a global scaling factor $\alpha=1$, referred to as PBE-SIC, and $\alpha=1/2$, referred to as PBE-SIC/2. All calculations are performed in the gas phase and in $C_1$ symmetry within the spin-unrestricted formalism. The calculations are performed in a cubic simulation box with at least 10.5~{\AA} of vacuum around the atoms, using a uniform real-space grid spacing of 0.16~{\AA} to represent the valence electrons. Numerical tests were performed for the most diffuse excited states considered in this work, specifically the S$_3$ and S$_4$ of methanol. The tests (see Table~S1 of the \gls*{si}) show that increasing the simulation box to include 15~{\AA} of vacuum around the atoms changes the excitation energy and the components of the dipole moment by less than $0.5\cdot 10^{-2}$~eV and 0.07~D, respectively.

The OO excited-state calculations are carried out using a direct optimization method employing a limited-memory symmetric rank 1 (L-SR1) algorithm~\cite{Ivanov2021, Levi2020-nz}. For the PW calculations, the nested optimization scheme presented by Ivanov \textit{et al.}~\cite{Ivanov2021} is used. There, an outer loop minimizes the energy in the tangent space of the orbitals, and an inner loop locates the excited-state saddle point using the exponential transformation with L-SR1. For water, the summed elapsed optimization time of the PW/PBE calculations is approximately 1.8 times that of the d-aug/PBE calculations over excited state calculations performed using 28 cores (Table~S3 of the \gls*{si}).

The initial orbitals for the excited-state calculations are taken as the ground-state optimized orbitals with nonaufbau occupation reflecting an excitation of one electron within one spin channel for the mixed-spin solutions or across spin channels for the triplet solutions. The identity of the occupied and virtual orbitals was monitored throughout the SCF procedure by inspection of their orbital characters using the maximum overlap method~\cite{Gilbert2008-ye}. The orbitals involved in the excitations obtained in PBE calculations are shown in Figures~S1, S3, S5, and S7 of the \gls*{si}. All singlet excited states are open-shell, and their energy and dipole moment are obtained using the spin purification formulas, eqs\ \eqref{eq:energy_spin_purification} and \eqref{eq:dipole_spin_purification}.

In some cases, when the target excitation involves a pair of degenerate orbitals, complex-valued orbitals need to be used to preserve the spatial symmetry of the electron density, otherwise occupation of only one component of a pair of real degenerate orbitals can artificially lower the symmetry~\cite{Barreiro-Lage2026, Ivanov2021, Sigurdarson2023-do}. For example, ammonia belongs to the $C_{3v}$ point group and has degenerate orbitals belonging to the $E$ irreducible representation, such as the nitrogen 3p lone-pair orbitals (unoccupied in the ground state). In this work, to describe a single-electron excitation into the degenerate pair while preserving the rotational symmetry of the density, complex $\mathrm{p}_{\pm}$ orbitals are used corresponding to the following linear combination of real $\mathrm{p}_x$ and $\mathrm{p}_y$ orbitals
\begin{equation}
    3\mathrm{p}_{\pm} = 3\mathrm{p}_x \pm i\,3\mathrm{p}_y .
\end{equation}

For methanol and water, additional calculations are performed using equation-of-motion coupled-cluster singles and doubles (EOM-CCSD)~\cite{Krylov2008-nk,Kowalski2000-gw,Nooijen1995-zu,Stanton1993-ze,Geertsen1989-ib,Rowe1968-hu} with a d-aug-cc-pVTZ basis set, as implemented in ORCA v.~6.1.0~\cite{Neese2025-cc,Neese2012-bk}. For these calculations, the EOM-CCSD excited-state energy is expanded in a Taylor series with respect to an external electric field of 0.00005 atomic units, and the corresponding excited-state dipole moment is determined using a finite-difference approach:
\begin{equation}
\mu_i^k = -\dfrac{E^k(+F_i)-E^k(-F_i)}{2F_i}\, ,
\end{equation}
where the subscript $i$ indicates the components of the dipole moment and the electric field. The excited-state dipole moments of water obtained with this procedure reproduce those calculated at the same level of theory from the relaxed CCSD density in ref.~\citen{Chrayteh2021-ib}. As an additional test of the finite-difference procedure, the Hartree--Fock ground-state dipole moment obtained from finite differences of the Hartree--Fock energy differs by only 0.002\% from that evaluated directly from the Hartree--Fock electron density.

\begin{figure}[hbt]
    \centering
    \includegraphics[width=0.35\linewidth]{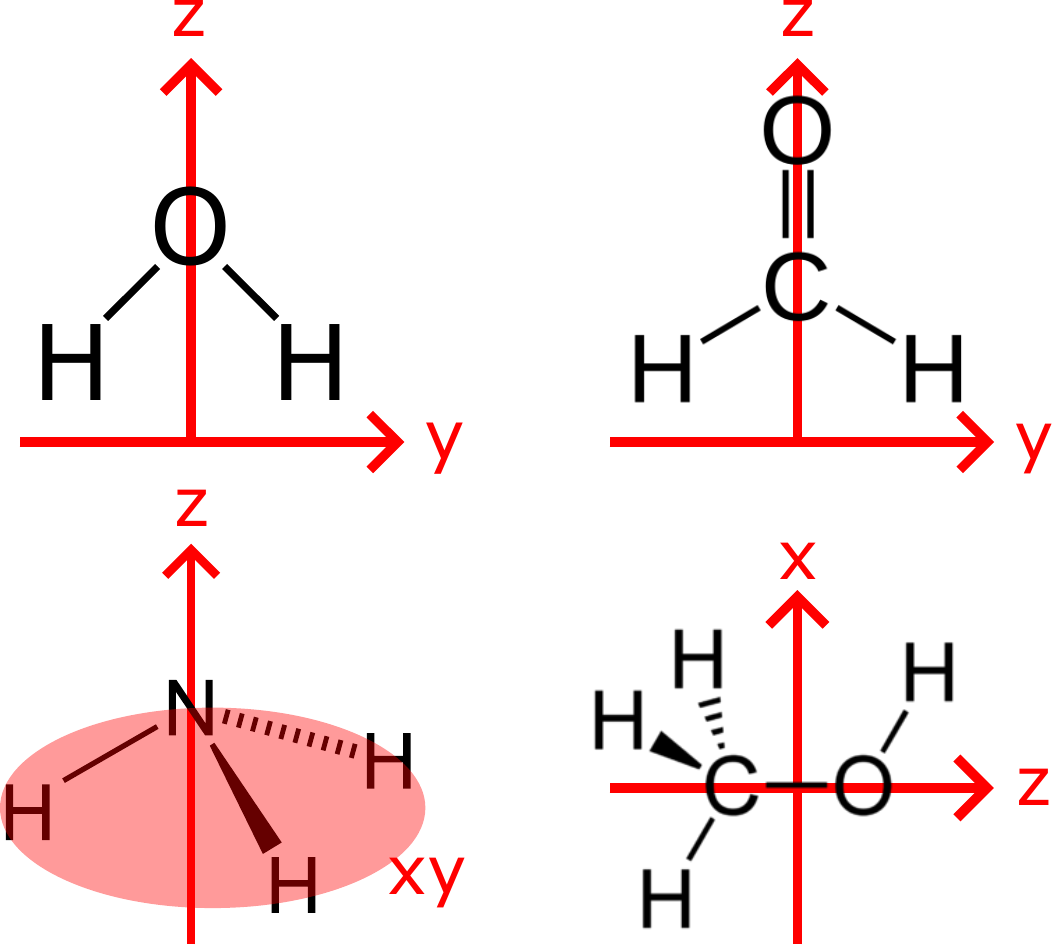}
    \caption{Cartesian coordinate frames used for the molecules investigated in this manuscript.}
    \label{fig:cartesian_frame}
\end{figure}

The molecular geometries are ground-state geometries optimized using high-level coupled-cluster methods as reported in the literature. The geometries of water, formaldehyde, and ammonia are taken from ref.~\citen{Loos2018-ww}, where they were computed at the CC3/aug-cc-pVTZ level of theory. The geometry of methanol is taken from ref.~\citen{Lange2020-sy}, where it was calculated at the CCSD(T)/aug-cc-pVQZ level of theory. The molecular Cartesian coordinate frame adopted for each system is shown in Figure~\ref{fig:cartesian_frame}. All vector quantities are reported with respect to these coordinate frames unless otherwise stated.

\section{Results}
\subsection{Basis set effect}
\begin{figure}[hbt]
    \centering
    \includegraphics[width=\linewidth]{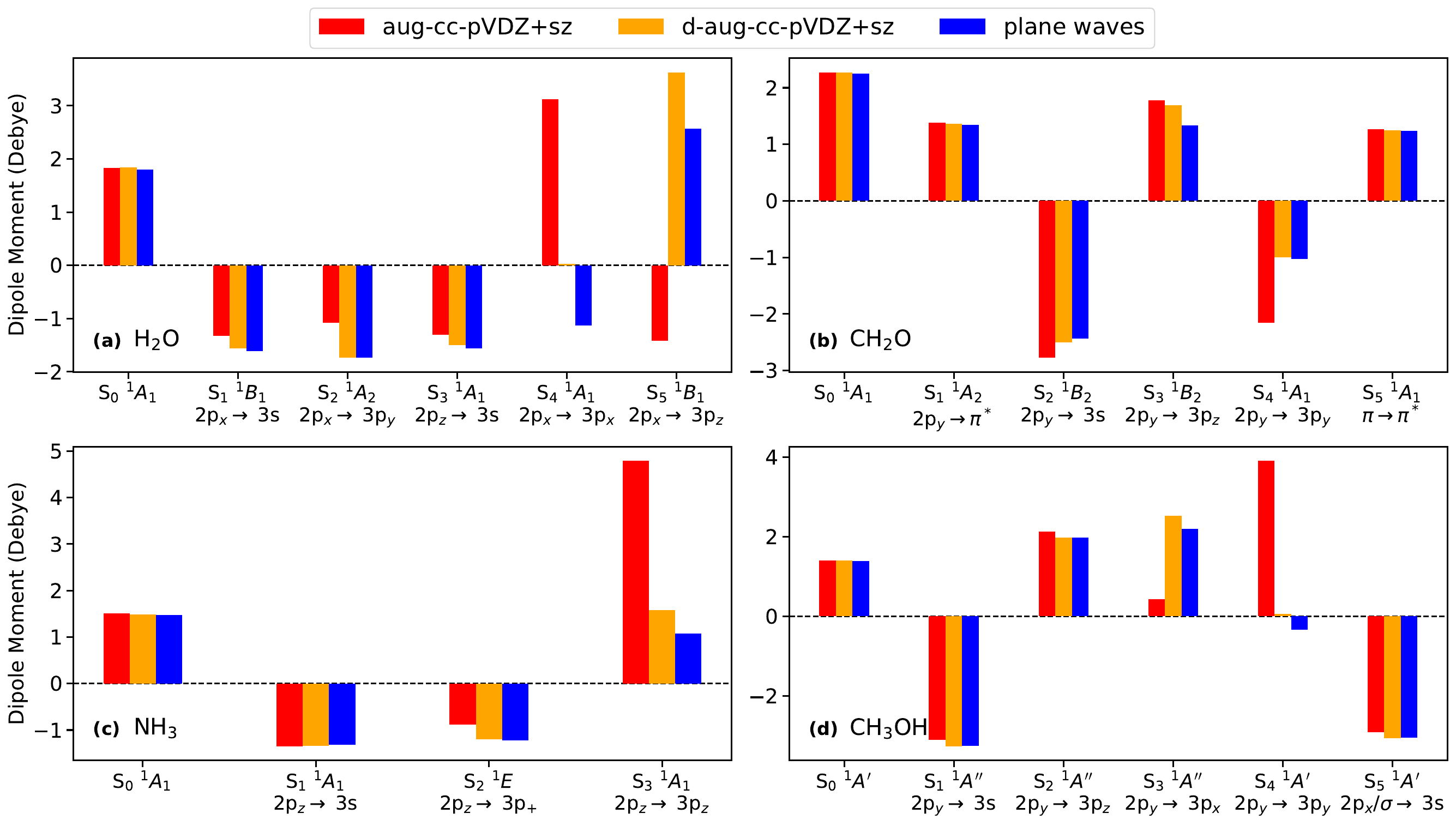}
    \caption{Effect of the basis set on the dipole moments of ground and singlet excited states of water (a), formaldehyde (b), ammonia (c), and methanol (d) obtained with orbital-optimized calculations with the PBE functional. For water, formaldehyde, and ammonia, the shown value is the single nonzero component of the dipole moment vector along the molecular axis of rotation ($\mu _z$). For methanol (d), which has two nonzero components, the reported value is along the $x$ direction of the molecular coordinate frame ($\mu _x$, see Figure~\ref{fig:cartesian_frame}). The excited-state dipole moments are spin-purified according to eq\ \ref{eq:dipole_spin_purification}. The aug-cc-pVDZ+sz basis set gives significant deviations from the plane waves for most cases because it overconfines the density (see Figure\ \ref{fig:3px_water_1D_cut}). For d-aug-cc-pVDZ+sz, which includes extra diffuse functions, deviations remain for the most diffuse states.}
    \label{fig:DM_basis_effect}
\end{figure}

Figure~\ref{fig:DM_basis_effect} compares the excited-state dipole moments of water, formaldehyde, ammonia, and methanol obtained in OO calculations with the PBE functional and different basis sets. The corresponding dipole moment values are reported in Table~\ref{tab:basis_effect_table} together with the values of vertical excitation energy and variance of the electronic position operator,
\begin{equation}
    \sigma(\mathbf r) = \ev{{\mathbf r}^2} - \ev{\mathbf r}^2 \, ,
\end{equation}
which measures the overall spatial extent of the variationally optimized
excited-state density.
Valence excitations redistribute electron density within the
molecular valence region, whereas Rydberg excitations populate spatially
diffuse orbital extending beyond the molecular region. We assign the character of the excited state by comparing $\sigma(\mathbf r)$ with the ground-state value.
Valence excitations have a variance
similar to that of the ground state, whereas Rydberg excitations have
a significantly larger variance. The assignments reported in
Table~\ref{tab:basis_effect_table} combine this measure of spatial extent
with the character of the orbitals involved.

The variation of the excitation energy and variance with respect to the
basis set is also visualized in Figures~S2, S4, S6, and S8 of the \gls{si}.

\begin{table}[htb!]
\centering
\caption{%
Spin-purified vertical excitation energy, $\Delta E$ (eV), variance, $\sigma$ (bohr$^2$), and dipole moment, $\mu$ (Debye), of the singlet excited states of water, formaldehyde, ammonia, and methanol computed with orbital-optimized calculations with the PBE functional and several basis sets. The excited states are identified as Rydberg (R) or valence (V) based on the orbital character and the value of the variance compared to the variance of the ground state. Dipole moments from high-level coupled-cluster calculations using atomic orbitals basis sets including diffuse functions are also shown. For water, formaldehyde, and ammonia, $\mu$ is the single nonzero component of the dipole moment vector along the molecular axis of rotation. For methanol, both nonzero components $x$ and $z$ are reported (see Figure \ref{fig:cartesian_frame}). Available coupled-cluster reference dipole moments are reported in the last column.
}
\label{tab:basis_effect_table}
\scriptsize
\begin{tabular}{
l
S[table-format=2.2]
S[table-format=2.1]
S[table-format=-1.2]
S[table-format=2.2]
S[table-format=2.1]
S[table-format=-1.2]
S[table-format=2.2]
S[table-format=2.1]
S[table-format=-1.2]
c
}
\hline
 & \multicolumn{3}{c}{{aug-cc-pVDZ+sz}} 
 & \multicolumn{3}{c}{{d-aug-cc-pVDZ+sz}} 
 & \multicolumn{3}{c}{{plane waves}}
 & \multicolumn{1}{c}{{Ref.}} \\
\cmidrule(lr){2-4}
\cmidrule(lr){5-7}
\cmidrule(lr){8-10}
\cmidrule(lr){11-11}
State 
& {$\Delta E$} & {$\sigma$} & {$\mu$} 
& {$\Delta E$} & {$\sigma$} & {$\mu$}
& {$\Delta E$} & {$\sigma$} & {$\mu$}
& {$\mu$} \\
\hline

\multicolumn{11}{l}{{H$_2$O}} \\
S$_0$ $A_1$ &   & 13.3 &  1.83 &  & 13.3 &  1.84 &  & 13.3 &  1.80 & 1.87$^\P$\\
S$_1$ $B_1$ (R) & 7.47  & 26.7 & -1.33 & 7.47 & 26.7 & -1.56 & 7.44 & 26.6 & -1.61 & -1.77$^\P$ \\
S$_2$ $A_2$ (R) & 8.95  & 33.5 & -1.07 & 8.94 & 37.3 & -1.74 & 8.90 & 37.0 & -1.73 & -1.62$^\P$ \\
S$_3$ $A_1$ (R) & 9.79  & 28.2 & -1.30 & 9.80 & 28.4 & -1.50 & 9.73 & 28.4 & -1.56 & -1.16$^\P$ \\
S$_4$ $A_1$ (R) & 11.24 & 30.6 &  3.12 & 9.90 & 56.1 &  0.03 & 9.84 & 57.9 & -1.13 & 0.21$^\P$ \\
S$_5$ $B_1$ (R) & 11.55 & 37.3 & -1.41 & 9.90 & 55.0 &  3.63 & 9.83 & 57.8 &  2.56 & 4.89$^\P$ \\[0.6em]

\multicolumn{11}{l}{{CH$_2$O}} \\
S$_0$ $A_1$ &  & 23.4 &  2.27 &  & 23.5 &  2.27 &  & 23.4 &  2.25 & 2.41 $\pm$ 0.01$^\dagger$\\
S$_1$ $A_2$ (V) & 3.58 & 24.2 &  1.38 & 3.58 & 24.2 &  1.36 & 3.54 & 24.1 &  1.35 & 1.36 $\pm$ 0.01$^\dagger$ \\
S$_2$ $B_2$ (R) & 6.91 & 50.5 & -2.77 & 6.90 & 52.6 & -2.51 & 6.88 & 52.5 & -2.44 & -2.15 $\pm$ 0.03$^\dagger$ \\
S$_3$ $B_2$ (R) & 7.75 & 50.9 &  1.78 & 7.62 & 62.6 &  1.69 & 7.59 & 62.4 &  1.34 & 0.21 $\pm$ 0.20$^\dagger$ \\
S$_4$ $A_1$ (R) & 7.72 & 61.7 & -2.16 & 7.68 & 68.3 & -1.00 & 7.68 & 67.8 & -1.03 & -0.69 $\pm$ 0.43$^\dagger$ \\
S$_5$ $A_1$ (V) & 8.57 & 28.0 &  1.27 & 8.57 & 28.1 &  1.25 & 8.56 & 27.9 &  1.24 & 2.46 $\pm$ 1.36$^\dagger$ \\[0.6em]

\multicolumn{11}{l}{{NH$_3$}} \\
S$_0$ $A_1$ &  & 16.9 &  1.51 &  & 16.9 &  1.48 &  & 16.9 &  1.48 & 1.52$^\ddagger$\\
S$_1$ $A_1$ (R) & 6.45 & 36.2 & -1.35 & 6.45 & 37.2 & -1.34 & 6.42 & 37.2 & -1.32 & -1.02$^\ddagger$ \\
S$_2$ $E$ (R) & 7.82 & 51.2 & -0.88 & 7.80 & 55.0 & -1.20 & 7.83 & 54.7 & -1.22 & -0.92$^\ddagger$ \\
S$_3$ $A_1$ (R) & 9.20 & 40.0 &  4.79 & 8.31 & 70.2 &  1.58 & 8.26 & 71.6 &  1.07 & 1.13$^\ddagger$ \\[0.6em]

\multicolumn{11}{l}{{CH$_3$OH}} \\
S$_0$ $A^{\prime}$       &      & 27.6 &
\multicolumn{1}{c}{\begin{tabular}{@{}S[table-format=-1.2]@{\;}S[table-format=-1.2]@{}} 1.40 & -0.92 \end{tabular}} &
     & 27.6 &
\multicolumn{1}{c}{\begin{tabular}{@{}S[table-format=-1.2]@{\;}S[table-format=-1.2]@{}} 1.40 & -0.90 \end{tabular}} &
     & 27.6 &
\multicolumn{1}{c}{\begin{tabular}{@{}S[table-format=-1.2]@{\;}S[table-format=-1.2]@{}} 1.39 & -0.89 \end{tabular}} & 
\multicolumn{1}{c}{\begin{tabular}{@{}S[table-format=-1.2]@{\;}S[table-format=-1.2]@{}} 1.42 & -1.03 \end{tabular}\makebox[0pt][l]{$^\P$}}
\\

S$_1$ $A^{\prime\prime}$ (R) & 6.36 & 44.7 &
\multicolumn{1}{c}{\begin{tabular}{@{}S[table-format=-1.2]@{\;}S[table-format=-1.2]@{}} -3.10 & -3.61 \end{tabular}} &
6.35 & 45.7 &
\multicolumn{1}{c}{\begin{tabular}{@{}S[table-format=-1.2]@{\;}S[table-format=-1.2]@{}} -3.26 & -3.56 \end{tabular}} &
6.34 & 45.5 &
\multicolumn{1}{c}{\begin{tabular}{@{}S[table-format=-1.2]@{\;}S[table-format=-1.2]@{}} -3.25 & -3.53 \end{tabular}} &
\multicolumn{1}{c}{\begin{tabular}{@{}S[table-format=-1.2]@{\;}S[table-format=-1.2]@{}} -3.72 & -3.72 \end{tabular}\makebox[0pt][l]{$^\P$}} \\

S$_2$ $A^{\prime\prime}$ (R) & 7.40 & 56.1 &
\multicolumn{1}{c}{\begin{tabular}{@{}S[table-format=-1.2]@{\;}S[table-format=-1.2]@{}} 2.13 & 7.06 \end{tabular}} &
7.38 & 59.7 &
\multicolumn{1}{c}{\begin{tabular}{@{}S[table-format=-1.2]@{\;}S[table-format=-1.2]@{}} 1.97 & 7.33 \end{tabular}} &
7.38 & 59.8 &
\multicolumn{1}{c}{\begin{tabular}{@{}S[table-format=-1.2]@{\;}S[table-format=-1.2]@{}} 1.98 & 7.24 \end{tabular}} &
\multicolumn{1}{c}{\begin{tabular}{@{}S[table-format=-1.2]@{\;}S[table-format=-1.2]@{}} 2.49 & 7.56 \end{tabular}\makebox[0pt][l]{$^\P$}} \\

S$_3$ $A^{\prime\prime}$ (R) & 7.90 & 67.7 &
\multicolumn{1}{c}{\begin{tabular}{@{}S[table-format=-1.2]@{\;}S[table-format=-1.2]@{}} 0.43 & 2.58 \end{tabular}} &
7.74 & 77.2 &
\multicolumn{1}{c}{\begin{tabular}{@{}S[table-format=-1.2]@{\;}S[table-format=-1.2]@{}} 2.53 & -0.04 \end{tabular}} &
7.80 & 76.7 &
\multicolumn{1}{c}{\begin{tabular}{@{}S[table-format=-1.2]@{\;}S[table-format=-1.2]@{}} 2.20 & -0.20 \end{tabular}} &
\multicolumn{1}{c}{\begin{tabular}{@{}S[table-format=-1.2]@{\;}S[table-format=-1.2]@{}} 2.38 & -0.98 \end{tabular}\makebox[0pt][l]{$^\P$}} \\

S$_4$ $A^{\prime}$ (R) & 8.06 & 70.1 &
\multicolumn{1}{c}{\begin{tabular}{@{}S[table-format=-1.2]@{\;}S[table-format=-1.2]@{}} 3.90 & 3.67 \end{tabular}} &
7.90 & 83.5 &
\multicolumn{1}{c}{\begin{tabular}{@{}S[table-format=-1.2]@{\;}S[table-format=-1.2]@{}} 0.06 & 0.50 \end{tabular}} &
7.89 & 83.0 &
\multicolumn{1}{c}{\begin{tabular}{@{}S[table-format=-1.2]@{\;}S[table-format=-1.2]@{}} -0.34 & 0.22 \end{tabular}} &
\multicolumn{1}{c}{\begin{tabular}{@{}S[table-format=-1.2]@{\;}S[table-format=-1.2]@{}} -1.22 & -1.18 \end{tabular}\makebox[0pt][l]{$^\P$}} \\

S$_5$ $A^{\prime}$ (R) & 7.94 & 46.5 &
\multicolumn{1}{c}{\begin{tabular}{@{}S[table-format=-1.2]@{\;}S[table-format=-1.2]@{}} -2.91 & -3.30 \end{tabular}} &
7.93 & 47.4 &
\multicolumn{1}{c}{\begin{tabular}{@{}S[table-format=-1.2]@{\;}S[table-format=-1.2]@{}} -3.06 & -3.24 \end{tabular}} &
7.89 & 47.3 &
\multicolumn{1}{c}{\begin{tabular}{@{}S[table-format=-1.2]@{\;}S[table-format=-1.2]@{}} -3.05 & -3.24 \end{tabular}} &
\multicolumn{1}{c}{\begin{tabular}{@{}S[table-format=-1.2]@{\;}S[table-format=-1.2]@{}} -2.90 & -2.45 \end{tabular}\makebox[0pt][l]{$^\P$}} \\

\hline
\multicolumn{11}{l}{\footnotesize $^\P$CCSD/d-aug-cc-pVTZ values from the present work; $^\ddagger$CCSDT/\mbox{d-aug-cc-pVQZ} values from ref.~\citen{perscorr_jacquemin}.} \\
\multicolumn{11}{l}{\footnotesize $^\dagger$Theoretical best estimate/complete basis set values from ref.~\citen{Chrayteh2021-ib}. } \\
\end{tabular}
\end{table}

\begin{figure}
    \centering
    \includegraphics[width=0.5\linewidth]{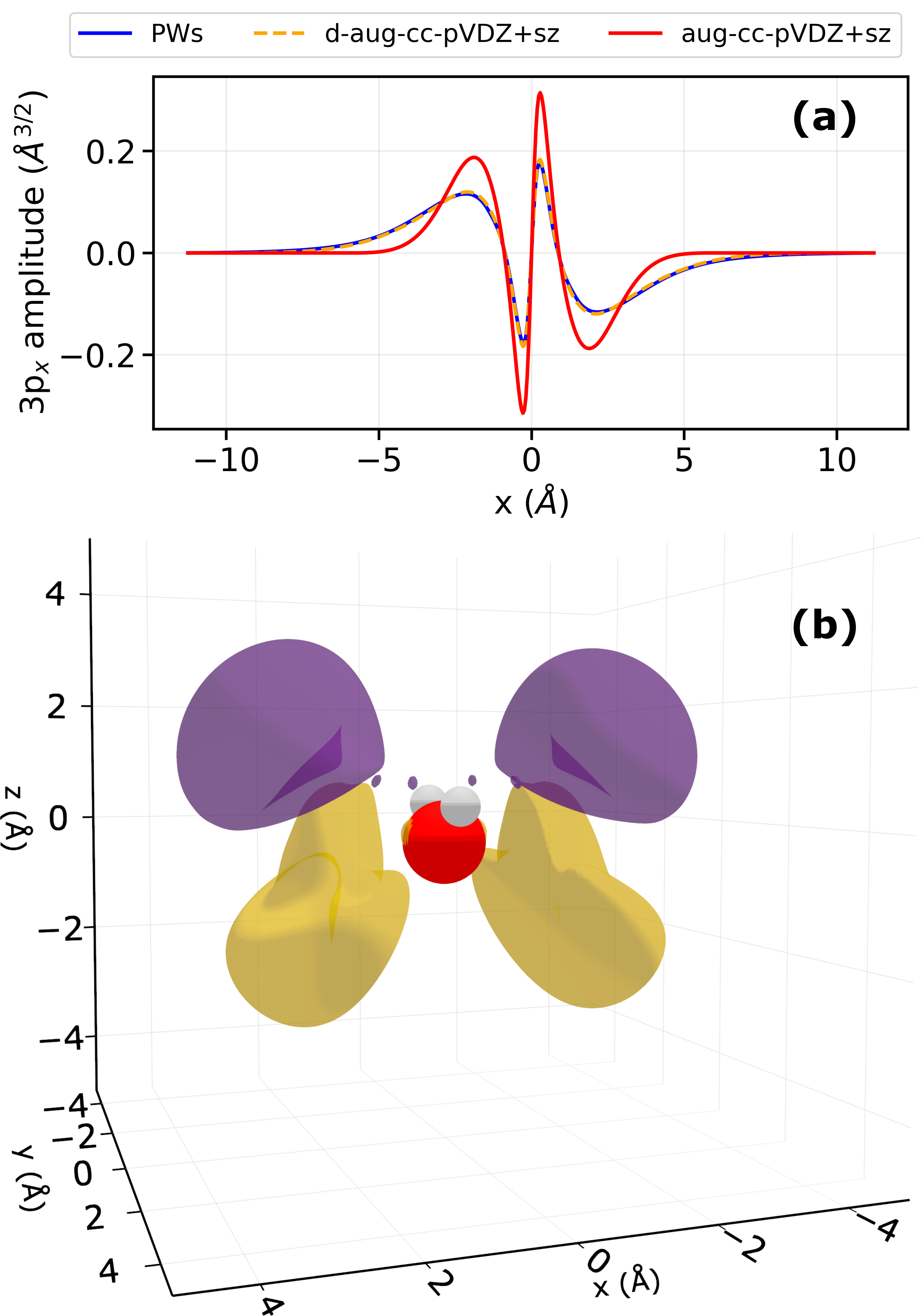}
    \caption{Comparison of the optimized 3p$_x$ orbital obtained in a mixed-spin calculation of the S$_4$ state of water obtained with different atomic orbitals basis sets and with plane waves (PWs). (a) 3p$_x$ amplitude along an axis perpendicular to the molecular plane and containing the oxygen atom calculated with aug-cc-pVDZ+sz (red), d-aug-cc-pVDZ+sz (orange), and PWs (blue). (b) Three-dimensional plot of the difference density, $\Delta n_{\mr{orb}} = n_{\mathrm{d\text{-}aug}} - n_{\mathrm{PW}}$, between the 3p$_x$ orbitals obtained with d-aug-cc-pVDZ+sz and PWs. Positive values of $\Delta n_{\mr{orb}}$ are shown in yellow, whereas negative values are shown in violet. The isosurfaces are plotted at $\pm 0.5 \times 10^{-3}$~{\AA}$^{-3}$. The d-aug basis set shifts density from the hydrogen atoms to the oxygen side, which makes the dipole moment slightly positive compared to PWs, where it is negative.}
    \label{fig:3px_water_1D_cut}
\end{figure}

For water, as also previously observed~\cite{Sigurdarson2023-do}, the excitation energy of S$_1$ (\mbox{2p$_x \rightarrow$ 3s}), S$_2$ (\mbox{2p$_x \rightarrow$ 3p$_y$}), and S$_3$ (\mbox{2p$_z \rightarrow$ 3s}) is only weakly affected by the basis set. This is consistent with the variance changing little with the basis set. In contrast, the excitation energy of S$_4$ (\mbox{2p$_x \rightarrow$ 3p$_x$}) and S$_5$ (\mbox{2p$_x \rightarrow$ 3p$_z$}), which are more diffuse based on the PW values of $\sigma(\bm r)$, shows a much stronger basis set dependence. It is overestimated by more than 1~eV with aug compared to \glspl*{pw}, and slightly overestimated (< 0.1~eV) with d-aug, which includes an additional set of diffuse functions. For S$_4$, $\sigma(\bm r)$ increases from 30.6 to 56.1 and 57.9~bohr$^2$ when going from aug to d-aug and then to \glspl*{pw}, while for S$_5$, $\sigma(\bm r)$ increases from 37.3 to 55.0 and 57.8~bohr$^2$. Therefore, the single-augmented basis set underestimates the spatial extent of these two Rydberg states, overly confining the electron density. This is analyzed in Figure~\ref{fig:3px_water_1D_cut}(a), which shows the 3p$_x$ orbital along an axis perpendicular to the molecular plane for all basis sets. The excited-state dipole moments of water show a much larger basis set dependence compared to the excitation energy. For S$_1$, S$_2$, and S$_3$, aug significantly underestimates the magnitude compared to \glspl*{pw}, despite the excitation energy being affected only marginally. For the more diffuse S$_4$ and S$_5$ states,  aug not only overestimates the magnitude of the dipole moment by more than 1~D compared to PWs, but also predicts an incorrect orientation. The d-aug basis set provides an improvement, but the dipole moment remains underestimated and overestimated compared to PWs for S$_4$ and S$_5$, respectively.

To assess whether the change in dipole orientation observed with OO/PBE is also found at the EOM-CCSD level, we additionally performed EOM-CCSD calculations with aug-cc-pVTZ, complementing the d-aug-cc-pVTZ results reported in Table~\ref{tab:basis_effect_table}. For S$_4$, EOM-CCSD shows the same trend as OO/PBE, with the dipole decreasing strongly from 3.52~D with aug-cc-pVTZ to 0.21~D with d-aug-cc-pVTZ. For S$_5$, however, the change of sign observed with OO/PBE is not reproduced, although the EOM-CCSD dipole still changes considerably, from 5.91 to 4.89~D.

At the OO/PBE level, the variances of S$_4$ and S$_5$ obtained with d-aug are nevertheless very close to the corresponding PW values. This suggests that the deviations in the dipole moment cannot be explained by the fact that d-aug is not diffuse enough. Figure~\ref{fig:3px_water_1D_cut}(b) shows the difference between the density of the 3p$_x$ orbital obtained in d-aug and \gls*{pw} calculations of the S$_4$ state. For this state, the \gls*{pw} dipole is negative, meaning that it points from the oxygen to the hydrogen atoms. Figure~\ref{fig:3px_water_1D_cut}(b) shows that d-aug shifts density from the hydrogen side toward the oxygen side compared to \glspl*{pw}, which makes the dipole moment slightly positive. Note that the differences between the \gls*{daug} and plane-wave representations cannot be attributed to the ``sz'' set of numerical atomic orbitals. These orbitals are confined to the vicinity of the nuclei and therefore cannot account for differences arising in regions of space far from the nuclei, such as those visible in Figure~\ref{fig:3px_water_1D_cut}(b).
The remaining differences in the dipole moments do not primarily reflect a failure to capture the overall spatial extent of the Rydberg state, but rather a limited flexibility of the atom-centered representation in describing the anisotropic density redistribution of the excitation.

For formaldehyde, the excitation energy changes only weakly with the basis set. This is expected for the valence states S$_1$ (\mbox{2p$_y \rightarrow \pi^*$}) and S$_5$ (\mbox{$\pi \rightarrow \pi^*$}), but it also holds for the Rydberg states S$_2$ (\mbox{2p$_y \rightarrow$ 3s}), S$_3$ (\mbox{2p$_y \rightarrow$ 3p$_z$}), and S$_4$ (\mbox{2p$_y \rightarrow$ 3p$_y$}). The largest deviations in the variance compared to PWs are observed for aug calculations of the S$_3$ and S$_4$ states. Aug gives a $\sigma(\bm r)$ of 50.9 and 62.4~bohr$^2$ for S$_3$ and S$_4$, respectively, while PWs give values of 62.4 and 67.8~bohr$^2$. For all states, the d-aug and \gls*{pw} variances are nearly identical. The dipole moments of formaldehyde also show smaller basis set effects compared to the other molecules, as shown in Figure~\ref{fig:DM_basis_effect}(b). The basis set dependence is largest for the Rydberg states, but remains moderate. In all cases, the LCAO basis sets provide the same dipole orientation as PWs. The largest deviation in the dipole moment magnitude compared to PWs ($\sim$1~D) is observed for S$_4$ calculated with aug, which is consistent with a relatively large deviation observed in the variance. The d-aug and \gls*{pw} dipole moments are close for all states. This indicates that the double-augmented atom-centered basis is sufficient to reproduce both the spatial extent and anisotropic density redistribution of the excited states for formaldehyde.

Ammonia shows a behavior similar to water. The variance and excitation energy of S$_1$ (\mbox{2p$_z \rightarrow$ 3s}) and S$_2$ (\mbox{2p$_z \rightarrow$ 3p$_+$}) are weakly affected by the basis set. In contrast, for S$_3$ (\mbox{2p$_z \rightarrow$ 3p$_z$}), which has a much larger variance and hence is more diffuse, the excitation energy computed with aug differs from the d-aug and PW values by more than 1~eV. For this state, aug significantly underestimates the variance, giving a value of 40.0~bohr$^2$, compared to d-aug and \glspl*{pw}, which instead give similar values of 70.2 and 71.6~bohr$^2$. The single-augmented atomic orbitals basis therefore underestimates the spatial extent of the most diffuse ammonia Rydberg state. The results further suggest that recovering the overall spatial extent, indicated by the close agreement in $\sigma(\bm r)$, is not sufficient to ensure convergence of the dipole moment. For S$_2$, aug gives a variance only 3~bohr$^2$ smaller than the PWs value, but the dipole moment is underestimated by about 30\%. For S$_3$, the magnitude of the dipole moment is overestimated by more than 4.5~D and by $\sim$0.5~D ($\sim$45\%) by aug and d-aug, respectively, despite both basis sets providing the correct orientation  (see Figure~\ref{fig:3px_water_1D_cut}(c)). 

For methanol, the largest differences in the excitation energy compared to PWs are exhibited by the S$_3$ (\mbox{2p$_y \rightarrow$ 3p$_x$}) and S$_4$ (\mbox{2p$_y \rightarrow$ 3p$_y$}) states calculated with aug, although the deviation remains below 0.3~eV. These two states have also the largest variance, consistent with their more diffuse Rydberg character. PW calculations provide $\sigma(\bm r)$ values of 76.7 and 83.0~bohr$^2$ for S$_3$ and S$_4$, respectively, while aug underestimates the variance, giving values of 67.7 and 70.1~bohr$^2$, respectively. The d-aug values of excitation energy and variance are close to those obtained with \glspl*{pw} for all states of methanol. The dipole moment of methanol has two components, $x$ and $z$ in the molecular frame shown in Figure \ref{fig:cartesian_frame}. Figure~\ref{fig:DM_basis_effect}(d) reports the $x$ component, while the $z$ component is shown in Figure~S9 of the \gls*{si}. For the more localized S$_1$, S$_2$, and S$_5$ states, the dipole moment shows only small changes with the basis set. The largest changes occur for the more diffuse S$_3$ and S$_4$ states. For S$_3$, $\mu_x$ changes from $-0.43$ D with aug to $-2.53$ D with d-aug and $-2.20$~D with \glspl*{pw}, while $\mu_z$ changes from $-2.58$ D with aug to 0.042 D with d-aug and 0.20~D with \glspl*{pw}. Thus, the aug basis set, which overly confines the density, rotates the dipole from the transverse $x$ direction to along the C--O bond, compared to PWs. For S$_4$, The $x$ component changes from $-3.90$ D with aug to $-0.06$ with d-aug and 0.34~D with \glspl*{pw}, while the $z$ component changes from $-3.67$ to $-0.50$ and $-0.22$~D. Thus, in this case, the aug basis set significantly overestimates the dipole moment magnitude compared to PWs. The d-aug basis set provides a significant improvement, but the deviation from PWs for S$_3$ and S$_4$ remains relatively large. Overall, the methanol results show that the basis set effect can involve a large change in the direction and magnitude of the dipole vector rather than a simple change in the magnitude of one component, and again illustrates the point that localized atomic orbitals calculations can yield a different dipole moment compared to PWs even when sufficiently diffuse functions are included.

To assess the use of PWs as a basis-set reference, additional OO/PBE calculations were performed using a real-space-grid (RSG) representation, in which the orbitals are represented directly on a three-dimensional grid rather than expanded in either plane waves or atom-centered basis functions. Methanol was chosen for this comparison because it contains the most diffuse Rydberg states among our set of molecules, as indicated by the largest values of the variance. Table~S2 of the \gls*{si} reports the differences in excitation energy and dipole-moment components obtained with PWs and d-aug for the mixed-spin solutions of the excited states of methanol relative to the RSG results. The PW and RSG results are in close agreement for all investigated states, including the most diffuse Rydberg states for which the d-aug representation shows the largest deviations in the dipole moment from the RSG results.

Table~\ref{tab:basis_effect_table} also includes available coupled-cluster reference dipole moments for comparison. These previous coupled-cluster calculations exhibited large basis-set dependence for diffuse Rydberg states of water and formaldehyde~\cite{Chrayteh2021-ib}. For these molecules, the maximum absolute changes in the CCSDT dipole moments upon going from aug-cc-pVDZ to d-aug-cc-pVDZ are 0.73 and 1.89~D, respectively~\cite{Chrayteh2021-ib}. With aug-cc-pVTZ and d-aug-cc-pVTZ, the corresponding maximum changes are 0.57 and 1.25~D. In both cases, the largest changes occur for the S$_2$ (\mbox{2p$_x \rightarrow$ 3p$_y$}) state of water and the S$_4$ (\mbox{2p$_y \rightarrow$ 3p$_y$}) state of formaldehyde.

\subsection{Assessment of Exchange--Correlation Functionals}
\begin{figure}[hbt]
    \centering
    \includegraphics[width=\linewidth]{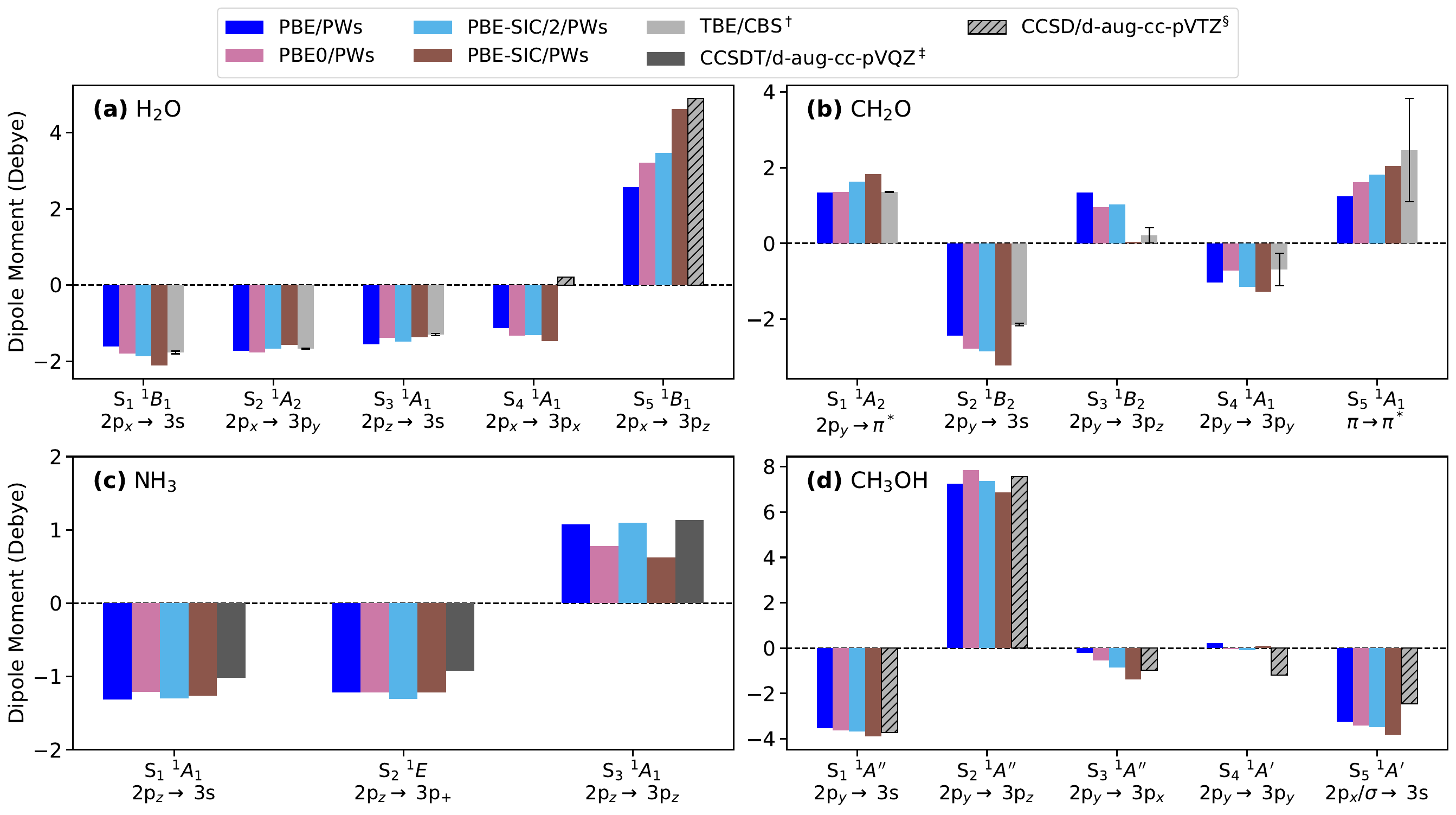}
    \caption{Spin-purified dipole moments of the singlet excited states of water (a), formaldehyde (b), ammonia (c), and methanol (d) obtained from orbital-optimized calculations with various exchange-correlation functionals and a plane-wave basis set. For methanol, which has two nonzero components, the reported value is along the z-direction of the molecular coordinate frame. Coupled-cluster results are shown in grey: $^\dagger$Theoretical best estimate (TBE)/complete basis set (CBS) values from ref.~\citen{Chrayteh2021-ib} obtained from high-order coupled cluster calculations and extrapolation to the complete basis set limit, with the corresponding CBS uncertainty shown as error bars; $^\ddagger$CCSDT/\mbox{d-aug-cc-pVQZ} values from ref.~\citen{perscorr_jacquemin}; $^\S$CCSD/\mbox{d-aug-cc-pVTZ} values from this work. In all cases, the OO calculations reproduce the dipole moment orientation of the coupled-cluster calculations. A strong functional dependence in the magnitude is observed only for the most diffuse Rydberg states and states with multi-configurational character}
    \label{fig:DM_xc_ref_comparison}
\end{figure}

Figure~\ref{fig:DM_xc_ref_comparison} compares the excited-state dipole moments of water, ammonia, formaldehyde, and methanol obtained from OO calculations with PBE, PBE0, PBE-SIC/2, and PBE-SIC using \glspl*{pw}. The corresponding numerical values are reported in Table~\ref{tab:xc_functional_table} together with results of coupled-cluster calculations.
For water and formaldehyde, \gls*{tbe}/\gls*{cbs} values from ref.~\citen{Chrayteh2021-ib} are reported, which were obtained from high-order coupled-cluster calculations up to CCSDTQP and extrapolation to the complete-basis-set limit using results with singly and doubly augmented basis sets. For methanol and for the two most diffuse Rydberg states of water, for which no \gls*{tbe}/\gls*{cbs} values are available, EOM-CCSD/d-aug-cc-pVTZ values calculated in the present work are shown. 
For ammonia, values from CCSDT/\mbox{d-aug-cc-pVTZ} calculations~\cite{perscorr_jacquemin} are shown. A comparison of the excitation energy and variance obtained with the various functionals is shown in Figures~S2, S4, S6, and S8 with corresponding values reported in Tables~S5, S8, S11 and S14.
\begin{table}[htb!]
\centering
\caption{%
Spin-purified dipole moment, $\mu$ (Debye), of singlet excited states of water, formaldehyde, ammonia, and methanol computed with orbital-optimized calculations using different exchange-correlation functionals with plane waves, alongside results of high-level coupled-cluster calculations using atomic orbitals basis sets including diffuse functions. For water, formaldehyde, and ammonia, $\mu$ is the single nonzero component of the dipole moment vector along the molecular axis of rotation. For methanol, each entry reports the $x$ and $z$ components of the dipole moment, respectively.
}
\label{tab:xc_functional_table}
\small
\begin{tabular}{
ll
S[table-format=-1.2]
S[table-format=-1.2]
S[table-format=-1.2,table-space-text-post=$^\mathsection$]
S[table-format=-1.2,table-space-text-post=$^\mathsection$]
c}
\hline
State & Character & {PBE} & {PBE0} & {SIC/2} & {SIC} & Coupled cluster\\
\hline

\multicolumn{7}{l}{H$_2$O} \\
S$_1$ $B_1$ & 2p$_x \rightarrow$ 3s
& -1.61 & -1.80 & -1.87 & -2.11
& \begin{tabular}{@{}S[table-format=-1.2]@{\;}l@{}}
-1.77 & {$\pm$ 0.04$^\dagger$} \\
-1.77$^\P$ & \phantom{$\pm$ 0.04$^\dagger$} 
\end{tabular} \\
S$_2$ $A_2$ & 2p$_x \rightarrow$ 3p$_y$
& -1.73 & -1.77 & -1.67 & -1.58
& \begin{tabular}{@{}S[table-format=-1.2]@{\;}l@{}}
-1.67 & {$\pm$ 0.01$^\dagger$} \\
-1.62$^\P$ & \phantom{$\pm$ 0.01$^\dagger$}
\end{tabular} \\
S$_3$ $A_1$ & 2p$_z \rightarrow$ 3s
& -1.56 & -1.38 & -1.48 & -1.37
& \begin{tabular}{@{}S[table-format=-1.2]@{\;}l@{}}
-1.30 & {$\pm$ 0.03$^\dagger$} \\
-1.16$^\P$ & \phantom{$\pm$ 0.03$^\dagger$}
\end{tabular} \\
S$_4$ $A_1$ & 2p$_x \rightarrow$ 3p$_x$
& -1.13 & -1.33 & -1.31 & -1.47
& \begin{tabular}{@{}S[table-format=-1.2]@{\;}l@{}}
0.21$^\P$ & \phantom{$\pm$ 0.04$^\dagger$}
\end{tabular} \\
S$_5$ $B_1$ & 2p$_x \rightarrow$ 3p$_z$
& 2.56 & 3.21 & 3.47 & 4.62
& \begin{tabular}{@{}S[table-format=-1.2]@{\;}l@{}}
4.89$^\P$ & \phantom{$\pm$ 0.04$^\dagger$}
\end{tabular} \\[0.6em]

\multicolumn{7}{l}{CH$_2$O} \\
S$_1$ $A_2$ & 2p$_y \rightarrow \pi^*$ &  1.35 &  1.36 &  1.63 &  1.83 &  1.36 $\pm$ 0.01$^\dagger$ \\
S$_2$ $B_2$ & 2p$_y \rightarrow$ 3s & -2.44 & -2.78 & -2.86 & -3.22 & -2.15 $\pm$ 0.03$^\dagger$ \\
S$_3$ $B_2$ & 2p$_y \rightarrow$ 3p$_z$ &  1.34 &  0.96 &  1.02 &  1.21 &  0.21 $\pm$ 0.20$^\dagger$ \\
S$_4$ $A_1$ & 2p$_y \rightarrow$ 3p$_y$ & -1.03 & -0.72 & -1.15 & 0.03 & -0.69 $\pm$ 0.43$^\dagger$ \\
S$_5$ $A_1$ & $\pi \rightarrow \pi^*$ &  1.24 &  1.62 &  1.81 &  2.05 &  2.46 $\pm$ 1.36$^\dagger$ \\[0.6em]

\multicolumn{7}{l}{NH$_3$} \\
S$_1$ $A_1$ & 2p$_z \rightarrow$ 3s
& -1.32 & -1.21 & -1.30 & -1.26
& \begin{tabular}{@{}S[table-format=-1.2]@{\;}l@{}}
-1.02$^\ddagger$ &
\phantom{$\pm$ 0.04$^\dagger$}
\end{tabular} \\
S$_2$ $E$ & 2p$_z \rightarrow$ 3p$_{+}$
& -1.22 & -1.22 & -1.30$^\mathsection$ & -1.22$^\mathsection$
& \begin{tabular}{@{}S[table-format=-1.2]@{\;}l@{}}
-0.92$^\ddagger$ &
\phantom{$\pm$ 0.04$^\dagger$}
\end{tabular} \\
S$_3$ $A_1$ & 2p$_z \rightarrow$ 3p$_z$
& 1.07 & 0.78 & 1.10$^\mathsection$ & 0.62$^\mathsection$
& \begin{tabular}{@{}S[table-format=-1.2]@{\;}l@{}}
1.13$^\ddagger$ &
\phantom{$\pm$ 0.04$^\dagger$}
\end{tabular} \\[0.6em]

\multicolumn{7}{l}{CH$_3$OH} \\
S$_1$ $A^{\prime\prime}$ & 2p$_y \rightarrow$ 3s &
\multicolumn{1}{c}{%
\begin{tabular}{@{}S[table-format=-1.2]@{\;}S[table-format=-1.2]@{}}
-3.25 & -3.53
\end{tabular}} &
\multicolumn{1}{c}{%
\begin{tabular}{@{}S[table-format=-1.2]@{\;}S[table-format=-1.2]@{}}
-3.74 & -3.64
\end{tabular}} &
\multicolumn{1}{c}{%
\begin{tabular}{@{}S[table-format=-1.2]@{\;}S[table-format=-1.2]@{}}
-3.86 & -3.68
\end{tabular}} &
\multicolumn{1}{c}{%
\begin{tabular}{@{}S[table-format=-1.2]@{\;}S[table-format=-1.2]@{}}
-4.45 & -3.90
\end{tabular}} &
\multicolumn{1}{c}{%
\begin{tabular}{@{}S[table-format=-1.2]@{\;}S[table-format=-1.2]@{}}
-3.72 & -3.72
\end{tabular}\makebox[0pt][l]{$^\P$}} \\

S$_2$ $A^{\prime\prime}$ & 2p$_y \rightarrow$ 3p$_z$ &
\multicolumn{1}{c}{%
\begin{tabular}{@{}S[table-format=-1.2]@{\;}S[table-format=-1.2]@{}}
1.98 & 7.24
\end{tabular}} &
\multicolumn{1}{c}{%
\begin{tabular}{@{}S[table-format=-1.2]@{\;}S[table-format=-1.2]@{}}
2.28 & 7.86
\end{tabular}} &
\multicolumn{1}{c}{%
\begin{tabular}{@{}S[table-format=-1.2]@{\;}S[table-format=-1.2]@{}}
2.80 & 7.36
\end{tabular}} &
\multicolumn{1}{c}{%
\begin{tabular}{@{}S[table-format=-1.2]@{\;}S[table-format=-1.2]@{}}
2.70 & 6.87
\end{tabular}} &
\multicolumn{1}{c}{%
\begin{tabular}{@{}S[table-format=-1.2]@{\;}S[table-format=-1.2]@{}}
2.49 & 7.56
\end{tabular}\makebox[0pt][l]{$^\P$}} \\

S$_3$ $A^{\prime\prime}$ & 2p$_y \rightarrow$ 3p$_x$ &
\multicolumn{1}{c}{%
\begin{tabular}{@{}S[table-format=-1.2]@{\;}S[table-format=-1.2]@{}}
2.20 & -0.20
\end{tabular}} &
\multicolumn{1}{c}{%
\begin{tabular}{@{}S[table-format=-1.2]@{\;}S[table-format=-1.2]@{}}
2.41 & -0.54
\end{tabular}} &
\multicolumn{1}{c}{%
\begin{tabular}{@{}S[table-format=-1.2]@{\;}S[table-format=-1.2]@{}}
2.53 & -0.86
\end{tabular}} &
\multicolumn{1}{c}{%
\begin{tabular}{@{}S[table-format=-1.2]@{\;}S[table-format=-1.2]@{}}
2.68 & -1.39
\end{tabular}\makebox[0pt][l]{$^\mathsection$}} &
\multicolumn{1}{c}{%
\begin{tabular}{@{}S[table-format=-1.2]@{\;}S[table-format=-1.2]@{}}
2.38 & -0.98
\end{tabular}\makebox[0pt][l]{$^\P$}} \\

S$_4$ $A^{\prime}$ & 2p$_y \rightarrow$ 3p$_y$ &
\multicolumn{1}{c}{%
\begin{tabular}{@{}S[table-format=-1.2]@{\;}S[table-format=-1.2]@{}}
-0.34 & 0.22
\end{tabular}} &
\multicolumn{1}{c}{%
\begin{tabular}{@{}S[table-format=-1.2]@{\;}S[table-format=-1.2]@{}}
-0.54 & -0.04
\end{tabular}\makebox[0pt][l]{$^\mathsection$}} &
\multicolumn{1}{c}{%
\begin{tabular}{@{}S[table-format=-1.2]@{\;}S[table-format=-1.2]@{}}
-0.05 & -0.08
\end{tabular}} &
\multicolumn{1}{c}{%
\begin{tabular}{@{}S[table-format=-1.2]@{\;}S[table-format=-1.2]@{}}
-0.68 & 0.09
\end{tabular}} &
\multicolumn{1}{c}{%
\begin{tabular}{@{}S[table-format=-1.2]@{\;}S[table-format=-1.2]@{}}
-1.22 & -1.18
\end{tabular}\makebox[0pt][l]{$^\P$}} \\

S$_5$ $A^{\prime}$ & 2p$_x$/$\sigma \rightarrow$ 3s &
\multicolumn{1}{c}{%
\begin{tabular}{@{}S[table-format=-1.2]@{\;}S[table-format=-1.2]@{}}
-3.05 & -3.24
\end{tabular}} &
\multicolumn{1}{c}{%
\begin{tabular}{@{}S[table-format=-1.2]@{\;}S[table-format=-1.2]@{}}
-3.50 & -3.42
\end{tabular}} &
\multicolumn{1}{c}{%
\begin{tabular}{@{}S[table-format=-1.2]@{\;}S[table-format=-1.2]@{}}
-3.70 & -3.48
\end{tabular}} &
\multicolumn{1}{c}{%
\begin{tabular}{@{}S[table-format=-1.2]@{\;}S[table-format=-1.2]@{}}
-4.36 & -3.83
\end{tabular}} &
\multicolumn{1}{c}{%
\begin{tabular}{@{}S[table-format=-1.2]@{\;}S[table-format=-1.2]@{}}
-2.90 & -2.45
\end{tabular}\makebox[0pt][l]{$^\P$}} \\
\hline
\multicolumn{7}{l}{\footnotesize $^\dagger$Theoretical best estimate/complete basis set values from ref.~\citen{Chrayteh2021-ib}.} \\
\multicolumn{7}{l}{\footnotesize $^\ddagger$CCSDT/\mbox{d-aug-cc-pVQZ} values from ref.~\citen{perscorr_jacquemin}; $^\P$CCSD/d-aug-cc-pVTZ values from the present work.} \\
\multicolumn{7}{l}{\footnotesize $^\mathsection$Evaluated from the mixed-spin solution (no spin-purification applied).}\\
\end{tabular}
\end{table}

For water, the different \gls*{xc} functionals predict the same orientation of the dipole moment for all five excited states considered. 
For water, the functional dependence is relatively small for S$_2$ and S$_3$, while somewhat larger variations are observed for S$_1$ and S$_4$. The largest functional dependence is found for the S$_5$ state.
As discussed in the previous section, their dipole moments display a pronounced basis-set dependence when going from the d-aug basis to \glspl*{pw}. In particular, for S$_4$, the CCSD calculation predicts a positive dipole orientation, whereas all \gls*{oo} calculations with PWs yield a negative orientation. Moreover, S$_3$ and S$_4$ have a multi-configurational character~\cite{Rubio2008-lk}, and the corresponding CCSD results are affected by the absence of triple excitations, which are expected to be important for these strongly correlated states, as also noted in ref.~\citen{Chrayteh2021-ib}.
 PBE0 gives the closest value for S$_1$, with $-1.80$~D compared to the TBE/CBS value of $-1.77 \pm 0.04$~D. For S$_2$, PBE and PBE0 slightly overestimate the dipole moment, whereas PBE-SIC/2 gives nearly exact agreement with the reference value of $-1.67 \pm 0.01$~D. For S$_3$, PBE overestimates the magnitude, while PBE0 and PBE-SIC are closer to the reference value of $-1.30 \pm 0.03$~D. Previous MS-CASPT2 calculations~\cite{Rubio2008-lk} using a basis set of atomic natural orbitals with additional diffuse functions (ANO-L+R) indicate that the S$_3$ and S$_4$ states of water have a multi-configurational character, with a strong mixing of the \mbox{2p$_z \rightarrow$ 3s} and \mbox{2p$_z \rightarrow$ 3p$_x$} excitations. The OO calculations however provide a single-configurational (diabatic-like) character, with the optimized S$_3$ and S$_4$ states being pure \mbox{2p$_z \rightarrow$ 3s} and \mbox{2p$_z \rightarrow$ 3p$_x$} excitations, respectively. This leads to a relatively large error on the variance of S$_3$. The reference CASPT2 value~\cite{Rubio2008-lk} is approximately $45$~bohr$^2$, while the OO calculations give a variance below $32$~bohr$^2$, because they lack mixing with the more diffuse 3p$_x$ excitation. It may therefore appear surprising that the OO calculations give a good estimate of the S$_3$ dipole moment compared to the TBE/CBS value. The good agreement can be explained by the fact that the \mbox{2p$_z \rightarrow$ 3s} and \mbox{2p$_z \rightarrow$ 3p$_x$} excitations have similar dipole moment, as indicated by the similar values obtained from the OO calculations of S$_3$ and S$_4$, and therefore a lack of mixing does not significantly affect the results. A similar cancellation of errors is observed for the excitation energy~\cite{Levi2026-ts, Sigurdarson2023-do}. The last and most diffuse state of water, S$_5$, exhibits the largest functional dependence. There, the dipole moment increases from 2.56~D with PBE to 3.21~D with PBE0, and 3.47 and 4.62~D with PBE-SIC/2 and PBE-SIC, respectively. High-level calculations of the dipole moment of this state beyond d-aug are currently not available.

For the excited states of formaldehyde, the functional dependence is more varied. The valence S$_1$ state is accurately described by PBE and PBE0, with errors on the dipole moment with respect to TBE/CBS of less than 0.01~D. In contrast, SIC worsens the agreement for this state, giving an overestimation of about 0.2 and 0.5~D with PBE-SIC/2 and PBE-SIC, respectively. A similar trend is observed for S$_2$. This state has a Rydberg character, but is less diffuse than the other Rydberg states of formaldehyde. PBE overestimates the magnitude of the dipole moment relative to \gls*{tbe}/\gls*{cbs} by $\sim0.3$ D ($\sim$14\%), and the deviation increases monotonically from PBE0 to PBE-SIC/2 and PBE-SIC. For the more diffuse Rydberg S$_3$ and S$_4$ states, all functionals give the correct sign, but the magnitude depends non-monotonically on the functional. For these states, the relative errors appear to be larger, but the reference dipole moments are small and have relatively large uncertainties. Finally, the S$_5$ valence state shows a strong dependence on the functional. There, the dipole moment increases from $1.24$~D with PBE to $1.62$~D with PBE0, $1.81$~D with PBE-SIC/2, and $2.05$~D with PBE-SIC. This state is known to have significant multi-configurational character in multireference calculations~\cite{Gomez-Carrasco2010-me,Schreiber2008-ui,Muller2001-hq,Merchan1995-gc}. In particular, the S$_5$ state is located in the proximity of two conical intersections at the ground-state geometry~\cite{Gomez-Carrasco2010-me}, therefore its description is especially challenging. This is also indicated by the fact that the \gls*{tbe}/\gls*{cbs} uncertainty is large. Thus, the large functional dependence may reflect the difficulty of OO calculations to describe the multi-configurational character, rather than a sensitivity to the treatment of self-interaction.

For ammonia, the SIC calculations tend to break the spatial symmetry of the density, and symmetric triplet solutions required for spin purification could not be found. Therefore, the PBE-SIC/2 and PBE-SIC values reported in Figure~\ref{fig:DM_xc_ref_comparison}(c) and Table \ref{tab:xc_functional_table} for S$_2$ and S$_3$ lack spin purification. Keeping that in mind, for ammonia the functional dependence is moderate for S$_1$ and S$_2$, but more pronounced for S$_3$, which is the most diffuse state. For S$_1$ and S$_2$, all functionals overestimate the magnitude of the dipole moment relative to the CCSDT/\mbox{d-aug-cc-pVTZ} values, with PBE0 giving the lowest errors. For S$_3$, the functional dependence is not monotonic. The dipole moment changes from $1.07$~D with PBE to $0.78$~D with PBE0, $1.10$~D with PBE-SIC/2, and $0.62$~D with PBE-SIC, compared with the CCSDT value of $1.30$~D. Thus, for this state of ammonia, neither PBE0 nor SIC give a dipole moment closer to the CCSDT value compared to PBE. The pronounced basis-set dependence observed for this state at the PBE level, with the d-aug dipole moment differing by about 45\% from the PW result, indicates that its dipole moment is particularly sensitive to the spatial representation. Although this difference cannot be used to quantify the basis-set error of the CCSDT calculation, it suggests that the CCSDT/d-aug-cc-pVTZ reference value should be interpreted with some caution.

Figure~\ref{fig:DM_xc_ref_comparison}(d) shows the variation of the $z$ component of the dipole moment of methanol with the functional, while the corresponding $x$ component is shown in Figure~S10 of the \gls*{si}. The magnitude of the $z$ component generally increases upon going from PBE to PBE0 and the SIC functionals for S$_1$, S$_2$, S$_3$, and S$_5$. This increase is monotonic for S$_1$, S$_3$, and S$_5$, whereas for S$_2$ the PBE-SIC value is slightly smaller than the PBE-SIC/2 value. For the first three states, PBE0 generally gives good agreement with the CCSD/d-aug-cc-pVTZ results, with deviations below 9\% for most components. The notable exception is the $z$ component of S$_3$, for which the PBE0 deviation is approximately 45\%. For the S$_5$ state, PBE provides the most accurate results, while the other functionals overestimate the magnitude of the dipole moment. The S$_4$ state is weakly polarized in the $x$ direction for all functionals, with values between $0.05$ and $0.68$~D. For this state, as for the S$_2$ and S$_3$ states of ammonia, the triplet calculation converges to a solution with spatial symmetry breaking when PBE-SIC is employed, so only a mixed-spin value of the dipole moment is available for PBE-SIC. All functionals exhibit a large deviation with respect to the CCSD/d-aug-cc-pVTZ results for the dipole moment of S$_4$. However, the dipole moment of this state shows a pronounced sensitivity to the spatial representation at the PBE level, with significant differences between the d-aug and PW results (see Table~\ref{tab:basis_effect_table}). As in the case of ammonia, this basis-set sensitivity warrants some caution when using the CCSD/d-aug-cc-pVTZ value as a reference.

\begin{figure}[hbt]
    \centering
    \includegraphics[width=0.5\linewidth]{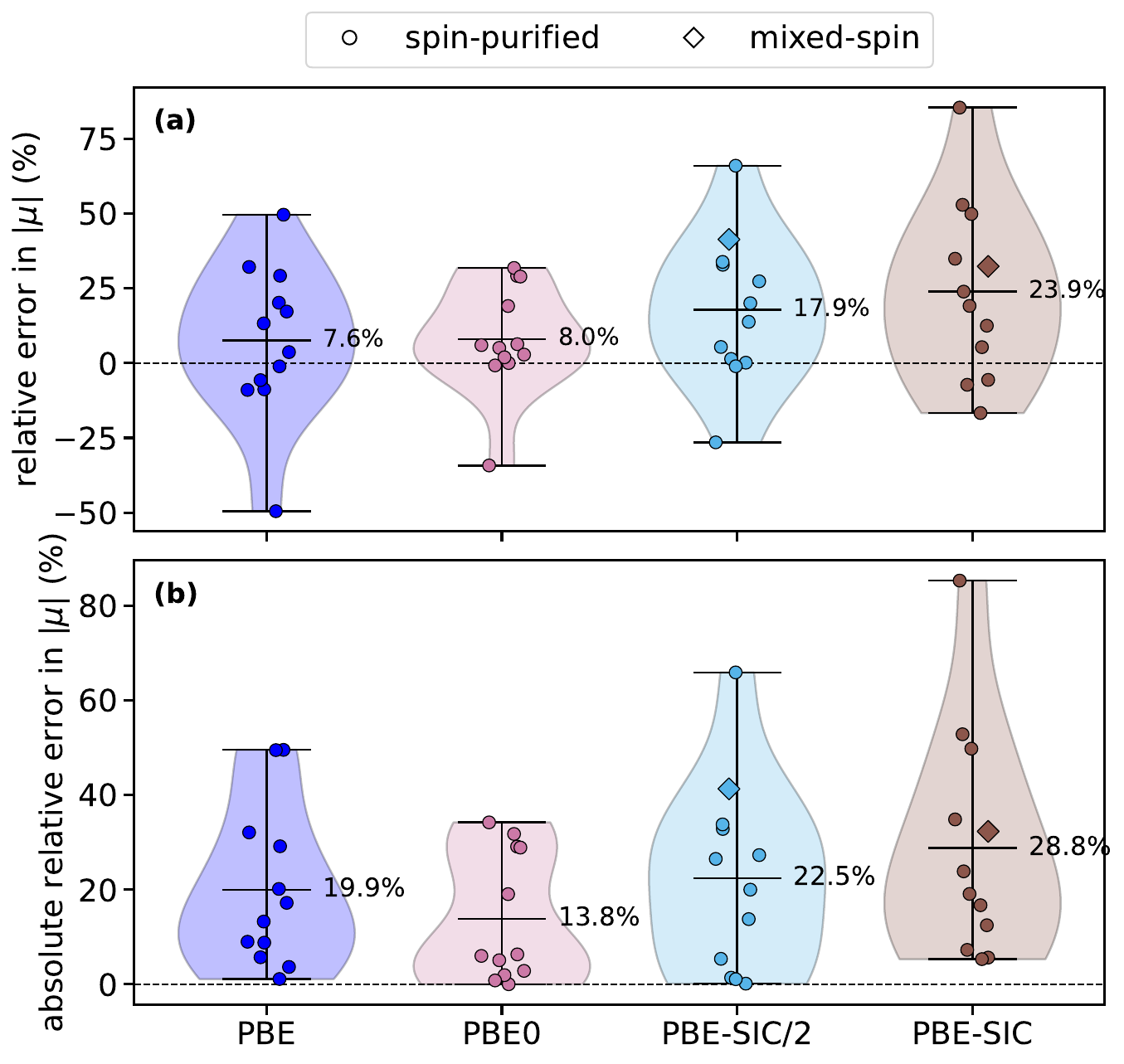}
    \caption{
    Signed (a) and absolute (b) relative errors in the magnitude of singlet excited-state dipole moments of water, formaldehyde, ammonia and methanol obtained with orbital-optimized calculations with different \gls*{xc} functionals and a plane-wave (PW) basis set. The errors are computed with respect to theoretical best estimates (see Table~\ref{tab:xc_functional_table}). Only states for which the difference between d-aug and PW PBE results is less than 5\% are included. The percentages reported next to each error distribution correspond to the mean error. Dipole moments evaluated from a mixed-spin solution only (no spin purification) are marked with diamonds, while all other points correspond to spin-purified dipole moments. The PBE0 functional provides the lowest mean error and spread, while inclusion of self-interaction correction with global scaling does not lead to an improvement over the PBE results.}
    \label{fig:relative_error}
\end{figure}
Figure~\ref{fig:relative_error} compares the signed and absolute relative errors in the magnitude of the dipole moment with respect to reference coupled-cluster values,
\begin{equation}
\text{relative error}_{|\boldsymbol{\mu}|}
=
\dfrac{|\boldsymbol{\mu}_{\mathrm{OO}}|-|\boldsymbol{\mu}_{\mathrm{ref}}|}
{|\boldsymbol{\mu}_{\mathrm{ref}}|}
\cdot 100\% ,
\end{equation}
for the different \gls*{xc} functionals.
For this statistical comparison, only states for which the relative difference between the d-aug and PW dipole moment magnitudes,
\begin{equation}
\Delta_{\mathrm{basis}} =  \frac{  |\boldsymbol{\mu}_{\mathrm{d\text{-}aug}}| - |\boldsymbol{\mu}_{\mathrm{PW}}|}{ |\boldsymbol{\mu}_{\mathrm{PW}}| }\times 100\%,
\end{equation}
is below 5\% in the OO/PBE calculations are included.
The percentages shown next to each error distribution correspond to the signed and absolute mean values, and therefore measure the typical error in the magnitude of the dipole moment for each functional.

PBE gives a mean absolute error of $\sim$20\%. PBE0 gives smaller errors and a narrower distribution, with a mean absolute relative error of only $\sim$14\%. The PBE and PBE0 signed error distributions are closely centered around $\sim$8\%, indicating a rather small systematic bias in the dipole moment magnitude. 
\Gls*{lrdft} calculations with the PBE0 functional have been found to give a larger mean absolute error of $\sim$60\% in the magnitude of the excited-state dipole moment for similar molecules~\cite{Sarkar2021-wg}. These calculations were performed using the aug-cc-pVTZ basis set, and the errors were evaluated with respect to coupled-cluster theoretical best estimates for the same basis set, constructed using CCSDTQ or CCSDTQP corrections. The different basis representations and the larger molecular sample considered by Sarkar \textit{et al.} prevent a direct quantitative comparison with the present OO/PBE0 results.

Inclusion of SIC, both in its full and $1/2$-scaled versions, seems to deteriorate the results compared to PBE. PBE-SIC/2 gives a mean absolute relative error of $\sim$22\%, while PBE-SIC gives $\sim$29\% and shows the broadest distribution, including the largest maximum error. The PBE-SIC/2 and PBE-SIC signed error distributions are shifted toward more positive values compared to PBE, showing that the SIC calculations tend to overestimate the dipole magnitude. Dipole moments computed with SIC that could not be spin-purified due to convergence of the triplet calculations on symmetry-broken solutions, are marked by diamonds in Figure~\ref{fig:relative_error}. Such points contribute to the spread of the error distribution for SIC, but they are not the only source of the larger errors. Large deviations from the reference values are also present for the spin-purified states.

\section{Discussion}
The present work shows that dipole moments are a stricter test for \gls*{oo} calculations of excited Rydberg states than the excitation energy. 

The analysis of the effect of the basis set shows that the variance, $\sigma(\bm r)$, and the dipole moment are much more sensitive to the choice of basis set than the excitation energy. Even when the excitation energy is weakly affected by the basis set, the variance and the dipole moment can change significantly. This is exemplified by the PBE calculations with the aug atomic basis set, which includes one set of diffuse functions. For the less diffuse Rydberg states investigated, aug calculations give small deviations in the excitation energy but large deviations in the variance and dipole moment compared to the PW results. The S$_3$ (\mbox{2p$_z \rightarrow$ 3s}) state of methanol is an illustrative case. There, the aug basis set provides an excitation energy close to the PW result, but a dipole dominated by the $z$ component, whereas the \gls*{pw} calculations give a dipole mainly along the $x$ direction. 

The limitation of the \gls*{lcao} representation is not only the number of diffuse functions, but also its atom-centered form. Adding diffuse functions improves the radial extent of the Rydberg density, as shown by a closer agreement in the variance from calculations with the d-aug basis set with respect to PWs, but it does not always provide enough flexibility to reproduce the anisotropic redistribution of the density of a given excitation. The S$_4$ (\mbox{2p$_x \rightarrow$ 3p$_x$}) and S$_5$ (\mbox{2p$_x \rightarrow$ 3p$_z$}) states of water provide a clear example. There, d-aug gives an excitation energy and variance close to the \gls*{pw} values, but the dipole moment still differs noticeably. To provide insights into the limitations of an atom-centered representation, OO PBE calculations are carried out using a composite basis set, consisting of the aug basis functions on oxygen and d-aug basis functions on hydrogen. Compared with the d-aug basis, the dipole moment of S$_4$ changes from 0.03 to $-0.41$~D, while that of S$_5$ changes from 3.62 to 3.07~D. Both values are closer to the corresponding \gls*{pw} results of $-1.13$ and 2.56~D, respectively, than for the d-aug basis set, despite such composite basis set having fewer diffuse functions than d-aug. This shows that the error of the \gls*{lcao} representation is not simply due to insufficiently diffuse basis functions, but also to the limited spatial flexibility of the atom-centered form. PWs do not have this constraint and provide a more flexible representation for diffuse Rydberg states. The close agreement between PW and RSG calculations for methanol (see Table~S2 in the \gls*{si}), including its most diffuse Rydberg states, further supports the use of PWs as a basis-set reference.

The smaller basis set sensitivity of the excitation energy can be rationalized on the basis of the variational nature of the OO calculations. In \gls*{oo} excited-state methods, each excited state corresponds to a stationary point of the electronic energy landscape. At convergence, the first-order contribution to the error on the energy with respect to orbital variations vanishes. Consequently, the excited-state energy is relatively insensitive to errors in the wave function compared to other properties, such as the dipole moment, which are not stationary with respect to orbital variations. 

The dependence of the calculated excited-state dipole moments on the \gls*{xc} functional is found to be smaller than the basis set dependence. This is in contrast to \gls*{lrdft}, where excited-state dipole moments show a strong dependence on the functional. For example, Sarkar \textit{et al.} report \gls*{lrdft} mean absolute relative errors of about 60\% with PBE0 and about 30\% with CAM-B3LYP for small molecules, including water and formaldehyde~\cite{Sarkar2021-wg}. As discussed above, however, this comparison should be made with caution, since the \gls*{lrdft} assessment was based on a larger sample of molecules and different basis set. 
For most states, the \gls*{oo}/PW calculations reproduce the dipole moment orientation of the coupled-cluster calculations, with few exceptions discussed in the text. PBE and PBE0 give mean absolute relative errors of about 20\% and 14\%. These results suggest that state-specific orbital relaxation, which is missing in practical implementations of \gls*{lrdft}, is important for accurate prediction of dipole moments of Rydberg excited states. 

The OO PW calculations also highlight an important limitation of current high-level wave function methods, such as coupled-cluster approaches. Reference calculations are typically performed using basis sets of localized atomic orbitals, such as d-aug. In the present work, a quantitative comparison with excited-state dipole moments obtained in coupled-cluster calculations could only be done for states where the d-aug basis set provides results in agreement with PW values. For highly diffuse Rydberg states, reference dipole moments are often unavailable, or are obtained with atom-centered basis sets that may not be sufficiently flexible to reach the PW limit. This highlights the need for more reliable benchmark data for excited-state dipole moments of diffuse Rydberg states, and for high-level wave function approaches that can use more flexible basis representations. Such developments are currently being pursued by some of the authors in the context of selected configuration interaction methods~\cite{Levi2026-ts, Schmerwitz2025}.

The results obtained here with \gls*{sic} are particularly interesting and show that improving the asymptotic form of the effective potential does not necessarily improve the dipole moments of Rydberg states. Previous work has shown that full Perdew--Zunger \gls*{sic} improves the excitation energy of diffuse Rydberg states, an effect attributed to the recovery of the correct asymptotic $-1/r$ dependence of the effective potential~\cite{Sigurdarson2023-do}. For the dipole moment however, the same correction does not seem to improve the results. Instead, PBE-SIC tends to overestimate the dipole moment magnitude and gives larger errors than PBE. Thus, restoring the correct long-range potential is not sufficient to improve the excited-state dipole moments. The half-scaled correction reduces the overestimation but still gives larger mean errors than PBE and PBE0. This behavior is consistent with the known tendency of full PZ-SIC to overcorrect approximate functionals in regions where occupied orbital densities overlap. Such regions remain important even for diffuse Rydberg excited states, because the total density contains contributions from several compact valence orbitals. The dipole moment depends on the full density redistribution, not only on the asymptotic tail of the Rydberg orbital. A globally scaled \gls*{sic} applies the same correction in regions where it is needed and in regions where it is too strong. Locally scaled \gls*{sic} methods have recently been developed~\cite{John2026arXiv, Shahi2026} and are a promising route to address this issue, because they can retain a strong correction in isolated orbital regions while reducing it in regions of high density overlap.

The \gls*{sic} calculations also introduce practical complications. Due to the orbital-density dependence, SIC can introduce multiple local solutions, with some breaking the spatial symmetry of the density. For some states of ammonia and methanol, only a triplet solution that breaks the symmetry could be found in the PBE-SIC calculations, preventing the application of spin purification (see Table~\ref{tab:xc_functional_table}). In formaldehyde, two triplet and two mixed-spin solutions with the same \mbox{2p$_y \rightarrow$ 3p$_z$} character are obtained with PBE-SIC. The corresponding spin-purified S$_3$ dipole moments differ substantially, 0.03 vs. 0.72~D. The former is closer to the \gls*{tbe}/\gls*{cbs} value and is therefore reported in Figure~\ref{fig:DM_xc_ref_comparison} and Table~\ref{tab:xc_functional_table}.

\section{Conclusions}
Orbital-optimized density functional calculations with a plane-wave basis set have been used to compute the electric dipole moment of several Rydberg excited states of water, formaldehyde, ammonia, and methanol. The results show that excited-state dipole moments are a more demanding test of OO approaches than the excitation energy. 

Commonly used atomic basis sets can give inaccurate dipole moments for Rydberg states even when the excitation energy is insensitive to the basis set choice. A single-augmented atomic basis set often overconfines the excited-state density, leading to large errors in both the magnitude and direction of the dipole moment. Adding an extra set of diffuse functions improves the results, but relatively large deviations with respect to PW calculations remain for the most diffuse Rydberg states. The limitation is not only the radial extent of the basis, but also the reduced spatial flexibility of atom-centered functions, indicated by the fact that large errors can affect the dipole moment even when the LCAO calculations provide a variance in agreement with PW calculations. On the contrary, PWs are not affected by confinement effects and provide a flexible representation of diffuse Rydberg densities.

The comparison with high-level coupled-cluster results, when available, shows that, unlike \gls*{lrdft}, OO calculations of excited-state dipole moments depend only moderately on the exchange-correlation functional. PBE already provides relatively good results, with a mean absolute relative error on the magnitude of the dipole moment of about 20\%. PBE0 performs better, reducing the mean absolute relative error to about 14\% and giving a narrower error distribution. In contrast, globally scaled Perdew--Zunger self-interaction correction does not improve the dipole moments. Although full SIC improves the excitation energy of Rydberg states by restoring the correct asymptotic form of the effective potential, it tends to overestimate the dipole moment magnitude and gives larger errors than PBE. The half-scaled SIC reduces this overestimation but still does not improve over PBE. Thus, correcting the long-range potential does not seem to be sufficient to improve excited-state dipole moments.

Looking forward, several developments would be beneficial. First, high-level reference dipole moments are still missing for the most diffuse Rydberg states, and existing values from coupled-cluster calculations are often obtained with atom-centered basis sets that may not be sufficiently flexible for these states. Thus,wave function methods with more flexible basis set representations are needed. Second, a restricted open-shell Kohn--Sham formulation~\cite{Frank1998, Filatov1998} would simplify the treatment of open-shell singlet excited states in OO density functional calculations by avoiding the separate mixed-spin and triplet calculations required for spin purification, which is especially critical when SIC is employed. Finally, further development of SIC-based approaches, including locally scaled SIC~\cite{John2026arXiv, Shahi2026}, may improve their accuracy and reduce the symmetry-breaking problems observed here.

\appendix

\section{Dipole moment in the projector augmented wave approach}\label{sec:AppendixA}
In the \gls*{paw} method~\cite{Blochl1994-gk}, the variational optimization is performed for smooth pseudo orbitals, $|\tilde{\psi}_i^k\rangle$, which are related to the corresponding orbitals, $|\psi_i^k\rangle$, through a linear transformation,
\begin{equation}
    |\psi_i^k\rangle = \hat{\mathcal T} |\tilde{\psi}_i^k\rangle ,
\end{equation}
with
\begin{equation}
    \hat{\mathcal T} = 1 + \sum_a \sum_n \left( |\phi_n^a\rangle - |\tilde{\phi}_n^a\rangle \right) \langle \tilde p_n^a | .
\end{equation}
Here, $a$ labels atoms and $n$ labels so-called all-electron, $|\phi_n^a\rangle$, and pseudo, $|\tilde{\phi}_n^a\rangle$, atom-centered partial waves, and $|\tilde p_n^a\rangle$ are smooth projector functions. This transformation restores the rapid oscillations and cusps of the wave function near the nuclei while allowing the smooth pseudo orbitals to be represented efficiently on a coarse real-space grid. In real space, the orbitals can be written as
\begin{equation}
    \psi_i^k(\mathbf r) = \tilde{\psi}_i^k(\mathbf r) + \sum_a \sum_n \left[ \phi_n^a(\mathbf r) - \tilde{\phi}_n^a(\mathbf r) \right] P_{in}^{a,k} ,
\end{equation}
where
\begin{equation}
    P_{in}^{a,k} = \langle \tilde p_n^a | \tilde\psi_i^k \rangle
\end{equation}
are projector overlaps. Using the projector overlaps, an atom-centered density matrix can be defined as 
\begin{equation}
    D_{nm}^{a,k} = \sum_i f_i^k P_{in}^{a,k*} P_{im}^{a,k},
\end{equation}
where $f_i^k$ are the orbital occupation numbers.

For a one-electron operator $\hat O$, the expectation value in the PAW formalism can be evaluated by applying the PAW transformation, giving
\begin{equation}\label{eq:paw_operator}
    \langle \hat O \rangle_k = \sum_i f_i^k \langle \psi_i^k | \hat O | \psi_i^k \rangle + O_{\mathrm{core}} = \sum_i f_i^k \langle \tilde{\psi}_i^k | \hat{\mathcal T}^{\dagger} \hat O \hat{\mathcal T} | \tilde{\psi}_i^k \rangle + O_{\mathrm{core}},
\end{equation}
which can be written as a smooth pseudo contribution plus atom-centered corrections~\cite{Blochl2002},
\begin{equation}
    \langle \hat O \rangle_k = \sum_i f_i^k \langle \tilde\psi_i^k | \hat O | \tilde\psi_i^k \rangle + \sum_a \sum_{nm} D_{nm}^{a,k} \left[ \langle \phi_n^a | \hat O | \phi_m^a \rangle - \langle \tilde\phi_n^a | \hat O | \tilde\phi_m^a \rangle \right] + O_{\mathrm{core}}.
\end{equation}
GPAW uses the frozen-core approximation for the orbitals and the $O_{\mathrm{core}}$ contribution is given by
\begin{equation}
    O_{\mathrm{core}} = \sum_c
    \left\langle \phi_{c}^{\mathrm{core}} \middle| \hat O \middle| \phi_{c}^{\mathrm{core}} \right\rangle ,
\end{equation}
where $\phi_{c}^{\mathrm{core}}$ are atomic core orbitals localized within the augmentation regions and obtained from calculations of the isolated atoms. 

Using eq\ \eqref{eq:dipole_density} and the expression of a one-electron operator in PAW formalism, eq\ \eqref{eq:paw_operator}, the electric dipole moment of a state $k$ obtained from OO calculations in the PAW formalism can be calculated as
\begin{equation}
    \boldsymbol{\mu}^k = -\left[\sum_i f_i^k \langle \tilde\psi_i^k | \mathbf r | \tilde\psi_i^k \rangle + \sum_a \sum_{nm} D_{nm}^{a,k} \left[ \langle \phi_n^a | \mathbf r | \phi_m^a \rangle - \langle \tilde\phi_n^a | \mathbf r | \tilde\phi_m^a \rangle \right] + \mathbf r_{\mathrm{core}} \right] + \sum_a \mathcal{Z}_a \mathbf R_a ,
\end{equation}
where 
\begin{equation}
    \mathbf r_{\mathrm{core}} = \sum_c
    \left\langle \phi_{c}^{\mathrm{core}} \middle| \mathbf r \middle| \phi_{c}^{\mathrm{core}} \right\rangle .
\end{equation}
In GPAW, the frozen-core densities are spherical and centered on the atoms. Therefore, the dipole moment around each atomic center is zero, and the core contribution to the electronic dipole moment reduces to
\begin{equation}
    \mathbf r_{\mathrm{core}} = \sum_a N_{\mathrm{core}}^a \mathbf R_a ,
\end{equation}
where $N_{\mathrm{core}}^a$ is the number of frozen core electrons of atom $a$. Combining this term with the nuclear contribution gives the final expression
\begin{equation}
    \boldsymbol{\mu}^k = - \left[ \sum_i f_i^k \langle \tilde\psi_i^k | \mathbf r | \tilde\psi_i^k \rangle + \sum_a \sum_{nm} D_{nm}^{a,k} \left( \langle \phi_n^a | \mathbf r | \phi_m^a \rangle - \langle \tilde\phi_n^a | \mathbf r | \tilde\phi_m^a \rangle \right) \right] + \sum_a \left( \mathcal{Z}_a - N_{\mathrm{core}}^a \right) \mathbf R_a .
\end{equation}
The first term is evaluated from the smooth pseudo-wave functions in the representation used in the calculation, i.e., in a \gls*{pw} basis set or from the LCAO coefficients and basis functions. The PAW correction terms are efficiently evaluated on atomic radial grids inside the augmentation spheres using the all-electron and pseudo partial waves from the atomic setups.

\begin{acknowledgement}
The authors thank Denis Jacquemin and Pierre-François Loos for providing the results of the CCSDT/\mbox{d-aug-cc-pVTZ} calculations of ammonia. The authors thank Aleksei V. Ivanov, Hannes Jónsson, and Philipp Hansmann for useful discussions. Y.L.A.S acknowledges support by the Max Planck Society. G.L. and D.L.P. acknowledge support by the Icelandic Research Fund (grant no. 2511544). E. Ö. J. acknowledges support by the Icelandic Research Fund (grant no. 2611846). G.L. and L.R. acknowledge support from the ERC under the European Union's Horizon Europe research and innovation programme (grant no. 101166044, project NEXUS). Views and opinions expressed are however those of the author(s) only and do not necessarily reflect those of the European Union or ERC Executive Agency. Neither the European Union nor the granting authority can be held responsible for them. The authors acknowledge computer resources, data storage, and user support by the Icelandic Research e-Infrastructure (IREI), funded by the Icelandic Infrastructure Fund.
\end{acknowledgement}

\begin{tocentry}
   \includegraphics[width=\textwidth]{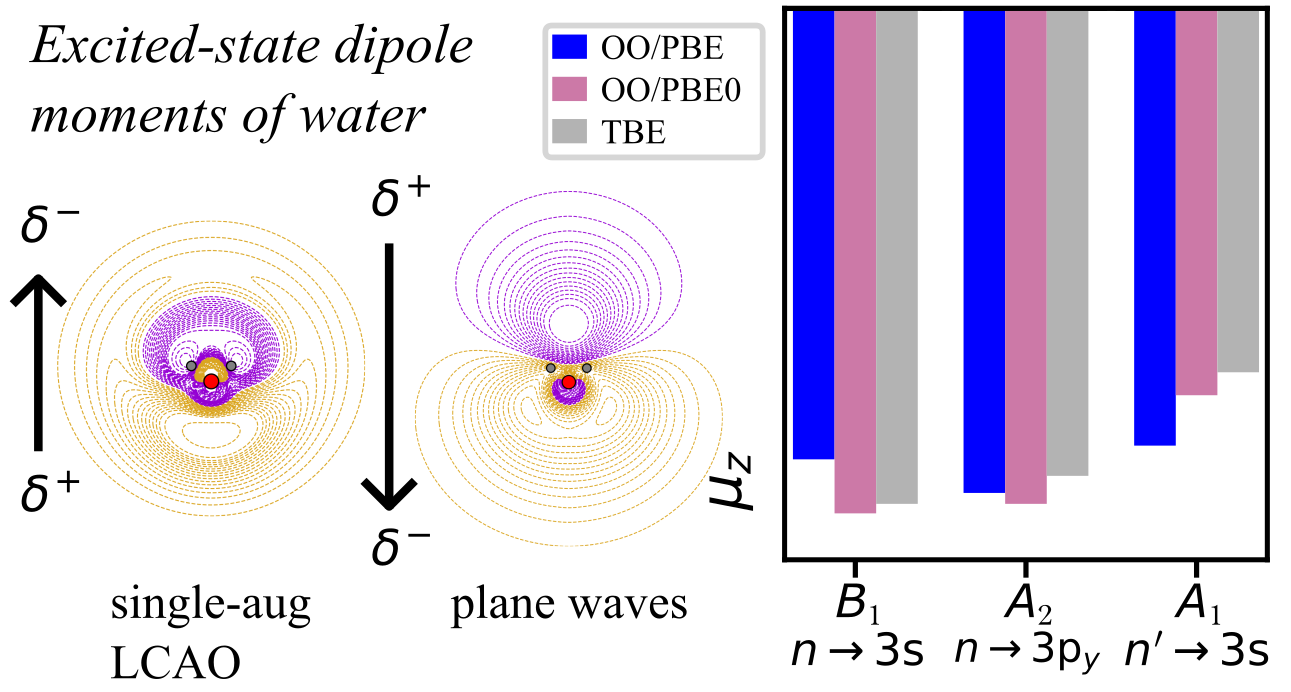}
\end{tocentry}

\begin{suppinfo}
The Supporting Information includes numerical tests on the size of the simulation box; calculations with a real-space grid representation of the orbitals; computational timings for orbital-optimized calculations of the mixed-spin and triplet excited states of water using LCAO and plane waves; graphical depictions of the molecular orbitals involved in the excitations examined in the article; tables listing the computed vertical transition energy, variance, and excited-state dipole moments for the mixed-spin and triplet states; and additional bar charts showing the effect of basis set and exchange-correlation functionals on the excitation energy and variance.
\end{suppinfo}

\section*{Data and Software Availability}
The data underlying this study are available in the published article and the Supporting Information. Scripts and code used to generate the results are publicly available at \url{https://gitlab.com/lore_94/OO-excited-state-properties}.

\bibliography{biblio.bib}
\end{document}


\tableofcontents
\clearpage

\section{Numerical Tests}
\begin{table}[hbt]
\caption{Differences in the excitation energy (eV) and components of the dipole moment (Debye) upon increasing the vacuum around the atoms from 10.5 to 15~\AA. Values are shown for the orbital-optimized mixed-spin solutions of the two most diffuse excited states of methanol calculated using different basis sets and the PBE functional (see values of excitation energy and dipole moment in Table S12).
}
\label{tab:vacum_size}
\begin{tabular}{lcccc}
\hline
State & Basis            & $\Delta \Delta E$  & $\Delta \mu _x$ & $\Delta \mu _z$ \\
\hline
M$_2$    & aug-cc-pVDZ+sz   & -0.001                 & -0.01               & -0.01               \\
M$_2$    & d-aug-cc-pVDZ+sz & -0.002                 & -0.06               & 0.01                \\
M$_2$    & plane waves      & 0.002                  & -0.06               & 0.00               \\
M$_3$    & aug-cc-pVDZ+sz   & -0.002                 & -0.04               & -0.03               \\
M$_3$    & d-aug-cc-pVDZ+sz & -0.001                 & -0.01               & -0.01               \\
M$_3$    & plane waves      &  0.002                 & 0.00                & 0.00    \\           \hline
\end{tabular}
\end{table}

\begin{table}[hbt]
\centering
\caption{Differences in excitation energy (eV) and dipole-moment components (D) of the mixed-spin excited states of methanol obtained from orbital-optimized PBE calculations using plane waves (PWs) and the d-aug-cc-pVDZ+sz (d-aug) basis set, relative to results obtained with a real-space-grid (RSG) representation.}
\label{tab:methanol_FD}
\begin{tabular}{llccc|ccc}
\hline
 & & \multicolumn{3}{c|}{PWs $-$ RSG} & \multicolumn{3}{c}{d-aug $-$ RSG} \\
\cline{3-8}
State & Character
& $\Delta\Delta E$ & $\Delta\mu_x$ & $\Delta\mu_z$
& $\Delta\Delta E$ & $\Delta\mu_x$ & $\Delta\mu_z$ \\
\hline
M$_0$ & $n \rightarrow 3s$
& 0.031 & -0.019 & -0.008
& 0.043 & -0.011 &  0.027 \\

M$_1$ & $n \rightarrow 3p_z$
& 0.033 &  0.008 &  0.040
& 0.036 &  0.019 & -0.037 \\

M$_2$ & $n \rightarrow 3p_x$
& 0.034 &  0.003 & -0.009
& 0.047 & -0.318 & -0.165 \\

M$_3$ & $n \rightarrow 3p_y$
& 0.036 &  0.006 &  0.003
& 0.048 & -0.358 & -0.244 \\

M$_4$ & $n' \rightarrow \sigma^{*}/3s$
& 0.035 & -0.018 & -0.005
& 0.073 & -0.022 & -0.006 \\
\hline
\end{tabular}
\end{table}

\clearpage

\section{Computational timings}

\begin{table}[htb]
\centering
\caption{Computational timings for orbital-optimized PBE calculations of the mixed-spin and triplet excited states of water using d-aug-cc-pVDZ+sz (d-aug) and plane waves (PWs). All reported calculations used 28 cores. The d-aug M$_2$ results are omitted because that calculation used a different number of cores. $N_{\mathrm{it}}$ is the number of outer optimization iterations, $T_{\mathrm{opt}}$ is the elapsed time between the first and last recorded outer iteration, and $T_{\mathrm{avg}}$ is the average elapsed time between consecutive outer iterations. Both $T_{\mathrm{opt}}$ and $T_{\mathrm{avg}}$ are in seconds. $T_{\mathrm{opt}}$ excludes initialization and operations outside the recorded iteration interval.}
\label{tab:timing_comparison}
\begin{tabular}{lccc|ccc}
\hline
& \multicolumn{3}{c|}{d-aug} & \multicolumn{3}{c}{PWs} \\
\cline{2-7}
State
& $N_{\mathrm{it}}$
& $T_{\mathrm{opt}}$ (s)
& $T_{\mathrm{avg}}$ (s)
& $N_{\mathrm{it}}$
& $T_{\mathrm{opt}}$ (s)
& $T_{\mathrm{avg}}$ (s) \\
\hline
M$_0$ & 26 & 170 &  6.8 &  31 &  314 & 10.5 \\
M$_1$ & 25 & 159 &  6.6 &  23 &  259 & 11.8 \\
M$_2$ & -- & --  & --    &  13 &  190 & 15.8 \\
M$_3$ & 15 & 594 & 42.4 &  52 & 1129 & 22.1 \\
M$_4$ & 28 & 800 & 29.6 &  26 &  744 & 29.8 \\
T$_0$ & 15 & 103 &  7.4 &  13 &  135 & 11.2 \\
T$_1$ & 22 &  96 &  4.6 &  65 &  308 &  4.8 \\
T$_2$ & 31 & 650 & 21.7 &  74 &  203 &  2.8 \\
T$_3$ & 16 &  48 &  3.2 &  72 &  691 &  9.7 \\
T$_4$ & 34 & 690 & 20.9 & 151 & 2140 & 14.3 \\
\hline
\end{tabular}
\end{table}

\begin{table}[htbp]
\centering
\caption{Computational timings for orbital-optimized calculations of the mixed-spin and triplet excited states of water using PBE and PBE0 with plane waves (PWs). Only states for which both calculations used 28 cores are included. $N_{\mathrm{it}}$ is the number of outer optimization iterations, $T_{\mathrm{opt}}$ is the elapsed time between the first and last recorded outer iteration, and $T_{\mathrm{avg}}$ is the average elapsed time between consecutive outer iterations. Both $T_{\mathrm{opt}}$ and $T_{\mathrm{avg}}$ are in seconds. $T_{\mathrm{opt}}$ excludes initialization and operations outside the recorded iteration interval.}
\label{tab:pbe_pbe0_pw_timings}
\begin{tabular}{lccc|ccc}
\hline
& \multicolumn{3}{c|}{PBE} & \multicolumn{3}{c}{PBE0} \\
\cline{2-7}
State
& $N_{\mathrm{it}}$
& $T_{\mathrm{opt}}$ (s)
& $T_{\mathrm{avg}}$ (s)
& $N_{\mathrm{it}}$
& $T_{\mathrm{opt}}$ (s)
& $T_{\mathrm{avg}}$ (s) \\
\hline
M$_1$ & 23 & 259  & 11.77 & 22 & 1272 & 60.57 \\
M$_2$ & 13 & 190  & 15.83 & 13 &  795 & 66.25 \\
M$_4$ & 26 & 744  & 29.76 & 28 & 3210 & 118.89 \\
T$_0$ & 13 & 135  & 11.25 & 13 &  822 & 68.50 \\
T$_1$ & 65 & 308  &  4.81 & 21 & 1261 & 63.05 \\
T$_2$ & 74 & 203  &  2.78 & 13 &  809 & 67.42 \\
T$_3$ & 72 & 691  &  9.73 & 23 & 2040 & 92.73 \\
\hline
\end{tabular}
\end{table}

\clearpage

\section{Water}
\subsection{Molecular orbitals}
\begin{figure}[hbt]
    \centering
    \includegraphics[width=0.5\linewidth]{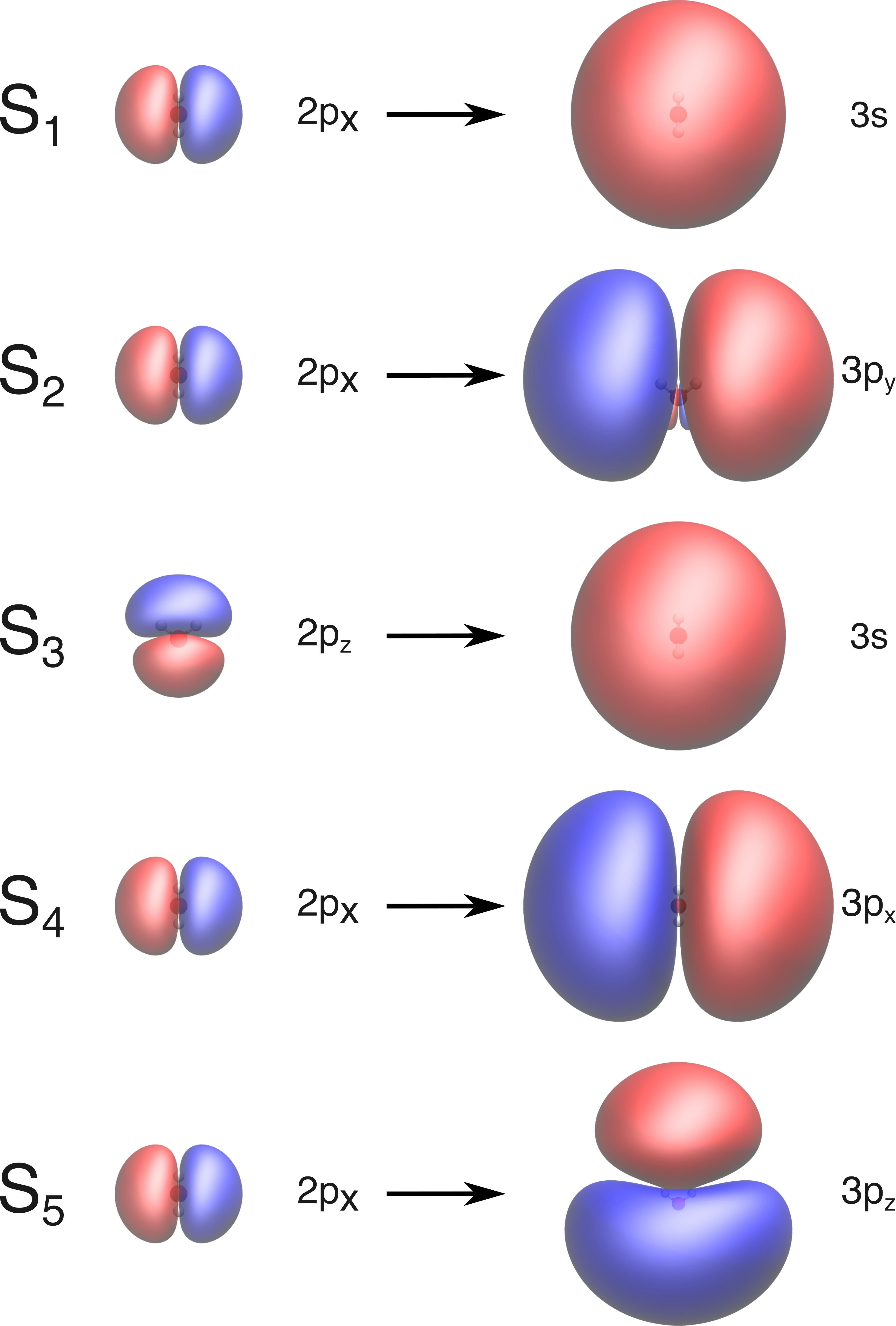}
    \caption{Molecular orbitals associated with the first five singlet Rydberg excitations of water at the Franck–Condon geometry. The orbitals shown on the left are obtained from a ground-state calculation at the PBE/\gls{pw} level of theory, while those on the right are obtained from orbital-optimized PBE/\gls{pw} calculations of the corresponding excited state.}
    \label{water_MOs}
\end{figure}

\subsection{Tables of excitation energy, dipole moment, and variance}
\begin{sidewaystable}[hbt]
\centering
\caption{Excitation energy, permanent dipole moment, and variance of the spin-purified singlet excited states of water computed using different basis set representations and exchange-correlation functionals: excitation energy, $\Delta E$ (eV), permanent dipole moment along the $z$ direction, $\mu _z$ (Debye), and variance of the position operator, $\sigma(\bm{r})$ (bohr$^2$).}
\label{tab:si_pure_singlets_dm_var}
\begin{tabular}{ll ccc ccc ccc}
\hline
& &
\multicolumn{3}{c}{\begin{tabular}{c} S$_1$ $B_1$ \\ 2p$_x \rightarrow$ 3s \end{tabular}}
&
\multicolumn{3}{c}{\begin{tabular}{c} S$_2$ $A_2$\\ 2p$_x \rightarrow$ 3p$_y$ \end{tabular}}
&
\multicolumn{3}{c}{\begin{tabular}{c} S$_3$ $A_1$\\ 2p$_z \rightarrow$ 3s \end{tabular}} \\
\cline{3-5}
\cline{6-8}
\cline{9-11}
Basis set & XC
& $\Delta E$ & $\mu _z$ & $\sigma(\bm{r})$
& $\Delta E$ & $\mu _z$ & $\sigma(\bm{r})$
& $\Delta E$ & $\mu _z$ & $\sigma(\bm{r})$ \\
\hline
plane waves      & PBE0      & 7.37 & -1.80 & 26.6 & 8.89 & -1.77 & 37.8 & 9.76 & -1.38 & 29.3 \\
plane waves      & PBE-SIC/2 & 7.38 & -1.87 & 27.1 & 9.03 & -1.67 & 40.1 & 9.69 & -1.48 & 29.9 \\
plane waves      & PBE-SIC   & 7.38 & -2.11 & 27.5 & 9.15 & -1.58 & 44.0 & 9.70 & -1.37 & 31.6 \\
Reference$^\dagger$ &        & $7.71\pm0.02$ & $-1.77\pm0.04$ & & $9.49\pm0.02$ & $-1.67\pm0.01$ & & $9.99\pm0.01$ & $-1.30\pm0.03$ & \\
\hline

\\[-1.4ex]

\hline
& &
\multicolumn{3}{c}{\begin{tabular}{c} S$_4$ $A_1$\\ 2p$_x \rightarrow$ 3p$_x$ \end{tabular}}
&
\multicolumn{3}{c}{\begin{tabular}{c} S$_5$ $B_1$\\ 2p$_x \rightarrow$ 3p$_z$ \end{tabular}}
&
\multicolumn{3}{c}{} \\
\cline{3-5}
\cline{6-8}
Basis set & XC
& $\Delta E$ & $\mu _z$ & $\sigma(\bm{r})$
& $\Delta E$ & $\mu _z$ & $\sigma(\bm{r})$
& \multicolumn{3}{c}{} \\
\cline{1-8}
plane waves      & PBE0      & 9.82 & -1.33 & 62.6 & 9.72 & 3.21 & 58.9 & \multicolumn{3}{c}{} \\
plane waves      & PBE-SIC/2 & 9.88 & -1.31 & 65.2 & 9.78 & 3.47 & 61.4 & \multicolumn{3}{c}{} \\
plane waves      & PBE-SIC   & 9.89 & -1.47 & 75.1 & 9.80 & 4.62 & 65.5 & \multicolumn{3}{c}{} \\
Reference$^\dagger$ &        & & & & & & & \multicolumn{3}{c}{} \\
\cline{1-8}
\multicolumn{11}{l}{\footnotesize{$^\dagger$CBS/TBE from ref.~\cite{Chrayteh2021-ib}.}} \\
\end{tabular}
\end{sidewaystable}

\begin{table}[hbt]
\centering
\caption{Excitation energy, permanent dipole moment, and variance of the mixed-spin excited states of water computed using different basis set representations and exchange--correlation functionals: excitation energy, $\Delta E$ (eV), permanent dipole moment along the $z$ direction, $\mu _z$ (Debye), and variance of the position operator, $\sigma(\bm{r})$ (bohr$^2$).}
\label{tab:si_mixed_spin_states_dm_var}

\begin{tabular}{ll ccc ccc ccc}
\hline
& &
\multicolumn{3}{c}{\begin{tabular}{c} M$_0$ \\ 2p$_x \rightarrow$ 3s \end{tabular}}
&
\multicolumn{3}{c}{\begin{tabular}{c} M$_1$ \\ 2p$_x \rightarrow$ 3p$_y$ \end{tabular}}
&
\multicolumn{3}{c}{\begin{tabular}{c} M$_2$ \\ 2p$_z \rightarrow$ 3s \end{tabular}} \\
\cline{3-5}
\cline{6-8}
\cline{9-11}
Basis set & XC
& $\Delta E$ & $\mu _z$ & $\sigma(\bm{r})$
& $\Delta E$ & $\mu _z$ & $\sigma(\bm{r})$
& $\Delta E$ & $\mu _z$ & $\sigma(\bm{r})$ \\
\hline
aug-cc-pVDZ+sz   & PBE       & 7.29 & -1.19 & 29.0 & 8.87 & -1.05 & 35.9 & 9.57 & -1.34 & 26.8 \\
d-aug-cc-pVDZ+sz & PBE       & 7.29 & -1.39 & 25.9 & 8.86 & -1.67 & 36.2 & 9.57 & -1.55 & 26.8 \\
plane waves      & PBE       & 7.26 & -1.44 & 25.8 & 8.82 & -1.68 & 36.0 & 9.50 & -1.62 & 26.8 \\
plane waves      & PBE0      & 7.19 & -1.61 & 25.7 & 8.81 & -1.71 & 36.6 & 9.50 & -1.63 & 27.1 \\
plane waves      & PBE-SIC/2 & 7.23 & -1.72 & 26.3 & 8.94 & -1.63 & 38.8 & 9.47 & -1.74 & 27.9 \\
plane waves      & PBE-SIC   & 7.22 & -1.99 & 26.7 & 9.04 & -1.55 & 42.3 & 9.45 & -1.85 & 29.0 \\
\hline

\\[-1.4ex]

\hline
& &
\multicolumn{3}{c}{\begin{tabular}{c} M$_3$ \\ 2p$_x \rightarrow$ 3p$_x$ \end{tabular}}
&
\multicolumn{3}{c}{\begin{tabular}{c} M$_4$ \\ 2p$_x \rightarrow$ 3p$_z$ \end{tabular}}
&
\multicolumn{3}{c}{} \\
\cline{3-5}
\cline{6-8}
Basis set & XC
& $\Delta E$ & $\mu _z$ & $\sigma(\bm{r})$
& $\Delta E$ & $\mu _z$ & $\sigma(\bm{r})$
& \multicolumn{3}{c}{} \\
\cline{1-8}
aug-cc-pVDZ+sz   & PBE       & 10.96 &  2.93 & 30.0 & 11.44 & -1.32 & 36.9 & \multicolumn{3}{c}{} \\
d-aug-cc-pVDZ+sz & PBE       & 9.77  &  0.08 & 53.1 & 9.87  &  3.48 & 54.5 & \multicolumn{3}{c}{} \\
plane waves      & PBE       & 9.72  & -0.93 & 54.7 & 9.80  &  2.44 & 57.1 & \multicolumn{3}{c}{} \\
plane waves      & PBE0      & 9.65  & -1.04 & 57.5 & 9.69  &  3.05 & 58.2 & \multicolumn{3}{c}{} \\
plane waves      & PBE-SIC/2 & 9.72  & -1.01 & 59.9 & 9.75  &  3.45 & 60.6 & \multicolumn{3}{c}{} \\
plane waves      & PBE-SIC   & 9.71  & -1.09 & 66.7 & 9.74  &  4.90 & 64.3 & \multicolumn{3}{c}{} \\
\cline{1-8}
\end{tabular}
\end{table}

\begin{table}[hbt]
\centering
\caption{Properties of the triplet states of water computed using different basis set representations and exchange--correlation functionals: excitation energy, $\Delta E$ (eV), permanent dipole moment along the $z$ direction, $\mu _z$ (Debye), and variance of the position operator, $\sigma(\bm{r})$ (bohr$^2$).}
\label{tab:si_triplet_states}

\begin{tabular}{ll ccc  ccc  ccc}
\hline
& &
\multicolumn{3}{c}{\begin{tabular}{c} T$_0$ \\ 2p$_x \rightarrow$ 3s \end{tabular}}
&
\multicolumn{3}{c}{\begin{tabular}{c} T$_1$ \\ 2p$_x \rightarrow$ 3p$_y$ \end{tabular}}
&
\multicolumn{3}{c}{\begin{tabular}{c} T$_2$ \\ 2p$_z \rightarrow$ 3s \end{tabular}} \\
\cline{3-5}
\cline{6-8}
\cline{9-11}
Basis set & XC
& $\Delta E$ & $\mu _z$ & $\sigma(\bm{r})$
& $\Delta E$ & $\mu _z$ & $\sigma(\bm{r})$
& $\Delta E$ & $\mu _z$ & $\sigma(\bm{r})$ \\
\hline
aug-cc-pVDZ+sz   & PBE       & 7.11 & -1.06 & 25.1 & 8.80 & -1.03 & 38.3 & 9.34 & -1.39 & 25.4 \\
d-aug-cc-pVDZ+sz & PBE       & 7.11 & -1.21 & 25.1 & 8.79 & -1.61 & 35.2 & 9.35 & -1.61 & 25.3 \\
plane waves      & PBE       & 7.08 & -1.26 & 25.0 & 8.74 & -1.62 & 35.0 & 9.28 & -1.68 & 25.2 \\
plane waves      & PBE0      & 7.00 & -1.42 & 24.8 & 8.74 & -1.65 & 35.5 & 9.24 & -1.87 & 25.0 \\
plane waves      & PBE-SIC/2 & 7.07 & -1.58 & 25.5 & 8.85 & -1.58 & 37.5 & 9.24 & -2.01 & 25.9 \\
plane waves      & PBE-SIC   & 7.06 & -1.87 & 25.9 & 8.93 & -1.53 & 40.6 & 9.21 & -2.32 & 26.4 \\
\hline

\\[-1.4ex]

\multicolumn{8}{l}{} \\
\cline{1-8}
& &
\multicolumn{3}{c}{\begin{tabular}{c} T$_3$ \\ 2p$_x \rightarrow$ 3p$_x$ \end{tabular}}
&
\multicolumn{3}{c}{\begin{tabular}{c} T$_4$ \\ 2p$_x \rightarrow$ 3p$_z$ \end{tabular}}
&
\multicolumn{3}{c}{} \\
\cline{3-5}
\cline{6-8}
Basis set & XC
& $\Delta E$ & $\mu _z$ & $\sigma(\bm{r})$
& $\Delta E$ & $\mu _z$ & $\sigma(\bm{r})$
& \multicolumn{3}{c}{} \\
\cline{1-8}
aug-cc-pVDZ+sz   & PBE       & 10.67 &  2.73 & 29.5 & 11.33 & -1.23 & 36.6 & \multicolumn{3}{c}{} \\
d-aug-cc-pVDZ+sz & PBE       & 9.35  &  0.14 & 50.1 & 9.84  &  3.33 & 53.9 & \multicolumn{3}{c}{} \\
plane waves      & PBE       & 9.59  & -0.72 & 51.5 & 9.77  &  2.32 & 56.5 & \multicolumn{3}{c}{} \\
plane waves      & PBE0      & 9.49  & -0.75 & 52.4 & 9.67  &  2.88 & 57.5 & \multicolumn{3}{c}{} \\
plane waves      & PBE-SIC/2 & 9.57  & -0.71 & 54.6 & 9.73  &  3.43 & 59.8 & \multicolumn{3}{c}{} \\
plane waves      & PBE-SIC   & 9.52  & -0.71 & 58.2 & 9.67  &  5.19 & 63.0 & \multicolumn{3}{c}{} \\
\cline{1-8}
\end{tabular}

\end{table}

\clearpage
\newpage

\subsection{Bar plots of excitation energy and variance}
\begin{figure}[hbt]
    \centering
    \includegraphics[width=\linewidth]{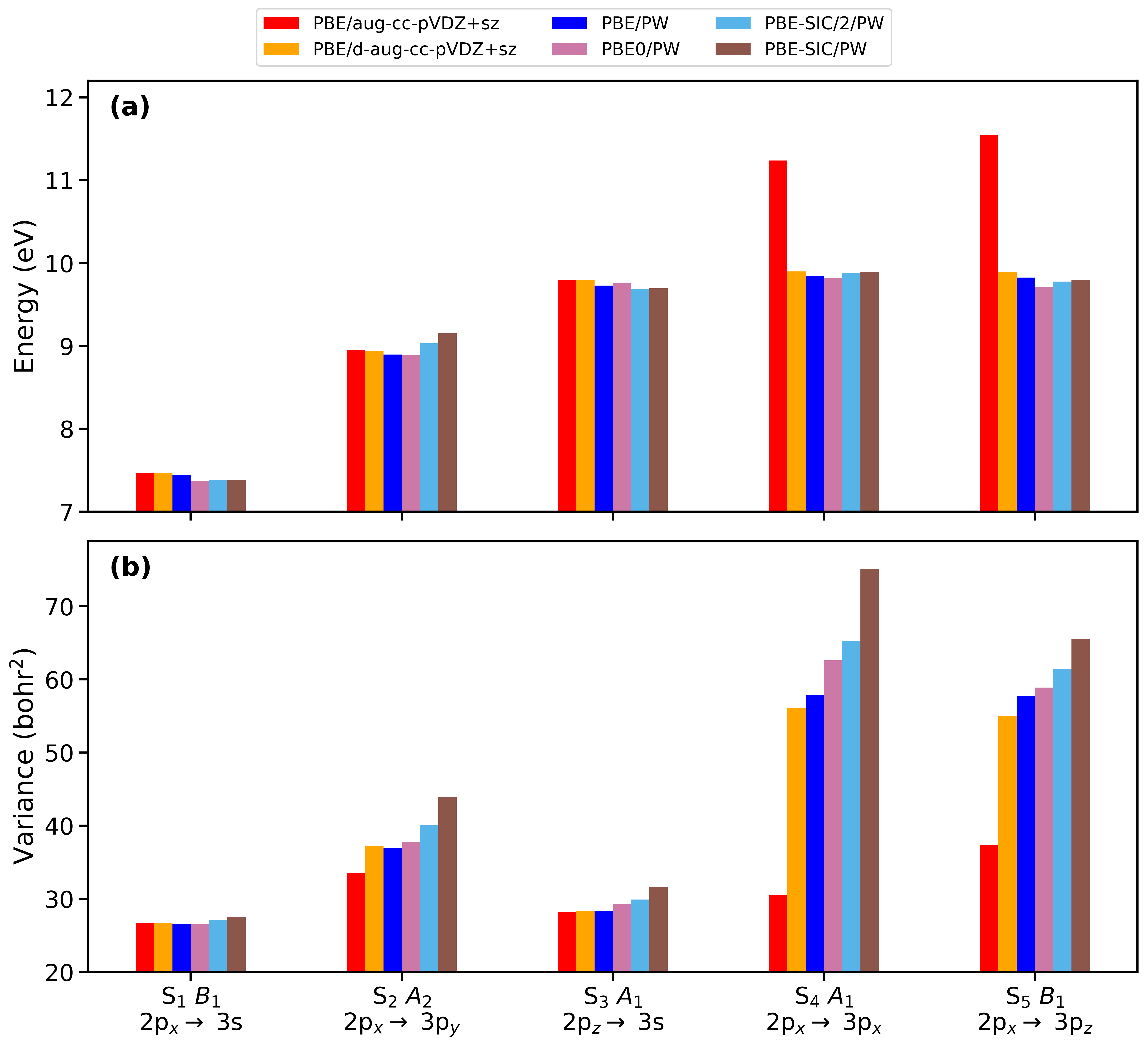}
    \caption{Basis set and exchange-correlation functional dependence of (a) the vertical excitation energy, and (b) the variance of the electronic position operator for the first five singlet excited states of water.}
    \label{water_basis_effect_E}
\end{figure}

\clearpage
\newpage

\section{Formaldehyde}
\subsection{Molecular orbitals}
\begin{figure}[hbt]
    \centering
    \includegraphics[width=0.5\linewidth]{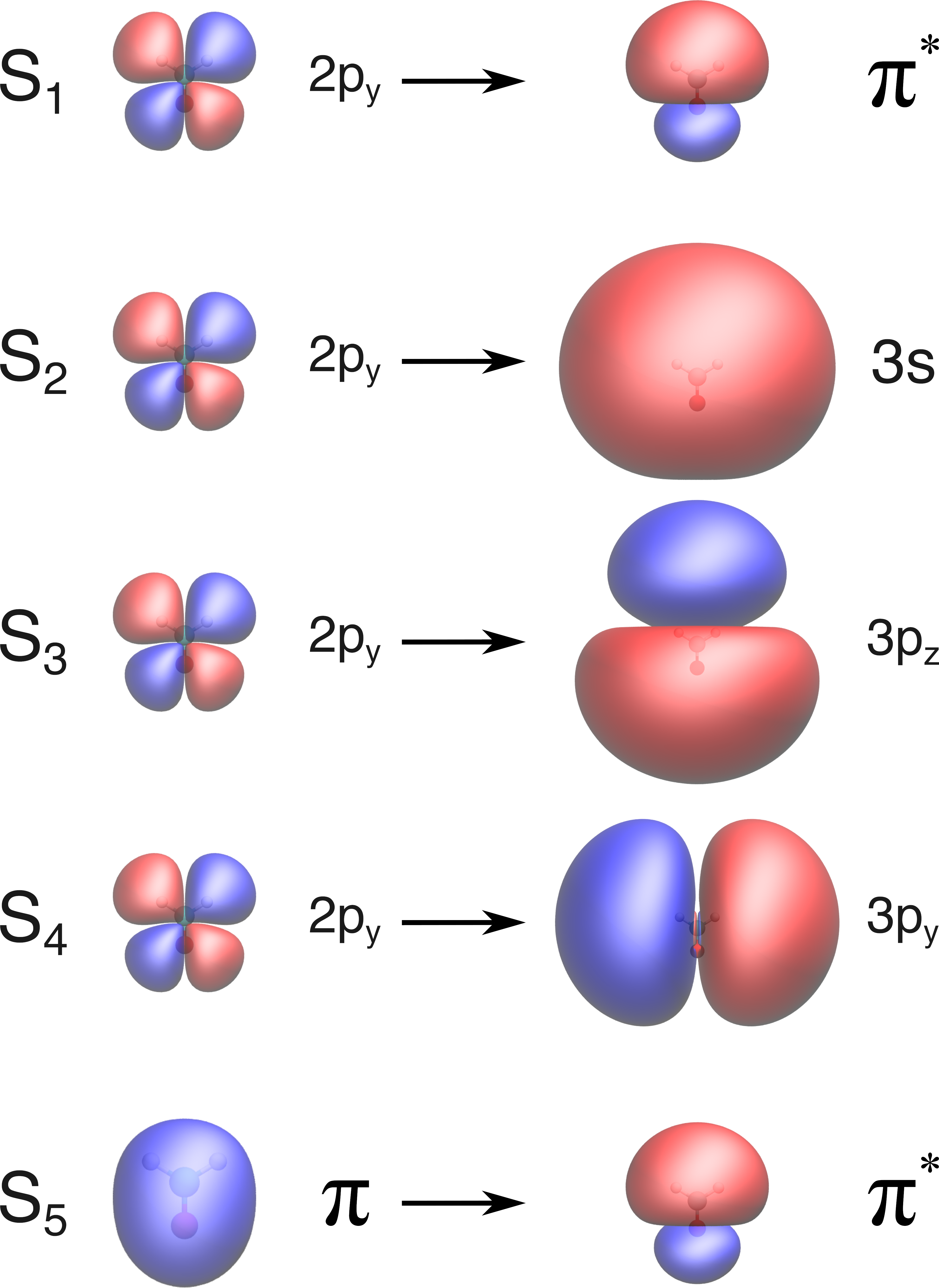}
    \caption{Molecular orbitals associated with the first five singlet excitations of formaldehyde at the Franck–Condon geometry. The orbitals shown on the left are obtained from a ground-state calculation at the PBE/\gls{pw} level of theory, while those on the right are obtained from orbital-optimized PBE/\gls{pw} calculations of the corresponding excited state.}
    \label{formaldehyde_MOs}
\end{figure}

\subsection{Tables of excitation energy, dipole moment, and variance}
\begin{sidewaystable}[hbt]
\centering
\caption{Excitation energy, permanent dipole moment, and variance of the spin-purified singlet excited states of formaldehyde computed using different basis set representations and exchange--correlation functionals: excitation energy, $\Delta E$ (eV), permanent dipole moment along the $z$ direction, $\mu _z$ (Debye), and variance of the position operator, $\sigma(\bm{r})$ (bohr$^2$).}
\label{tab:si_formaldehyde_pure_singlets_dm_var}

\begin{tabular}{ll ccc ccc ccc}
\hline
& &
\multicolumn{3}{c}{\begin{tabular}{c} S$_1$ $A_2$\\ 2p$_y \rightarrow \pi ^*$ \end{tabular}}
&
\multicolumn{3}{c}{\begin{tabular}{c} S$_2$ $B_2$\\ 2p$_y \rightarrow$ 3s \end{tabular}}
&
\multicolumn{3}{c}{\begin{tabular}{c} S$_3$ $B_2$\\ 2p$_y \rightarrow$ 3p$_z$ \end{tabular}} \\
\cline{3-5}
\cline{6-8}
\cline{9-11}
Basis set & XC
& $\Delta E$ & $\mu _z$ & $\sigma(\bm{r})$
& $\Delta E$ & $\mu _z$ & $\sigma(\bm{r})$
& $\Delta E$ & $\mu _z$ & $\sigma(\bm{r})$ \\
\hline
plane waves      & PBE0      & 3.46 & 1.36 & 23.0 & 6.98 & -2.78 & 52.6 & 7.73 & 0.96 & 63.5 \\
plane waves      & PBE-SIC/2 & 3.38 & 1.63 & 22.8 & 7.08 & -2.86 & 54.4 & 7.82 & 1.02 & 65.8 \\
plane waves      & PBE-SIC   & 3.25 & 1.83 & 22.0 & 7.14 & -3.22 & 56.4 & 7.94 & 1.21 & 69.8 \\
Reference$^\dagger$   &      & $3.99 \pm 0.01$ & $1.36 \pm 0.01$ & & $7.34\pm 0.01$ & $-2.15 \pm 0.03$ & & $8.16\pm 0.02$ & $0.21 \pm 0.20$ & \\
\hline

\\[-1.4ex]

\hline
& &
\multicolumn{3}{c}{\begin{tabular}{c} S$_4$ $A_1$\\ 2p$_y \rightarrow$ 3p$_y$ \end{tabular}}
&
\multicolumn{3}{c}{\begin{tabular}{c} S$_5$ $A_1$ \\ $\pi \rightarrow \pi ^*$ \end{tabular}}
&
\multicolumn{3}{c}{} \\
\cline{3-5}
\cline{6-8}
Basis set & XC
& $\Delta E$ & $\mu _z$ & $\sigma(\bm{r})$
& $\Delta E$ & $\mu _z$ & $\sigma(\bm{r})$
& \multicolumn{3}{c}{} \\
\cline{1-8}
plane waves      & PBE0      & 7.81 & -0.72 & 69.7 & 9.40  & 1.62 & 28.0 & \multicolumn{3}{c}{} \\
plane waves      & PBE-SIC/2 & 7.90 & -1.15 & 70.4 & 9.47  & 1.81 & 27.3 & \multicolumn{3}{c}{} \\
plane waves      & PBE-SIC   & 8.00 & -1.28 & 74.5 & 10.25 & 2.05 & 26.6 & \multicolumn{3}{c}{} \\
Reference$^\dagger$   &      & $8.28 \pm 0.04$ & $-0.69 \pm 0.43$ & & $9.52\pm 0.12 $ & $2.46 \pm 1.36$ & & \multicolumn{3}{c}{} \\
\hline
\multicolumn{8}{l}{\footnotesize{$^\dagger$CBS/TBE from Ref.~\cite{Chrayteh2021-ib}.}} \\
\end{tabular}

\end{sidewaystable}

\begin{table}
\centering
\caption{Excitation energy, permanent dipole moment, and variance of the mixed-spin excited states of formaldehyde computed using different basis set representations and exchange--correlation functionals: excitation energy, $\Delta E$ (eV), permanent dipole moment along the $z$ direction, $\mu _z$ (Debye), and variance of the position operator, $\sigma(\bm{r})$ (bohr$^2$).}
\label{tab:si_formaldehyde_mixed_spin_states_dm_var}

\begin{tabular}{ll ccc ccc ccc}
\hline
& &
\multicolumn{3}{c}{\begin{tabular}{c} M$_0$ \\ 2p$_y \rightarrow \pi ^*$ \end{tabular}}
&
\multicolumn{3}{c}{\begin{tabular}{c} M$_1$ \\ 2p$_y \rightarrow$ 3s \end{tabular}}
&
\multicolumn{3}{c}{\begin{tabular}{c} M$_2$ \\ 2p$_y \rightarrow$ 3p$_z$ \end{tabular}} \\
\cline{3-5}
\cline{6-8}
\cline{9-11}
Basis set & XC
& $\Delta E$ & $\mu _z$ & $\sigma(\bm{r})$
& $\Delta E$ & $\mu _z$ & $\sigma(\bm{r})$
& $\Delta E$ & $\mu _z$ & $\sigma(\bm{r})$ \\
\hline
aug-cc-pVDZ+sz   & PBE       & 3.42 &  1.34 & 24.4 & 6.81 & -2.71 & 49.1 & 7.66 &  2.09 & 49.8 \\
d-aug-cc-pVDZ+sz & PBE       & 3.42 &  1.32 & 24.4 & 6.80 & -2.46 & 50.9 & 7.54 &  2.06 & 60.7 \\
plane waves      & PBE       & 3.39 &  1.30 & 24.2 & 6.78 & -2.39 & 50.9 & 7.51 &  1.75 & 60.5 \\
plane waves      & PBE0      & 3.31 &  1.30 & 23.2 & 6.90 & -2.76 & 51.0 & 7.66 &  1.40 & 61.7 \\
plane waves      & PBE-SIC/2 & 3.38 &  1.33 & 23.0 & 7.02 & -2.66 & 53.9 & 7.75 &  1.22 & 64.2 \\
plane waves      & PBE-SIC   & 3.35 &  1.38 & 22.2 & 7.11 & -2.90 & 56.5 & 7.87 &  0.63 & 68.4 \\
\hline

\\[-1.4ex]

\hline
& &
\multicolumn{3}{c}{\begin{tabular}{c} M$_3$ \\ 2p$_y \rightarrow$ 3p$_y$ \end{tabular}}
&
\multicolumn{3}{c}{\begin{tabular}{c} M$_4$ \\ $\pi \rightarrow \pi ^*$ \end{tabular}}
&
\multicolumn{3}{c}{} \\
\cline{3-5}
\cline{6-8}
Basis set & XC
& $\Delta E$ & $\mu _z$ & $\sigma(\bm{r})$
& $\Delta E$ & $\mu _z$ & $\sigma(\bm{r})$
& \multicolumn{3}{c}{} \\
\cline{1-8}
aug-cc-pVDZ+sz   & PBE       & 7.64 & -2.13 & 60.1 & 7.38 & 0.96 & 25.2 & \multicolumn{3}{c}{} \\
d-aug-cc-pVDZ+sz & PBE       & 7.62 & -1.12 & 66.2 & 7.38 & 0.95 & 25.2 & \multicolumn{3}{c}{} \\
plane waves      & PBE       & 7.60 & -1.15 & 65.8 & 7.34 & 0.94 & 25.0 & \multicolumn{3}{c}{} \\
plane waves      & PBE0      & 7.75 & -0.97 & 67.7 & 7.56 & 1.17 & 24.8 & \multicolumn{3}{c}{} \\
plane waves      & PBE-SIC/2 & 7.87 & -1.11 & 70.0 & 7.70 & 1.29 & 24.4 & \multicolumn{3}{c}{} \\
plane waves      & PBE-SIC   & 7.99 & -1.09 & 75.0 & 8.00 & 1.47 & 23.8 & \multicolumn{3}{c}{} \\
\cline{1-8}
\end{tabular}
\end{table}

\begin{table}
\centering
\caption{Properties of the triplet states of formaldehyde computed using different basis set representations and exchange--correlation functionals: excitation energy, $\Delta E$ (eV), permanent dipole moment along the $z$ direction, $\mu _z$ (Debye), and variance of the position operator, $\sigma(\bm{r})$ (bohr$^2$).}
\label{tab:si_formaldehyde_triplet_states}

\begin{tabular}{ll ccc  ccc  ccc}
\hline
& &
\multicolumn{3}{c}{\begin{tabular}{c} T$_0$ \\ 2p$_y \rightarrow \pi ^*$ \end{tabular}}
&
\multicolumn{3}{c}{\begin{tabular}{c} T$_1$ \\ 2p$_y \rightarrow$ 3s \end{tabular}}
&
\multicolumn{3}{c}{\begin{tabular}{c} T$_2$ \\ 2p$_y \rightarrow$ 3p$_z$ \end{tabular}} \\
\cline{3-5}
\cline{6-8}
\cline{9-11}
Basis set & XC
& $\Delta E$ & $\mu _z$ & $\sigma(\bm{r})$
& $\Delta E$ & $\mu _z$ & $\sigma(\bm{r})$
& $\Delta E$ & $\mu _z$ & $\sigma(\bm{r})$ \\
\hline
aug-cc-pVDZ+sz   & PBE       & 3.26 &  1.29 & 24.5 & 6.71 & -2.65 & 47.7 & 7.58 & 2.41 & 48.7 \\
d-aug-cc-pVDZ+sz & PBE       & 3.26 &  1.27 & 24.5 & 6.70 & -2.41 & 49.3 & 7.46 & 2.42 & 58.8 \\
plane waves      & PBE       & 3.23 &  1.25 & 24.3 & 6.68 & -2.35 & 49.3 & 7.43 & 2.17 & 58.7 \\
plane waves      & PBE0      & 3.16 &  1.24 & 23.5 & 6.83 & -2.74 & 49.5 & 7.59 & 1.84 & 59.8 \\
plane waves      & PBE-SIC/2 & 3.39 &  1.03 & 23.2 & 6.97 & -2.47 & 53.4 & 7.68 & 1.41 & 62.6 \\
plane waves      & PBE-SIC   & 3.44 &  0.92 & 22.5 & 7.08 & -2.57 & 56.5 & 7.80 & 0.06 & 66.9 \\
\hline

\\[-1.4ex]

\multicolumn{8}{l}{} \\
\cline{1-8}
& &
\multicolumn{3}{c}{\begin{tabular}{c} T$_3$ \\ 2p$_y \rightarrow$ 3p$_y$ \end{tabular}}
&
\multicolumn{3}{c}{\begin{tabular}{c} T$_4$ \\ $\pi \rightarrow \pi ^*$ \end{tabular}}
&
\multicolumn{3}{c}{} \\
\cline{3-5}
\cline{6-8}
Basis set & XC
& $\Delta E$ & $\mu _z$ & $\sigma(\bm{r})$
& $\Delta E$ & $\mu _z$ & $\sigma(\bm{r})$
& \multicolumn{3}{c}{} \\
\cline{1-8}
aug-cc-pVDZ+sz   & PBE       & 7.57 & -2.10 & 58.5 & 6.18 & 0.66 & 22.3 & \multicolumn{3}{c}{} \\
d-aug-cc-pVDZ+sz & PBE       & 7.55 & -1.24 & 64.1 & 6.18 & 0.64 & 22.3 & \multicolumn{3}{c}{} \\
plane waves      & PBE       & 7.53 & -1.27 & 63.7 & 6.13 & 0.63 & 22.1 & \multicolumn{3}{c}{} \\
plane waves      & PBE0      & 7.70 & -1.23 & 65.7 & 5.72 & 0.72 & 21.6 & \multicolumn{3}{c}{} \\
plane waves      & PBE-SIC/2 & 7.83 & -1.07 & 69.5 & 5.92 & 0.78 & 21.5 & \multicolumn{3}{c}{} \\
plane waves      & PBE-SIC   & 7.98 & -0.90 & 75.5 & 5.76 & 0.89 & 21.0 & \multicolumn{3}{c}{} \\
\cline{1-8}
\end{tabular}

\end{table}

\clearpage
\newpage

\subsection{Bar plots of excitation energy and variance}
\begin{figure}[hbt]
    \centering
    \includegraphics[width=\linewidth]{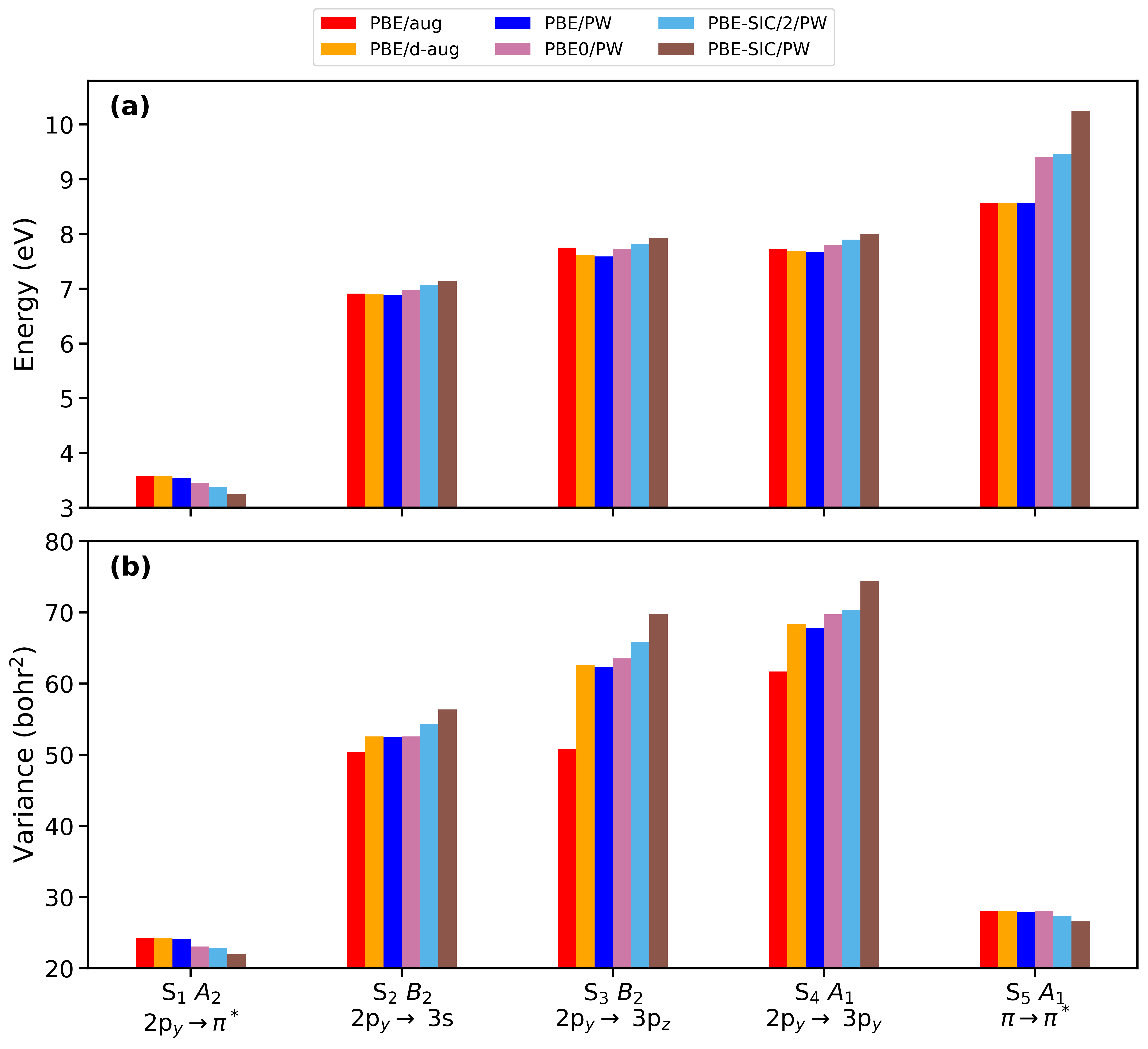}
    \caption{Basis set and exchange-correlation functional dependence of (a) the vertical excitation energy, and (b) the variance of the electronic position operator for the first five singlet excited states of formaldehyde.}
    \label{formaldehyde_basis_effect_E}
\end{figure}

\clearpage
\newpage

\section{Ammonia}
\subsection{Molecular orbitals}
\begin{figure}[hbt]
    \centering
    \includegraphics[width=0.5\linewidth]{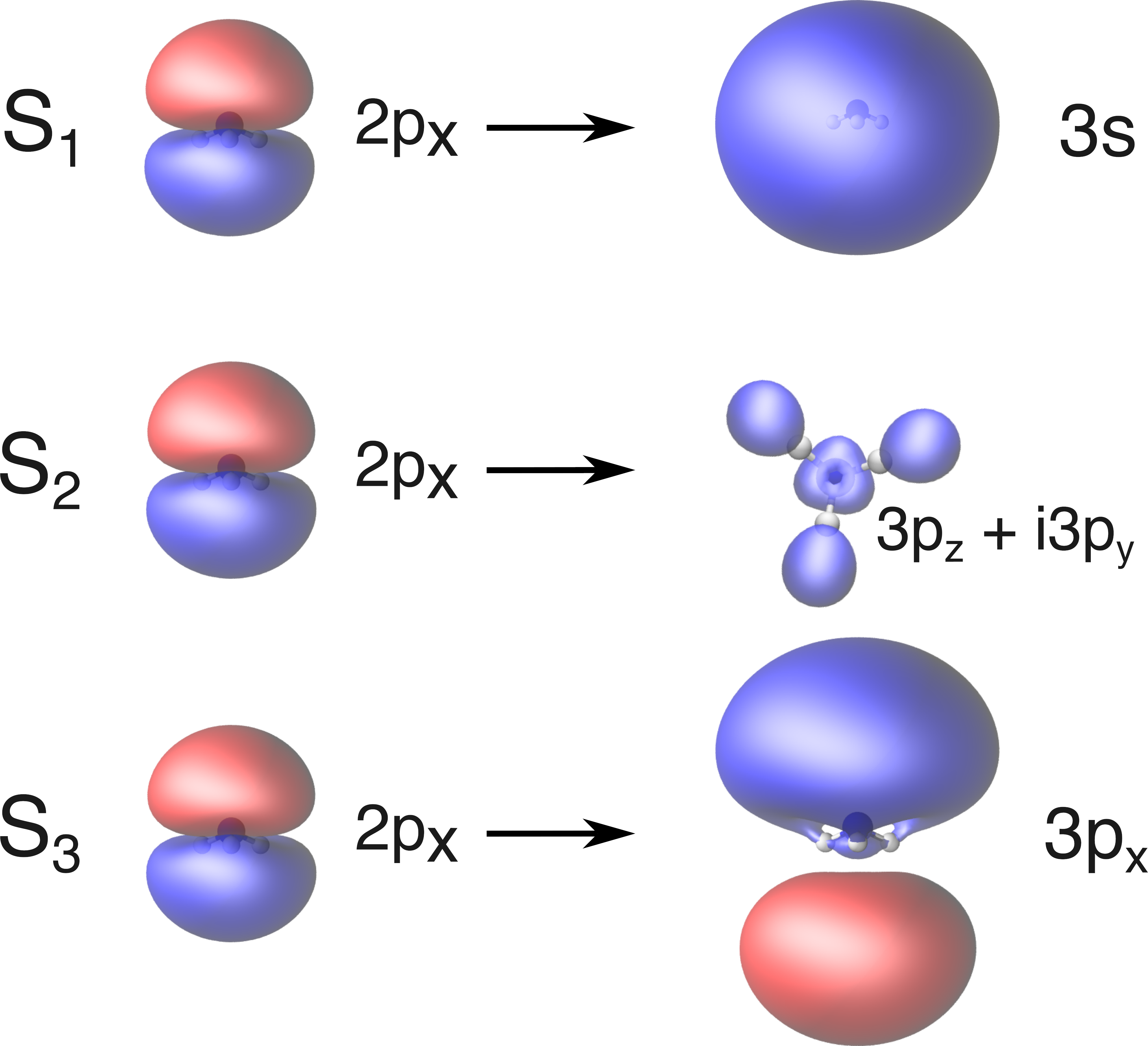}
    \caption{Molecular orbitals associated with the first three singlet excitations of ammonia at the Franck–Condon geometry. The orbitals shown on the left are obtained from a ground-state calculation at the PBE/\gls{pw} level of theory, while those on the right are obtained from orbital-optimized PBE/\gls{pw} calculations of the corresponding excited state. For S$_2$ with symmetry $E$, the orbital density of the linear combination $\mathrm{3p}_z + i3\mathrm{p}_y$ is shown instead.}
    \label{ammonia_MOs}
\end{figure}

\subsection{Tables of excitation energy, dipole moment, and variance}
\begin{table}[hbt]
\centering
\caption{Excitation energy, permanent dipole moment, and variance of the spin-purified singlet excited states of ammonia computed using different exchange--correlation functionals: excitation energy, $\Delta E$ (eV), permanent dipole moment along the $z$ direction, $\mu_z$ (Debye), and variance of the position operator, $\sigma(\bm{r})$ (bohr$^2$).}
\label{tab:si_ammonia_pure_singlets_dm_var}
\begin{tabular}{ll ccc ccc ccc}
\hline
& &
\multicolumn{3}{c}{\begin{tabular}{c} S$_1$ $A_1$\\ 2p$_z \rightarrow$ 3s \end{tabular}}
&
\multicolumn{3}{c}{\begin{tabular}{c} S$_2$ $E$\\ 2p$_z \rightarrow$ 3p$_+$ \end{tabular}}
&
\multicolumn{3}{c}{\begin{tabular}{c} S$_3$ $A_1$\\ 2p$_z \rightarrow$ 3p$_z$ \end{tabular}} \\
\cline{3-5}
\cline{6-8}
\cline{9-11}
Basis set & XC
& $\Delta E$ & $\mu_z$ & $\sigma(\bm{r})$
& $\Delta E$ & $\mu_z$ & $\sigma(\bm{r})$
& $\Delta E$ & $\mu_z$ & $\sigma(\bm{r})$ \\
\hline
plane waves      & PBE0      & 6.41 &  1.21 & 37.9 & 7.84 &  1.22 & 57.7 & 8.28 & -0.78 & 77.4 \\
plane waves      & PBE-SIC/2 & 6.37 & -1.30 & 39.5 &      &       &      &      &       &      \\
plane waves      & PBE-SIC   & 6.33 & -1.26 & 41.7 &      &       &      &      &       &      \\
Reference$^\dagger$ &        & 6.57 & -1.01 &      & 8.11 & -0.88 &      & 8.56 &  1.30 &      \\
\hline
\multicolumn{11}{l}{\footnotesize{$^\dagger$CCSDT/d-aug-cc-pVTZ values taken from Ref.~\cite{perscorr_jacquemin}.}}
\end{tabular}
\end{table}

\begin{table}
\centering
\caption{Excitation energy, permanent dipole moment, and variance of the mixed-spin excited states of ammonia computed using different basis set representations and exchange--correlation functionals: excitation energy, $\Delta E$ (eV), permanent dipole moment along the $z$ direction, $\mu_z$ (Debye), and variance of the position operator, $\sigma(\bm{r})$ (bohr$^2$).}
\label{tab:si_ammonia_mixed_spin_states_dm_var}

\begin{tabular}{ll ccc ccc ccc}
\hline
& &
\multicolumn{3}{c}{\begin{tabular}{c} M$_0$ \\ 2p$_z \rightarrow$ 3s \end{tabular}}
&
\multicolumn{3}{c}{\begin{tabular}{c} M$_1$ \\ 2p$_z \rightarrow$ 3p$_+$ \end{tabular}}
&
\multicolumn{3}{c}{\begin{tabular}{c} M$_2$ \\ 2p$_z \rightarrow$ 3p$_z$ \end{tabular}} \\
\cline{3-5}
\cline{6-8}
\cline{9-11}
Basis set & XC
& $\Delta E$ & $\mu_z$ & $\sigma(\bm{r})$
& $\Delta E$ & $\mu_z$ & $\sigma(\bm{r})$
& $\Delta E$ & $\mu_z$ & $\sigma(\bm{r})$ \\
\hline
aug-cc-pVDZ+sz   & PBE       & 6.31 &  1.21 & 35.1 & 7.82 &  0.93 & 49.2 & 8.95 & -4.68 & 39.0 \\
d-aug-cc-pVDZ+sz & PBE       & 6.31 &  1.23 & 35.9 & 7.79 &  1.29 & 52.6 & 8.18 & -1.71 & 65.8 \\
plane waves      & PBE       & 6.28 &  1.23 & 35.9 & 7.74 &  1.30 & 52.3 & 8.13 & -1.26 & 67.0 \\
plane waves      & PBE0      & 6.27 &  1.27 & 36.4 & 7.76 &  1.34 & 54.9 & 8.12 & -1.24 & 70.6 \\
plane waves      & PBE-SIC/2 & 6.23 & -1.36 & 37.8 & 7.70 & -1.30 & 57.7 & 8.08 &  1.10 & 73.6 \\
plane waves      & PBE-SIC   & 6.18 & -1.48 & 39.6 & 7.64 & -1.22 & 62.6 & 8.05 &  0.62 & 82.0 \\
\hline
\end{tabular}
\end{table}

\begin{table}
\centering
\caption{Excitation energy, permanent dipole moment, and variance of the triplet states of ammonia computed using different basis set representations and exchange--correlation functionals: excitation energy, $\Delta E$ (eV), permanent dipole moment along the $z$ direction, $\mu_z$ (Debye), and variance of the position operator, $\sigma(\bm{r})$ (bohr$^2$). Missing values indicate that the corresponding calculation did not converge.}
\label{tab:si_ammonia_triplet_states_dm_var}

\begin{tabular}{ll ccc ccc ccc}
\hline
& &
\multicolumn{3}{c}{\begin{tabular}{c} T$_0$ \\ 2p$_z \rightarrow$ 3s \end{tabular}}
&
\multicolumn{3}{c}{\begin{tabular}{c} T$_1$ \\ 2p$_z \rightarrow$ 3p$_+$ \end{tabular}}
&
\multicolumn{3}{c}{\begin{tabular}{c} T$_2$ \\ 2p$_z \rightarrow$ 3p$_z$ \end{tabular}} \\
\cline{3-5}
\cline{6-8}
\cline{9-11}
Basis set & XC
& $\Delta E$ & $\mu_z$ & $\sigma(\bm{r})$
& $\Delta E$ & $\mu_z$ & $\sigma(\bm{r})$
& $\Delta E$ & $\mu_z$ & $\sigma(\bm{r})$ \\
\hline
aug-cc-pVDZ+sz   & PBE       & 6.17 &  1.08 & 34.0 & 7.73 &  0.98 & 47.4 & 8.71 & -4.58 & 37.9 \\
d-aug-cc-pVDZ+sz & PBE       & 6.17 &  1.12 & 34.7 & 7.70 &  1.38 & 50.1 & 8.04 & -1.84 & 61.4 \\
plane waves      & PBE       & 6.14 &  1.14 & 34.6 & 7.66 &  1.38 & 49.9 & 8.00 & -1.44 & 62.4 \\
plane waves      & PBE0      & 6.12 &  1.32 & 34.8 & 7.68 &  1.47 & 52.0 & 7.96 & -1.69 & 63.7 \\
plane waves      & PBE-SIC/2 & 6.08 & -1.42 & 36.1 &       &       &      &      &       &      \\
plane waves      & PBE-SIC   & 6.03 & -1.69 & 37.5 &       &       &      &      &       &      \\
\hline
\end{tabular}
\end{table}

\clearpage
\newpage
\subsection{Bar plots of excitation energy and variance}
\begin{figure}[hbt]
    \centering
    \includegraphics[width=0.95\linewidth]{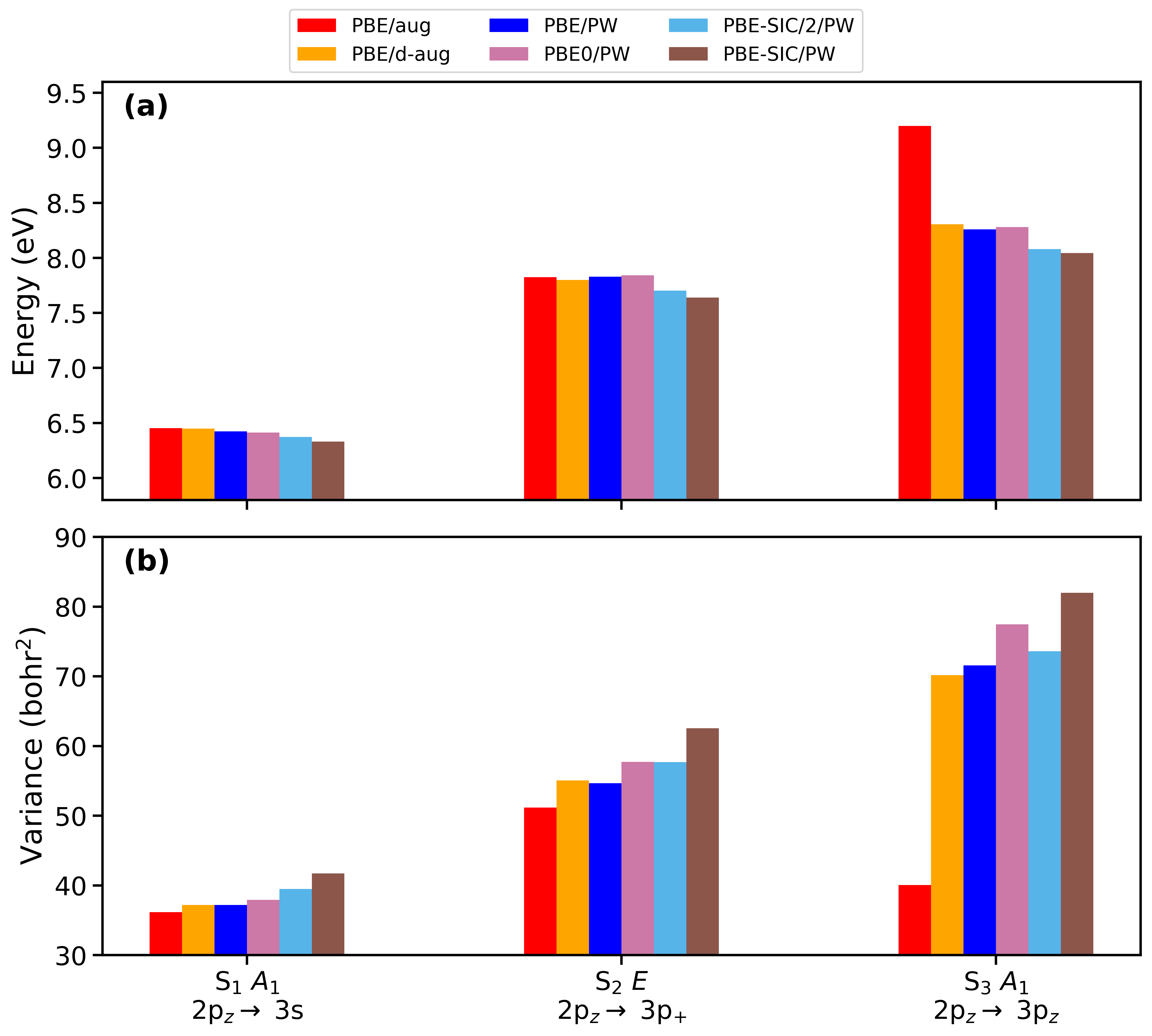}
    \caption{Basis set and exchange-correlation functional dependence of (a) the vertical excitation energy, and (b) the variance of the electronic position operator for the first three excited states of ammonia. For the PBE-SIC/2 and PBE-SIC functionals, the spin-purification formula could not be applied to S$_2$ and S$_3$ because the triplet calculations could not be converged to a solution with the correct spatial symmetry. Therefore, for these two functionals, mixed-spin values are shown instead.}
    \label{ammonia_basis_effect_E}
\end{figure}

\clearpage
\newpage

\section{Methanol}
\subsection{Molecular orbitals}
\begin{figure}[hbt]
    \centering
    \includegraphics[width=0.5\linewidth]{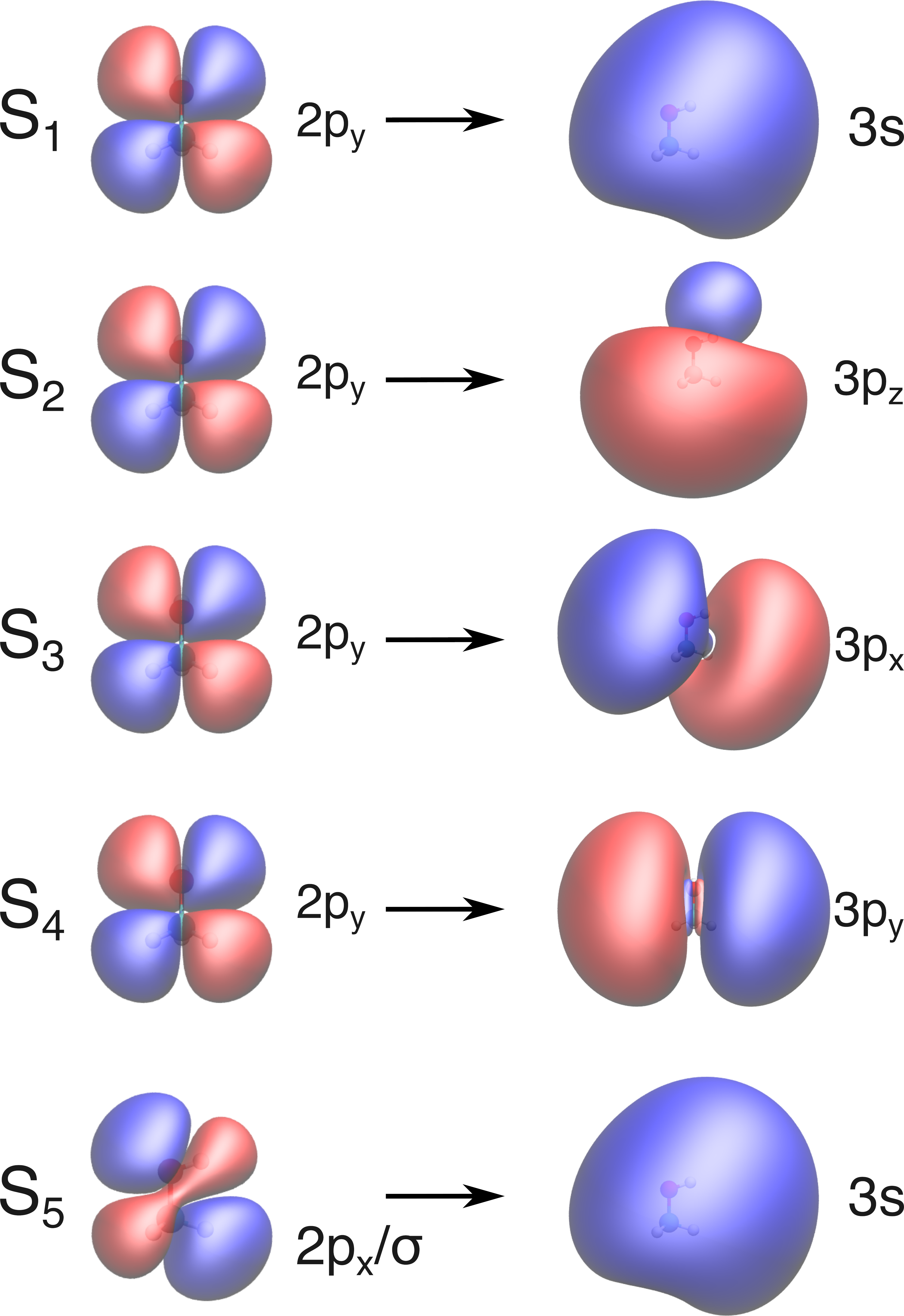}
    \caption{Molecular orbitals associated with the first five singlet excitations of methanol at the Franck–Condon geometry. The orbitals shown on the left are obtained from a ground-state calculation at the PBE/\gls{pw} level of theory, while those on the right are obtained from orbital-optimized PBE/\gls{pw} calculations of the corresponding excited state.}
    \label{ammonia_MOs}
\end{figure}

\begin{sidewaystable}[hbt]
\centering
\caption{Excitation energy, permanent dipole moment, and variance of the spin-purified singlet excited states of methanol computed using different exchange--correlation functionals: excitation energy, $\Delta E$ (eV), permanent dipole-moment components $\mu_x$ and $\mu_z$ (Debye), and variance of the position operator, $\sigma(\bm{r})$ (bohr$^2$).}
\label{tab:si_methanol_pure_singlets_dm_var}

\begin{tabular}{ll cccc cccc cccc}
\hline
& &
\multicolumn{4}{c}{\begin{tabular}{c} S$_1$ $A^{\prime\prime}$\\ 2p$_y \rightarrow$ 3s \end{tabular}}
&
\multicolumn{4}{c}{\begin{tabular}{c} S$_2$ $A^{\prime\prime}$\\ 2p$_y \rightarrow$ 3p$_z$ \end{tabular}}
&
\multicolumn{4}{c}{\begin{tabular}{c} S$_3$ $A^{\prime\prime}$\\ 2p$_y \rightarrow$ 3p$_x$ \end{tabular}} \\
\cline{3-6}
\cline{7-10}
\cline{11-14}
Basis set & XC
& $\Delta E$ & $\mu_x$ & $\mu_z$ & $\sigma(\bm{r})$
& $\Delta E$ & $\mu_x$ & $\mu_z$ & $\sigma(\bm{r})$
& $\Delta E$ & $\mu_x$ & $\mu_z$ & $\sigma(\bm{r})$ \\
\hline
plane waves      & PBE0      & 6.54 &  3.74 &  3.64 & 43.9 & 7.63 & -2.28 & -7.86 & 60.5 & 8.04 & -2.41 &  0.54 & 79.2 \\
plane waves      & PBE-SIC/2 & 6.66 &  3.86 &  3.68 & 43.8 & 7.92 & -2.80 & -7.36 & 61.6 & 8.23 & -2.54 &  0.86 & 82.0 \\
plane waves      & PBE-SIC   & 6.85 &  4.45 &  3.90 & 42.2 & 8.42 & -2.70 & -6.87 & 66.1 &       &       &       &      \\
\hline

\\[-1.4ex]

\hline
& &
\multicolumn{4}{c}{\begin{tabular}{c} S$_4$ $A^\prime$\\ 2p$_y \rightarrow$ 3p$_y$ \end{tabular}}
&
\multicolumn{4}{c}{\begin{tabular}{c} S$_5$ $A^\prime$\\ $2\mathrm{p}_x/\sigma \rightarrow$ 3s \end{tabular}}
&
\multicolumn{4}{c}{} \\
\cline{3-6}
\cline{7-10}
Basis set & XC
& $\Delta E$ & $\mu_x$ & $\mu_z$ & $\sigma(\bm{r})$
& $\Delta E$ & $\mu_x$ & $\mu_z$ & $\sigma(\bm{r})$
& \multicolumn{4}{c}{} \\
\cline{1-10}
plane waves      & PBE0      &       &       &       &      & 8.21 & 3.50 & 3.42 & 46.5 & \multicolumn{4}{c}{} \\
plane waves      & PBE-SIC/2 & 8.31 &  0.05 &  0.08 & 86.8 & 8.34 & 3.70 & 3.48 & 46.8 & \multicolumn{4}{c}{} \\
plane waves      & PBE-SIC   & 8.50 &  0.68 & -0.09 & 94.1 & 8.56 & 4.36 & 3.83 & 45.6 & \multicolumn{4}{c}{} \\
\cline{1-10}
\end{tabular}
\end{sidewaystable}

\begin{sidewaystable}[hbt]
\centering
\caption{Excitation energy, permanent dipole moment, and variance of the mixed-spin excited states of methanol computed using different basis set representations and exchange--correlation functionals: excitation energy, $\Delta E$ (eV), permanent dipole-moment components $\mu_x$ and $\mu_z$ (Debye), and variance of the position operator, $\sigma(\bm{r})$ (bohr$^2$). Missing values indicate that the corresponding calculation did not converge.}
\label{tab:si_methanol_mixed_spin_states_dm_var}

\begin{tabular}{ll cccc cccc cccc}
\hline
& &
\multicolumn{4}{c}{\begin{tabular}{c} M$_0$ \\ 2p$_y \rightarrow$ 3s \end{tabular}}
&
\multicolumn{4}{c}{\begin{tabular}{c} M$_1$ \\ 2p$_y \rightarrow$ 3p$_z$ \end{tabular}}
&
\multicolumn{4}{c}{\begin{tabular}{c} M$_2$ \\ 2p$_y \rightarrow$ 3p$_x$ \end{tabular}} \\
\cline{3-6}
\cline{7-10}
\cline{11-14}
Basis set & XC
& $\Delta E$ & $\mu_x$ & $\mu_z$ & $\sigma(\bm{r})$
& $\Delta E$ & $\mu_x$ & $\mu_z$ & $\sigma(\bm{r})$
& $\Delta E$ & $\mu_x$ & $\mu_z$ & $\sigma(\bm{r})$ \\
\hline
aug-cc-pVDZ+sz   & PBE       & 6.25 &  2.90 &  3.63 & 43.8 & 7.31 & -2.08 & -6.89 & 54.8 & 7.86 & -0.32 & -2.43 & 66.8 \\
d-aug-cc-pVDZ+sz & PBE       & 6.24 &  3.05 &  3.58 & 44.7 & 7.29 & -1.92 & -7.11 & 58.1 & 7.77 & -2.14 &  0.07 & 75.8 \\
plane waves      & PBE       & 6.23 &  3.04 &  3.55 & 44.6 & 7.29 & -1.93 & -7.04 & 58.2 & 7.76 & -1.82 &  0.23 & 75.3 \\
plane waves      & PBE0      & 6.42 &  3.48 &  3.65 & 42.8 & 7.56 & -2.14 & -7.73 & 58.7 & 8.01 & -2.26 &  0.54 & 78.0 \\
plane waves      & PBE-SIC/2 & 6.57 &  3.67 &  3.72 & 42.7 & 7.83 & -2.56 & -7.36 & 59.6 & 8.20 & -2.57 &  0.89 & 80.7 \\
plane waves      & PBE-SIC   & 6.75 &  4.20 &  3.93 & 41.4 & 8.33 & -2.48 & -7.25 & 62.6 & 8.41 & -2.68 &  1.39 & 86.8 \\
\hline

\\[-1.4ex]

\hline
& &
\multicolumn{4}{c}{\begin{tabular}{c} M$_3$ \\ 2p$_y \rightarrow$ 3p$_y$ \end{tabular}}
&
\multicolumn{4}{c}{\begin{tabular}{c} M$_4$ \\ $2\mathrm{p}_x/\sigma \rightarrow$ 3s \end{tabular}}
&
\multicolumn{4}{c}{} \\
\cline{3-6}
\cline{7-10}
Basis set & XC
& $\Delta E$ & $\mu_x$ & $\mu_z$ & $\sigma(\bm{r})$
& $\Delta E$ & $\mu_x$ & $\mu_z$ & $\sigma(\bm{r})$
& \multicolumn{4}{c}{} \\
\cline{1-10}
aug-cc-pVDZ+sz   & PBE       & 8.00 & -3.93 & -3.54 & 68.8 & 7.79 &  2.84 &  3.22 & 45.2 & \multicolumn{4}{c}{} \\
d-aug-cc-pVDZ+sz & PBE       & 7.85 & -0.32 & -0.48 & 81.8 & 7.78 &  2.99 &  3.16 & 46.0 & \multicolumn{4}{c}{} \\
plane waves      & PBE       & 7.84 &  0.04 & -0.23 & 81.5 & 7.74 &  2.99 &  3.16 & 45.8 & \multicolumn{4}{c}{} \\
plane waves      & PBE0      & 8.06 &  0.54 &  0.04 & 83.3 & 8.05 &  3.43 &  3.34 & 44.7 & \multicolumn{4}{c}{} \\
plane waves      & PBE-SIC/2 & 8.25 &  0.09 &  0.38 & 84.6 & 8.20 &  3.66 &  3.46 & 44.9 & \multicolumn{4}{c}{} \\
plane waves      & PBE-SIC   & 8.45 &  0.86 &  0.42 & 90.9 & 8.43 &  4.26 &  3.77 & 43.6 & \multicolumn{4}{c}{} \\
\cline{1-10}
\end{tabular}
\end{sidewaystable}

\begin{sidewaystable}[hbt]
\centering
\caption{Excitation energy, permanent dipole moment, and variance of the triplet states of methanol computed using different basis set representations and exchange--correlation functionals: excitation energy, $\Delta E$ (eV), permanent dipole-moment components $\mu_x$ and $\mu_z$ (Debye), and variance of the position operator, $\sigma(\bm{r})$ (bohr$^2$). Missing values indicate that the corresponding calculation did not converge.}
\label{tab:si_methanol_triplet_states_dm_var}

\begin{tabular}{ll cccc cccc cccc}
\hline
& &
\multicolumn{4}{c}{\begin{tabular}{c} T$_0$ \\ 2p$_y \rightarrow$ 3s \end{tabular}}
&
\multicolumn{4}{c}{\begin{tabular}{c} T$_1$ \\ 2p$_y \rightarrow$ 3p$_z$ \end{tabular}}
&
\multicolumn{4}{c}{\begin{tabular}{c} T$_2$ \\ 2p$_y \rightarrow$ 3p$_x$ \end{tabular}} \\
\cline{3-6}
\cline{7-10}
\cline{11-14}
Basis set & XC
& $\Delta E$ & $\mu_x$ & $\mu_z$ & $\sigma(\bm{r})$
& $\Delta E$ & $\mu_x$ & $\mu_z$ & $\sigma(\bm{r})$
& $\Delta E$ & $\mu_x$ & $\mu_z$ & $\sigma(\bm{r})$ \\
\hline
aug-cc-pVDZ+sz   & PBE       & 6.14 &  2.69 &  3.64 & 43.0 & 7.22 & -2.03 & -6.73 & 53.5 & 7.82 & -0.22 & -2.29 & 65.9 \\
d-aug-cc-pVDZ+sz & PBE       & 6.14 &  2.83 &  3.60 & 43.7 & 7.20 & -1.87 & -6.90 & 56.6 & 7.80 & -1.75 &  0.10 & 74.4 \\
plane waves      & PBE       & 6.12 &  2.83 &  3.57 & 43.6 & 7.20 & -1.88 & -6.83 & 56.6 & 7.71 & -1.44 &  0.26 & 73.9 \\
plane waves      & PBE0      & 6.31 &  3.23 &  3.67 & 41.7 & 7.48 & -2.00 & -7.59 & 56.9 & 7.99 & -2.11 &  0.55 & 76.7 \\
plane waves      & PBE-SIC/2 & 6.46 &  3.49 &  3.76 & 41.6 & 7.75 & -2.33 & -7.36 & 57.6 & 8.16 & -2.60 &  0.91 & 79.4 \\
plane waves      & PBE-SIC   & 6.64 &  3.96 &  3.96 & 40.5 & 8.24 & -2.27 & -7.64 & 59.1 &       &       &       &      \\
\hline

\\[-1.4ex]

\hline
& &
\multicolumn{4}{c}{\begin{tabular}{c} T$_3$ \\ 2p$_y \rightarrow$ 3p$_y$ \end{tabular}}
&
\multicolumn{4}{c}{\begin{tabular}{c} T$_4$ \\ $2\mathrm{p}_x/\sigma \rightarrow$ 3s \end{tabular}}
&
\multicolumn{4}{c}{} \\
\cline{3-6}
\cline{7-10}
Basis set & XC
& $\Delta E$ & $\mu_x$ & $\mu_z$ & $\sigma(\bm{r})$
& $\Delta E$ & $\mu_x$ & $\mu_z$ & $\sigma(\bm{r})$
& \multicolumn{4}{c}{} \\
\cline{1-10}
aug-cc-pVDZ+sz   & PBE       & 7.93 & -3.96 & -3.40 & 67.5 & 7.63 & 2.78 & 3.15 & 43.9 & \multicolumn{4}{c}{} \\
d-aug-cc-pVDZ+sz & PBE       & 7.80 & -0.58 & -0.46 & 80.1 & 7.62 & 2.92 & 3.09 & 44.5 & \multicolumn{4}{c}{} \\
plane waves      & PBE       & 7.79 & -0.26 & -0.24 & 79.9 & 7.58 & 2.93 & 3.09 & 44.3 & \multicolumn{4}{c}{} \\
plane waves      & PBE0      &      &       &       &      & 7.90 & 3.37 & 3.27 & 42.9 & \multicolumn{4}{c}{} \\
plane waves      & PBE-SIC/2 & 8.20 &  0.12 &  0.68 & 82.3 & 8.06 & 3.63 & 3.44 & 43.1 & \multicolumn{4}{c}{} \\
plane waves      & PBE-SIC   & 8.40 &  1.04 &  0.92 & 87.7 & 8.29 & 4.16 & 3.71 & 41.7 & \multicolumn{4}{c}{} \\
\cline{1-10}
\end{tabular}
\end{sidewaystable}

\clearpage
\newpage

\subsection{Bar plots of excitation energy and variance}
\begin{figure}[hbt]
    \centering
    \includegraphics[width=0.95\linewidth]{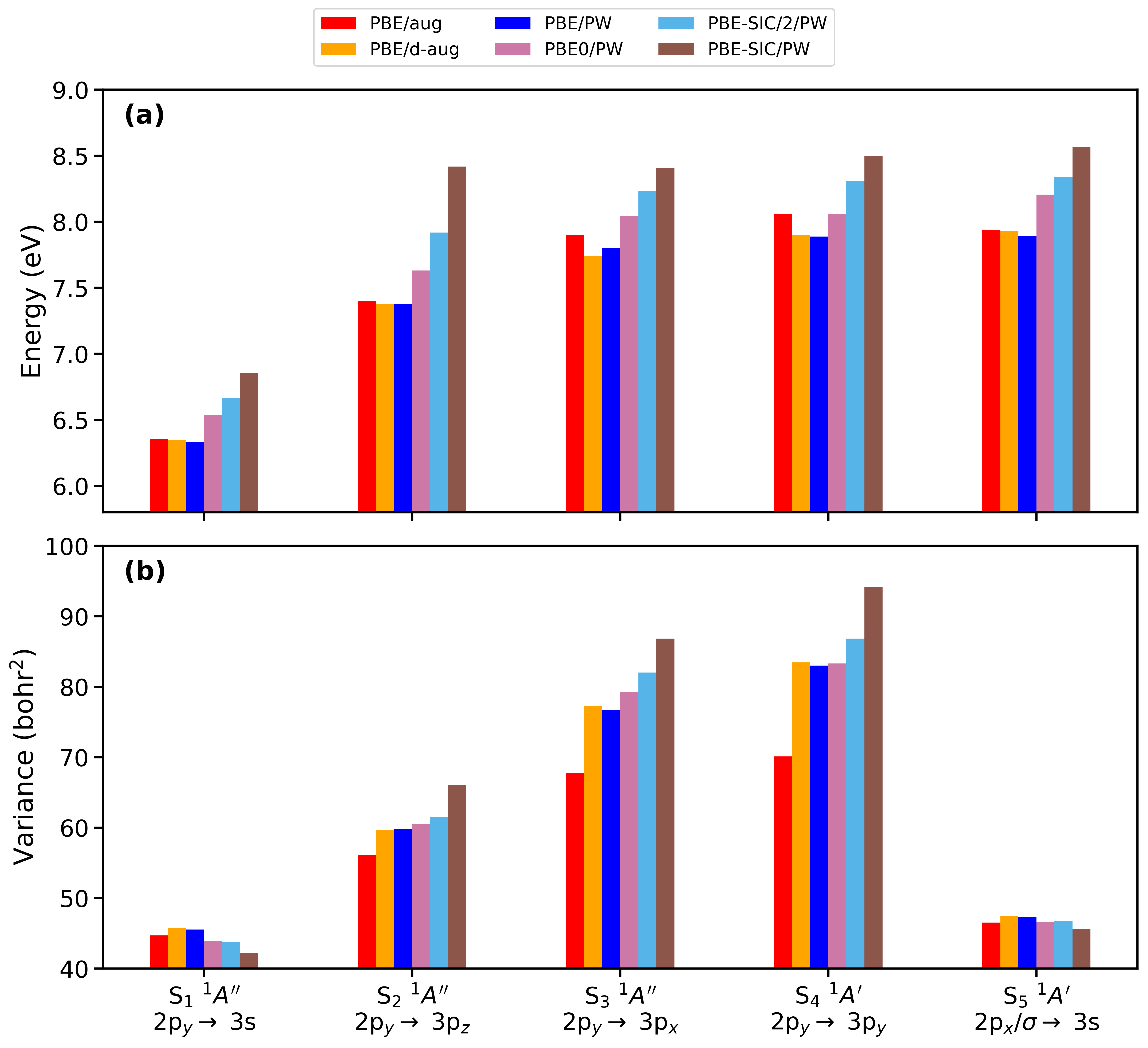}
    \caption{Basis set and exchange-correlation functional dependence of (a) the vertical excitation energy, and (b) the variance of the electronic position operator for the first five excited states of methanol. For PBE0 (state S$_4$) and PBE-SIC (state S$_3$), the spin-purification formula could not be applied, because the triplet calculations could not be converged to a solution with the correct spatial symmetry. Therefore, we instead report the mixed-spin values.}
    \label{ammonia_basis_effect_E}
\end{figure}

\clearpage
\newpage

\section{Additional plots}
\begin{figure}[hbt]
    \centering
    \includegraphics[width=\linewidth]{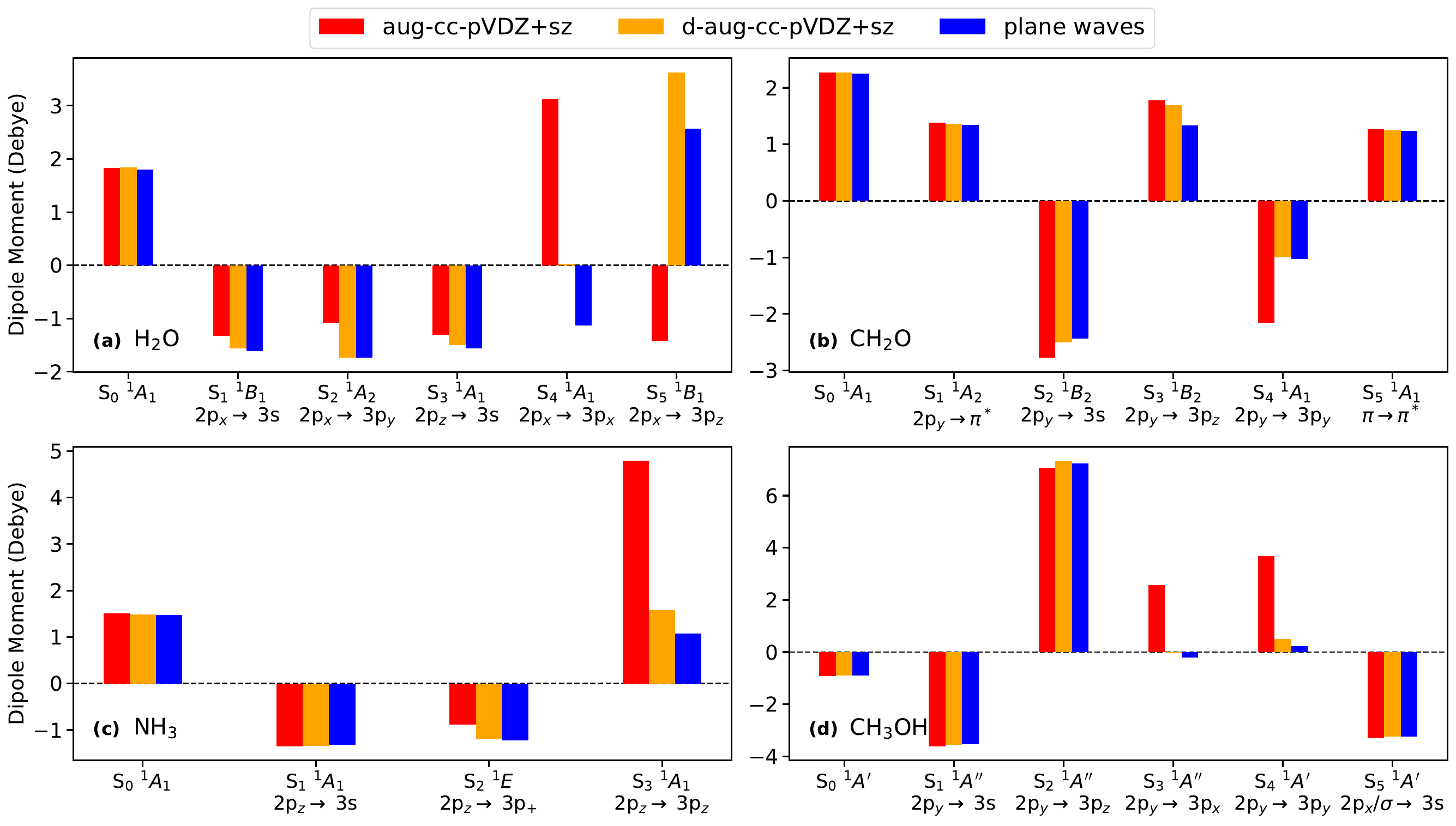}
    \caption{Effect of the basis representation on the dipole moments of water (a), formaldehyde (b), ammonia (c), and methanol (d). Dipole moments are computed at the \gls{oo}/PBE level of theory. The dominant character of each electronic transition is indicated below the corresponding state label. For water, formaldehyde, and ammonia, the reported value is the single nonzero component of the dipole-moment vector along the molecular axis of rotation. For methanol (d), which has two nonzero components, the reported value is along the z-direction of the molecular coordinate frame.}
    \label{fig:DM_basis_effect}
\end{figure}

\begin{figure}[hbt]
    \centering
    \includegraphics[width=\linewidth]{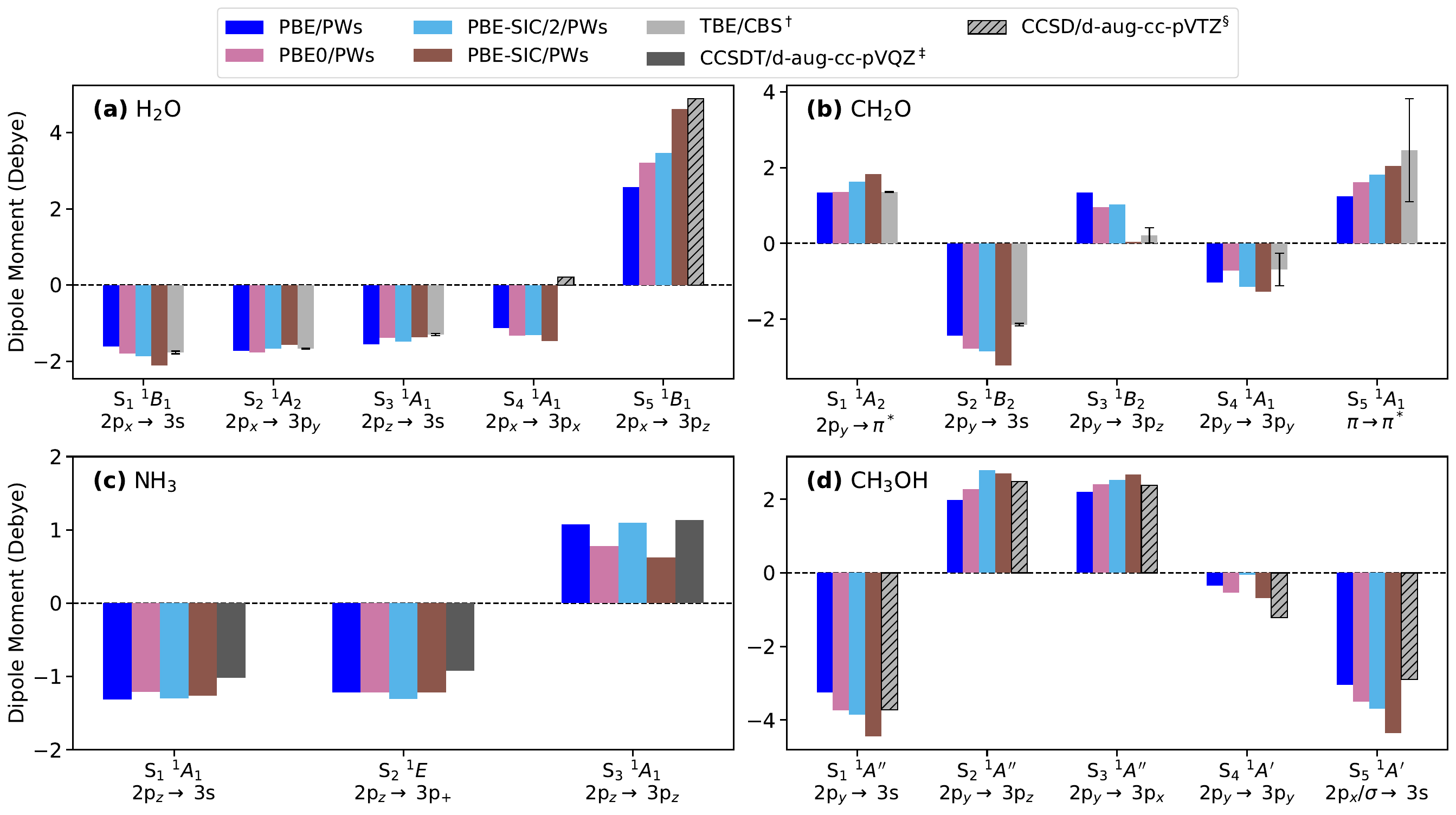}
    \caption{Spin-purified dipole moments of the singlet excited states of water (a), formaldehyde (b), ammonia (c), and methanol (d) obtained from orbital-optimized (OO) calculations with various exchange-correlation functionals and a plane-wave basis set. For methanol, which has two nonzero components, the reported value is along the x-direction of the molecular coordinate frame. Reference values are shown in gray: $^\dagger$Theoretical best estimate (TBE)/complete basis set (CBS) values from ref.~\citen{Chrayteh2021-ib} obtained from high-order coupled cluster calculations and extrapolation to the complete basis set limit, with the corresponding CBS uncertainty shown as error bars; $^\ddagger$CCSDT/\mbox{d-aug-cc-pVQZ} values from ref.~\citen{perscorr_jacquemin}; $^\S$CCSD/\mbox{d-aug-cc-pVTZ} values from this work. }
    \label{fig:DM_basis_effect}
\end{figure}

\clearpage
\newpage

\clearpage

\bibliography{biblio.bib}